\documentclass{article}
\usepackage{graphicx} 
\usepackage{amsmath}
\usepackage{geometry}
\usepackage{hyperref}
\usepackage{amssymb}
\usepackage{yhmath}
\usepackage{rotating}
\usepackage[affil-it]{authblk}

\title{Three-dimensional gravity-capillary standing waves: computation, resonance and instability}
\author{Xin Guan%
\thanks{\textit{E-mail address}: \texttt{x.guan@imperial.ac.uk}}}
\affil{Department of Mathematics, Imperial College London, United Kingdom}
\date{}

\begin{document}

\maketitle

\begin{abstract}
    \noindent We present a numerical study of three-dimensional gravity-capillary standing waves by using cubic and quintic truncated Hamiltonian formulations and the Craig–Sulem expansion of the Dirichlet–Neumann operator (DNO). The resulting models are treated as triply periodic boundary-value problems and solved via a spatio-temporal collocation method without executing initial-value calculations. This approach avoids the numerical stiffness associated with surface tension and numerical instabilities arising from time integration. We reduce the number of unknowns significantly by exploiting the spatio-temporal symmetries for three types of symmetric standing waves. Comparisons with existing asymptotic and numerical results illustrate excellent agreement between the models and the full potential-flow formulation. We investigate typical bifurcations and standing waves that feature square, hexagonal, and more complex flower-like patterns under the three-wave resonance. These solutions are generalisations of the classical Wilton ripples. Temporal simulations of the computed three-dimensional standing waves exhibit perfect periodicity and reveal an instability mechanism based on the reported oblique instability in two-dimensional standing waves\cite{zhu2003three}.
\end{abstract}

\section{Introduction}

Water waves are a ubiquitous natural phenomenon of both scientific and practical importance to mathematicians, physicists, and engineers. Due to their intrinsic nonlinearity, numerical simulations play an indispensable role in uncovering and understanding the underlying mechanisms. Although most real-world water waves are irregular and random, much insight can be gained by studying some simple and deterministic solutions such as travelling waves and standing waves, which serve as fundamental building blocks for more complex phenomena. 

Two-dimensional standing water waves have been extensively investigated through rigorous analysis, weakly nonlinear theory, fully nonlinear simulations, and laboratory experiments. Plotnikov \& Toland \cite{plotnikov2001nash} and Iooss \textit{et al.} \cite{iooss2005standing} proved the existence of small-amplitude gravity standing waves in finite and infinite depth, respectively. The Stokes-type expansions of gravity standing waves were derived by Rayleigh \cite{Rayleigh1915deep} and Penney \& Price \cite{penney1952some} for deep water, by Tadjbaksh \& Keller \cite{tadjbakhsh1960standing} for finite-depth water, and later extended to include surface tension by Concus \cite{concus1962standing} and Vanden-Broeck \cite{vanden1984nonlinear}. In recent years, direct numerical computations on two-dimensional standing waves have received more attention in studying the branching phenomenon \cite{okamura1985instability,mercer1994form,smith1999branching,wilkening2012overdetermined,shelton2023structure}, the mechanisms of instability \cite{mercer1992standing,bryant1994different,wilkening2019harmonic}, and the long-standing conjecture on the limiting configuration of gravity standing waves \cite{mercer1992standing,schultz1998highly,okamura1998enclosed,okamura2003standing,okamura2010almost,wilkening2011breakdown,wilkening2012overdetermined}. Three principal numerical approaches have been established: (i) spatio-temporal collocation methods based on the Fourier and other spectral expansions \cite{vanden1981numerical,tsai1987numerical,tsai1994numerical,bryant1994different,bryant1995water,okamura2003standing}, (ii) semi-analytic series-expansion methods \cite{schwartz1981semi,rottman1982steep,abassi2025semi}, and (iii) time-dependent methods (usually via the boundary-integral formulation) combined with Newton or nonlinear least-squares solvers \cite{mercer1992standing,schultz1998highly,wilkening2011breakdown,wilkening2012overdetermined,shelton2023structure}. The last approach is most commonly used owing to its relatively low computational cost, although the time integration may suffer from numerical instabilities, aliasing errors, and filtering errors. In contrast, the spatio-temporal collocation methods avoid these issues and guarantee exact temporal periodicity owing to the Fourier expansion in time. When the system is stiff or chaotic, this approach is particularly attractive, despite the heavy computational cost in solving large-scale algebraic systems. Such an issue had limited the application of spatio-temporal collocation methods in early years, but is largely alleviated over the past decades with a rapid growth of computational power.

In contrast, three-dimensional standing waves have received considerably less attention, and the existing works were primarily concerned with gravity waves based on weakly nonlinear theory in studying the Stokes-type expansion \cite{verma1962three}, bifurcation \cite{bridges1987secondary}, and linear stability \cite{okamura1985instability,bryant1995water}. Direct computation of nonlinear three-dimensional standing waves remains highly challenging, despite the recent progress \cite{tsai1987numerical,bryant1995water,zhu2003three,rycroft2013computation}. Of particular interest, Bryant \& Stiassnie \cite{bryant1995water} and Zhu \textit{et al.} \cite{zhu2003three} showed that two-dimensional standing waves may be unstable to transverse and oblique disturbances whose frequencies are close to that of the base standing waves. Note that, in the linear sense, a three-dimensional standing wave can be decomposed into two obliquely oriented standing waves
\begin{align}
    \cos(kx)\cos(ly)\cos(\omega t) = \frac{1}{2}\cos(kx+ly)\cos(\omega t) + \frac{1}{2}\cos(kx-ly)\cos(\omega t),\label{decom1}
\end{align}
where $k$ and $l$ denote the wavenumbers in the $x$- and $y$-directions, and $\omega$ is the nonlinear frequency. If one of the two components is slightly disturbed, then the instability found by Zhu \textit{et al.} suggests that energy could flow between these two components cyclically, leading to a recurrence phenomenon over long temporal periods. In this paper, we present some numerical evidence supporting this assertion. On the other hand, all these mentioned works neglected surface tension, which is known to give rise to various non-trivial effects to water waves \cite{perlin2000capillary} by modifying the dispersion relation. In particular, it changes the dominant resonance mechanism from four-wave to three-wave interactions, therefore accelerating energy cascade rate. Such resonances lead to the classical Wilton ripples, travelling gravity-capillary waves whose fundamental and second-order spatial harmonics have comparable amplitudes and propagate at the same speed. For standing waves, the effect of surface tension has been investigated only in two dimensions \cite{vanden1984nonlinear,schultz1998highly,wilkening2012overdetermined,shelton2023structure}, thereby motivating the present work to study more complex, non-collinear resonant modes in three dimensions. 

A central difficulty in simulating water waves lies within resolving the non-local relationship between the normal and tangential velocities on the free surface. This is commonly achieved by solving Laplace's equation subject to suitable boundary conditions. However, directly solving Laplace's equation in a three-dimensional fluid domain is computationally prohibitive \cite{rycroft2013computation}. In contrast, the boundary-integral method is widely employed owing to its ability to reduce the dimensionality of the problem. Despite its superior numerical efficiency, the three-dimensional boundary-integral formulation presents additional difficulty in handling the slowly converging lattice sums due to the absence of a closed-form integral kernel. To address these difficulties, we adopt a third approach based on Zakharov's Hamiltonian formulation of water waves \cite{zakharov1968stability} and the Dirichlet-Neumann operator (DNO). This method is computationally efficient because it provides explicit and systematic approximations to the non-local relationship of the full potential-flow problem. Therefore, it is widely used in water-wave problems \cite{craig1993numerical,craig2000traveling,craig2002traveling,nicholls1998traveling,xu2009numerical,wang2012dynamics,wang2014computation,wilkening2021quasi,guan2022interfacial,xu2025dynamics}. The key steps of this method are to truncate the DNO in the Hamiltonian of the system and then take variational derivatives with respect to the surface elevation $\eta$ and the surface velocity potential $\varphi$, which lead to reduced models of the full water-wave problem. Alternatively, one can first take the variational derivatives of the exact Hamiltonian, which gives rise to the kinematic and dynamic boundary conditions, and then truncate the DNO. Note that even when truncated to the same order, the two approaches usually give different high-order terms in their resulting models. In the present paper, we follow Wang \& Milewski \cite{wang2012dynamics} and Wang \textit{et al.} \cite{wang2014computation} who employed the third- and fifth-order truncations through the first approach to study gravity-capillary and flexural-gravity solitary waves, respectively. A novelty of their models is the fully nonlinear form of the surface-tension or flexural-tension terms, which provide considerably better agreement to the full potential-flow formulation. Because of the numerical stiffness introduced by surface tension, initial-value calculations for three-dimensional gravity-capillary standing waves are computationally intensive. Therefore, we develop a boundary-value algorithm based on the spatio-temporal collocation methods that were previously used in two dimensions. By exploiting a large amount of spatio-temporal symmetries, we reduce the number of unknowns by a factor of 64 and then solve the problem in physical space via Newton's method. Nevertheless, the numerical computation is still intensive and is therefore performed on a 128-core compute node equipped with AMD EPYC 7742 processors. Typically, executing one Newton iteration takes $1.5$ hours and requires $32$ GB memory for constructing Jacobian matrix on $128\times128\times128$ grid points. For most computations presented in this paper, this resolution is good enough to give highly accurate solutions.

The remainder of the paper is organised as follows. {\S} 2 gives the full potential-flow and Hamiltonian formulations for three-dimensional deep-water gravity-capillary waves. {\S} 3 details the numerical method based on a triply boundary-value calculation. {\S} 4 validates the numerical accuracy and exhibits typical standing-wave solutions and bifurcations under resonances. {\S} 5 exhibits temporal simulations for the computed standing waves and examine their stability. Finally, {\S} 6 provides concluding remarks.

\section{Formulations of deep-water gravity-capillary waves}
We consider the mathematical formulation of deep-water gravity-capillary waves, which requires solving the velocity potential $\phi(x,y,z,t)$ and surface elevation $\eta(x,y,t)$ from
\begin{align}
    &\Delta\phi+\phi_{zz} = 0,\quad\qquad\qquad\qquad\qquad\qquad\qquad\qquad\qquad\qquad-\infty<z<\eta(x,y,t),\\
    &\eta_t + \nabla \phi \cdot \nabla \eta = \phi_z,\qquad\qquad\qquad\qquad\qquad\qquad\qquad\quad\qquad \text{at }\,\,z = \eta(x,y,t),\\
    &\phi_t + \frac{1}{2}\Big(|\nabla \phi|^2 + \phi_z^2\Big) + \eta = \nabla\cdot \Bigg(\frac{\nabla\eta}{\sqrt{1+|\nabla\eta|^2}}\Bigg),\,\,\,\qquad\qquad \text{at }\,\,z = \eta(x,y,t),\\
    &\phi_x,\phi_y,\phi_z \rightarrow 0, \qquad\qquad\qquad\qquad\,\,\,\,\,\qquad\qquad\qquad\qquad\qquad \text{as }\,\, z\rightarrow -\infty,
\end{align}
where $x,y$ are the horizontal Cartesian coordinates, $z$ is the vertical one pointing opposite to the direction of gravity, $t$ denotes time, $\nabla = (\partial_x,\partial_y)$, and $\Delta = \partial_{xx}+\partial_{yy}$. For convenience, we put the $(x,y)$-plane on the mean water level. Without loss of generality, we have non-dimensionalised the system by choosing the typical scalings of length, time, and mass such that the gravitational acceleration, fluid density and surface tension coefficient become unit. For real water whose surface tension coefficient equals to $72$ dyn/cm, $2\pi$ spatial units and one temporal unit correspond to $1.7$cm and $0.0167$ seconds physically.

In 1968, Zakharov \cite{zakharov1968stability} proved that the above formulation is equivalent to Hamilton's canonical equations
\begin{align}
    \eta_t = \frac{\delta \mathcal H}{\delta \varphi},\quad \varphi_t = -\frac{\delta \mathcal H}{\delta \eta},\label{Hamilton}
\end{align}
where $\varphi(x,y,t) = \phi(x,y,\eta(x,y,t),t)$, and the Hamiltonain $\mathcal H$ is the total energy of the system
\begin{align}
    \mathcal H = \iint \bigg(\frac{1}{2}\varphi G(\eta) \varphi +\frac{1}{2}\eta^2  + \sqrt{1+|\nabla\eta|^2}-1\bigg) \,\mathrm dx\,\mathrm dy.\label{Hamiltonian}
\end{align}
Here $G(\eta)$ is the scaled Dirichlet-Neumann operator (DNO) and is formally defined by
\begin{align}
   G(\eta)\varphi = \phi_{\boldsymbol{n}}\sqrt{1+|\nabla\eta|^2},
\end{align}
where $\phi_{\boldsymbol{n}}$ is the normal derivative of $\phi$ in the outward direction to the surface. It is rigorously proved that the DNO has a convergent Taylor series 
\begin{align}
    G(\eta) = \sum_{i=0}^{\infty} G_i(\eta),
\end{align}
provided the norm of $\eta$ is small than a certain constant \cite{coifman1985nonlinear,craig2000traveling}. For the convenience of numerical computation, the expressions of $G_i$ are commonly written in a recursive form \cite{craig1993numerical,craig2002traveling} (see Appendix 1).
By truncating the DNO to a certain order $n$, one obtains an approximated Hamiltonian. Substituting it into \eqref{Hamilton} leads to a $n$-th order truncated model
\begin{align}
    \eta_t & = \sum_{i=0}^{n-1} G_i(\eta)\varphi,\label{eq1}\\
    \varphi_t &= \sum_{i=2}^{n}\mathcal N_i(\eta,\varphi) - \eta + \nabla\cdot \Bigg( \frac{\nabla\eta}{\sqrt{1+|\nabla\eta|^2}}\Bigg),\label{eq2}
\end{align}
where $\mathcal N_i$ represents the DNO-involved $i$-th order nonlinear term whose expression can be found in Appendix 1. Wang \& Milewski \cite{wang2012dynamics} and Wang \textit{et al.} \cite{wang2014computation} employed a cubic ($n=3$) and a quintic ($n=5$) truncations (denoted by \textit{\textbf{cubic model}} and \textit{\textbf{quintic model}} hereafter) to study gravity-capillary and flexural-gravity solitary waves, respectively. They showed that these models are quite accurate compared with the original water-wave formulation, even when solutions have large amplitudes. In the following sections, we investigate three-dimensional gravity-capillary standing waves in rectangular or square basins by searching for triply periodic (in the $x,y$ and $t$ directions) solutions of the cubic and quintic models.

\section{Spatio-temporal collocation method}
\subsection{Linear solutions}
By solving the linearised water-wave equations, one obtains the following linear standing-wave solution 
\begin{align}
    \widetilde{\eta}_{k,l}(x,y,t) = \cos(k x)\cos(l y)\cos(\omega t), \quad \widetilde{\varphi}_{k,l}(x,y,t) = -\frac{\omega}{\kappa} \cos(kx)\cos(l y)\sin(\omega t),\label{linear solution}
\end{align}
where $\kappa = \sqrt{k^2+l^2}$, and $\omega$ satisfies the dispersion relation for deep-water gravity-capillary waves
\begin{align}
    \omega = \sqrt{\kappa + \kappa^3}.\label{dispersion}
\end{align}
We use the following two assumptions when computing standing waves:
\begin{itemize}
    \item $\varphi = 0$ at $t = 0,T/2,T,3T/2\cdots$,
    \item $\eta$ and $\varphi$ are even functions of $x$ and $y$,
\end{itemize}
where $T$ is the smallest temporal period. The first assumption implies that the kinetic energy vanishes every half period. Although this also seems true for potential energy because $\widetilde{\eta}_{k,l}$ vanishes periodically in time, the potential energy of nonlinear standing waves never vanishes, as will be shown later. To construct standing waves in rectangular or square basins, we use the following three types of linear solutions
\begin{itemize}
    \item \textit{Case I} (rectangular basin): $\eta =  \epsilon\widetilde{\eta}_{k,l}, \varphi = \epsilon\widetilde{\varphi}_{k,l}$, $k,l\neq 0,k\neq l$
    \item \textit{Case II} (square basin): $\eta = \epsilon\widetilde{\eta}_{k,l} + \epsilon\widetilde{\eta}_{l,k},\varphi = \epsilon\widetilde{\varphi}_{k,l} + \epsilon\widetilde{\varphi}_{l,k}$, $l = rk, r = 1,3,5,\cdots$
    \item \textit{Case III} (square basin): $\eta = \epsilon\widetilde{\eta}_{k,l} + \epsilon\widetilde{\eta}_{l,k},\varphi = \epsilon\widetilde{\varphi}_{k,l} + \epsilon\widetilde{\varphi}_{l,k}$, $l = rk, r = 0,2,4,\cdots$
\end{itemize}
where $\epsilon$ is a small parameter. To support standing waves in square basins, it is required that $k$ and $l$ being rationally related, but we only study the case that $l$ is an integer multiple of $k$. In general, one can also consider the following form of linear solution 
\begin{align}
    \eta = \epsilon\widetilde{\eta}_{k,rk} - \epsilon\widetilde{\eta}_{rk,k},\varphi = \epsilon\widetilde{\varphi}_{k,rk} - \epsilon\widetilde{\varphi}_{rk,k}.\label{another initial guess}
\end{align}
For $r = 0,2,4,\cdots$ and $y = Y+\pi/k$, we have
\begin{align}
    \cos(kx)\cos(rk y)-\cos(rkx)\cos(k y) = \cos(kx)\cos(rk Y)+\cos(rkx)\cos(k Y),
\end{align}
thus \eqref{another initial guess} is equivalent to \textit{Case III} without specifying additional symmetries. On the other hand, for $r=1,3,5,\cdots$, we find no convergent solution using \eqref{another initial guess}. Note that
\begin{align}
    \widetilde{\eta}_{k,l}(x,y,t) = \frac{1}{2}\widetilde{\eta}_{\kappa,0}(\xi,\zeta,t) + \frac{1}{2}\widetilde{\eta}_{0,\kappa}(\xi,\zeta,t),\label{Case223}
\end{align}
where $\xi = (kx+ly)/\kappa$ and $\zeta = (kx-ly)/\kappa$. This means a three-dimensional standing wave can be decomposed into obliquely oriented two-dimensional standing waves, at least linearly. In particular, when $k=l$, a \textit{Case II} standing wave can be transformed to a \textit{Case III} standing wave by rotating the $(x,y)$-plane by $\pi/4$.

\subsection{Spatio-temporal symmetries}
We consider triply periodic solutions in a three-dimensional cube
\begin{align}
    \{(x,y,t)|-L_1\le x\le L_1,\, -L_2\le y\le L_2,\, 0\le t \le T\},
\end{align}
where $2L_1$ and $2L_2$ are the smallest periods in the $x$- and $y$-directions. Using the time-reversal argument, solutions must have a temporal symmetry about $t = T/2$
\begin{align}
    \eta(x,y,T/2-t') &= \eta(x,y,T/2+t'),\\
    \varphi(x,y,T/2-t') &= -\varphi(x,y,T/2+t'),
\end{align}
where $t'\in \mathbb R$. Since $\varphi$ vanishes at $t = T/2$, $\eta$ must returns to its initial state with a spatial shift (we do not consider other possibilities in this paper). Otherwise, $T/2$ becomes the smallest temporal period. To ensure the consistency with linear solutions, there exist two types of shift 
\begin{itemize}
    \item For \textit{Case I} and \textit{II}
    \begin{align}
        \eta(x,y,0) = \eta(x\pm L_1,y,T/2) = \eta(x,y\pm L_2,T/2).\label{timesymmetry5}
    \end{align}
    \item For \textit{Case III}
    \begin{align}
        \eta(x,y,0) = \eta(x\pm L_1,y\pm L_2,T/2).
    \end{align}
\end{itemize}
Using the time-reversal argument again, solutions must be symmetric about $t = T/4$
\begin{itemize}
    \item For \textit{Case I} and \textit{II}
    \begin{align}
        \eta(x,y,T/4-t') &= \eta(x\pm L_1,y,T/4+t') = \eta(x,y\pm L_2,T/4+t'),\label{timesymmetry1}\\
        \varphi(x,y,T/4-t') &= -\varphi(x\pm L_1,y,T/4+t') = -\varphi(x,y\pm L_2,T/4+t').\label{timesymmetry2}
    \end{align}
    \item For \textit{Case III}
    \begin{align}
        \eta(x,y,T/4-t') &= \eta(x\pm L_1,y\pm L_2,T/4+t'),\label{timesymmetry3}\\
        \varphi(x,y,T/4-t') &= -\varphi(x\pm L_1,y\pm L_2,T/4+t').\label{timesymmetry4}
    \end{align}
\end{itemize}
Consequently, we only need to consider a quarter temporal period, say, $0\le t\le T/4$, in computation. 

\begin{figure}[h!]
    \centering
    \includegraphics[width=0.7\linewidth]{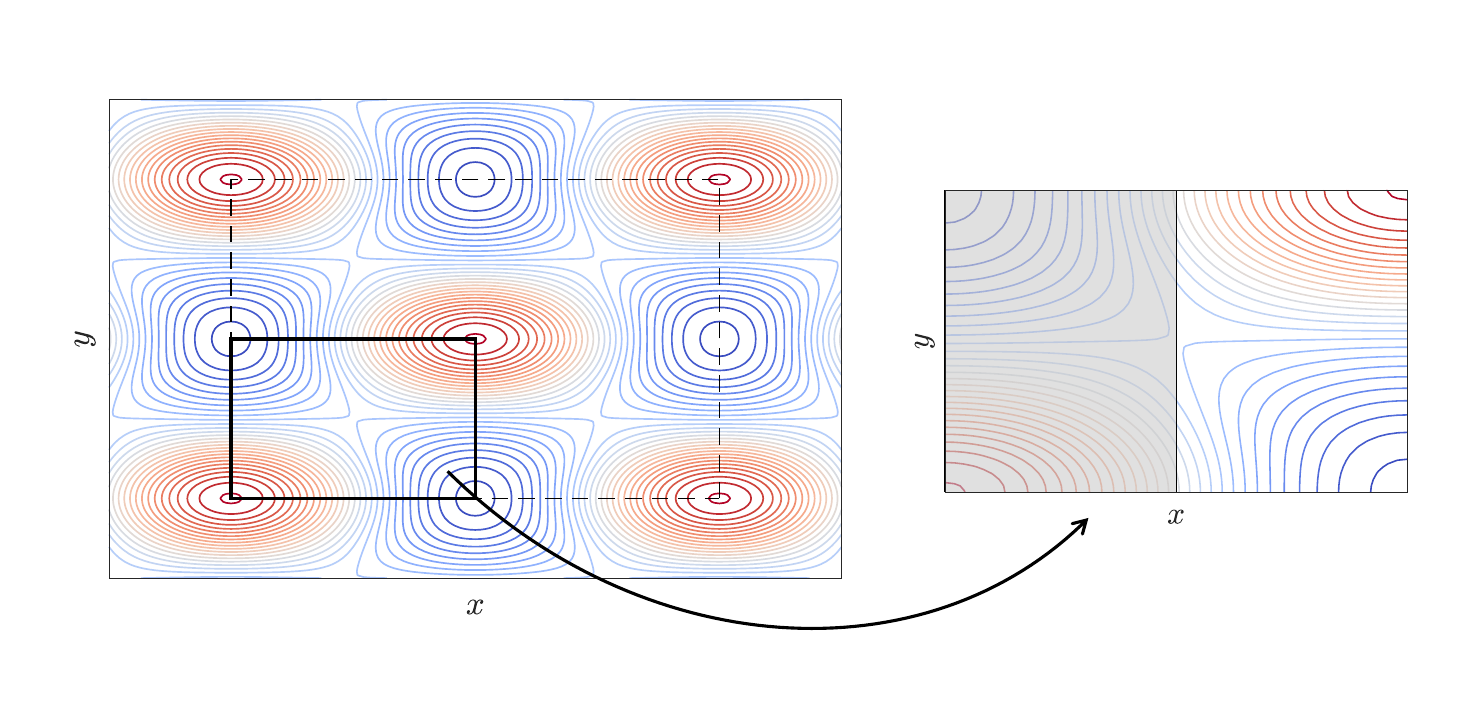}
    \caption{Typical contours of $\eta$ for a \textit{Case I} standing wave. The rectangles surround by the black dashed and solid lines represent a unit periodic cell and a quadrant on the $(x,y)$-plane, respectively. The shaded region denotes the real computational domain by using \eqref{symmetry_eta} and \eqref{symmetry_phi}.}
    \label{fig:sym}
\end{figure}

\begin{figure}[h!]
    \centering
    \includegraphics[width=0.7\linewidth]{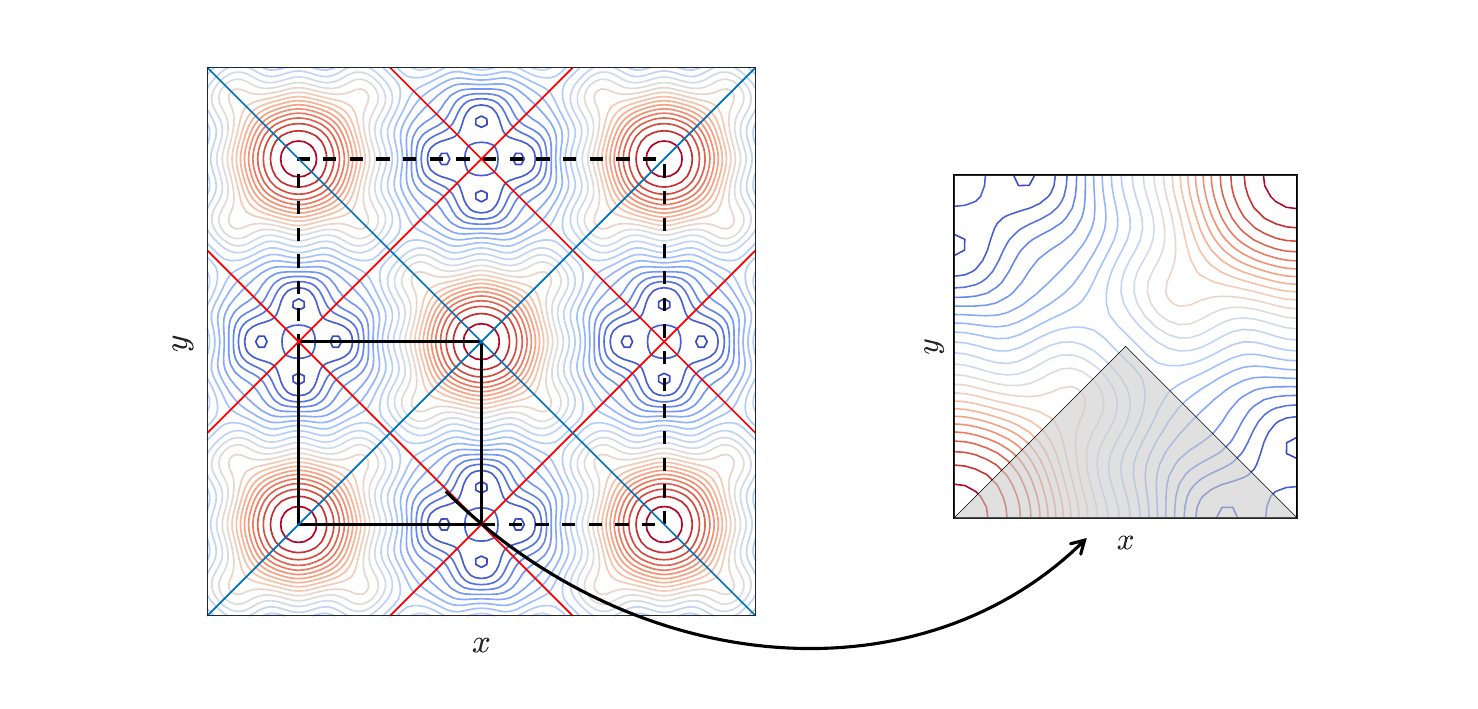}
    \caption{Typical contours of $\eta$ for a \textit{Case II} standing wave. The squares surround by the black dashed and solid lines represent a unit periodic cell and a quadrant on the $(x,y)$-plane, respectively. The blue lines stand for the two diagonals $x = \pm y$, and the four red lines denote $x+y = \pm L$ and $x-y = \pm L$. The shaded region denotes the real computational domain by using \eqref{symmetry_eta2} and \eqref{symmetry_phi2}.}
    \label{fig:sym2}
\end{figure}

\begin{figure}[h!]
    \centering
    \includegraphics[width=0.7\linewidth]{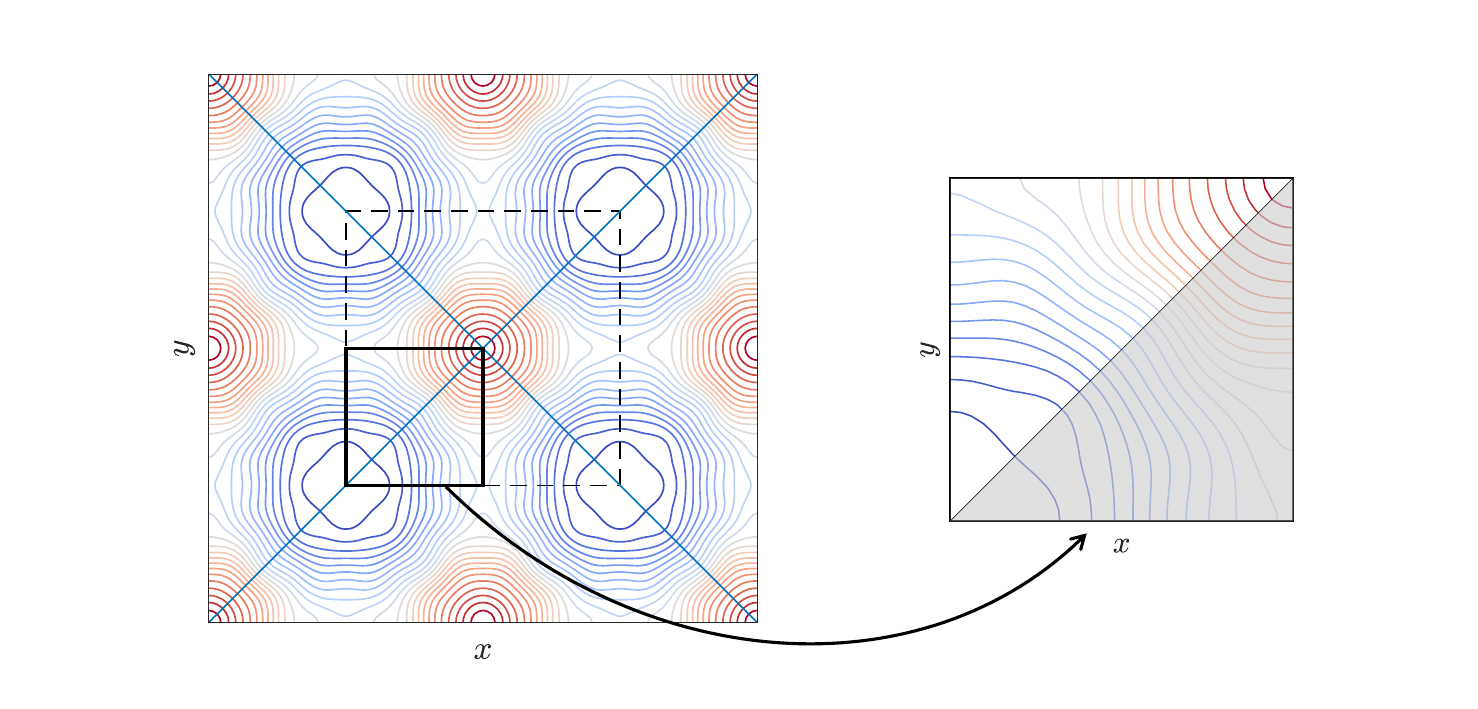}
    \caption{Typical contours of $\eta$ for a \textit{Case III} standing wave. The squares surround by the black dashed and solid lines represent a unit periodic cell and a quadrant on the $(x,y)$-plane, respectively. The blue lines stand for the two diagonals $x = \pm y$. The shaded region denotes the real computational domain by using \eqref{diagonal_eta} and \eqref{diagonal_phi}.}
    \label{fig:sym3}
\end{figure}

\begin{figure}[h!]
    \centering
    \includegraphics[width=0.9\linewidth]{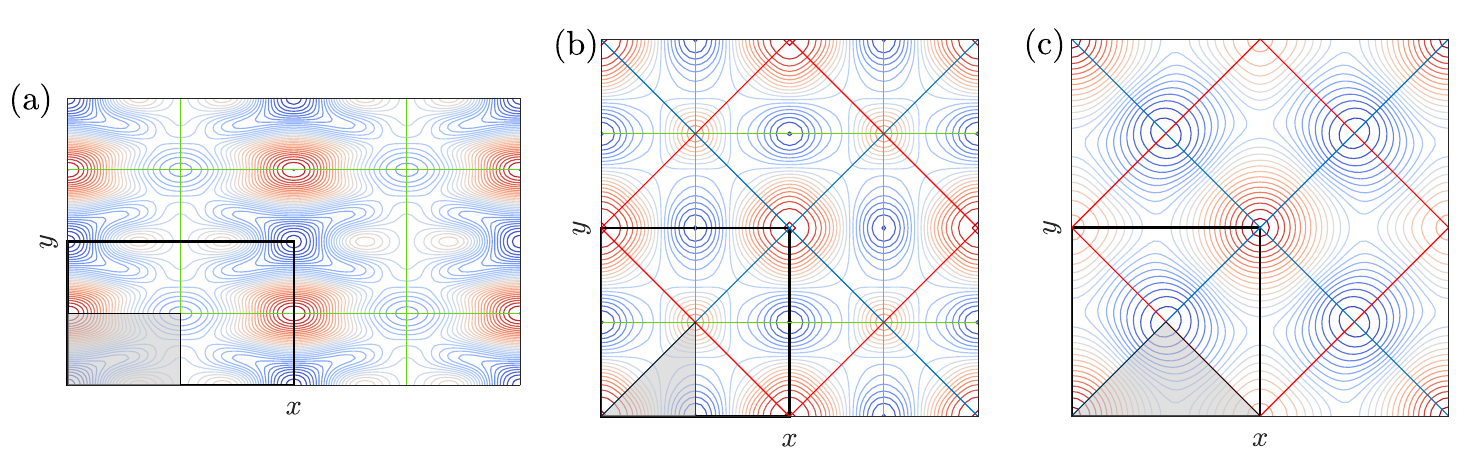}
    \caption{Typical contours of $\eta(x,y,T/4)$ for \textit{Case I} (a), \textit{Case II} (b) and \textit{Case III} (c) standing waves. The rectangle and squares surround by the solid lines represent a quadrant on the $(x,y)$-plane, respectively. The blue lines are the two diagonals $x = \pm y$. The red lines denote $x+y = \pm L$ and $x-y = \pm L$. The green lines stand for $x = \pm L_1/2$ and $y = \pm L_2/2$. The shaded regions denote the real computational domain.}
    \label{fig:sym4}
\end{figure}

Owing to the even symmetry with respect to $x$ and $y$ (denoted by \textit{\textbf{fourfold symmetry}} hereafter)
\begin{align}
    \eta(x,y,t) &= \eta(-x,y,t) = \eta(-x,-y,t) = \eta(x,-y,t),\\
    \varphi(x,y,t) &= \varphi(-x,y,t) = \varphi(-x,-y,t) = \varphi(x,-y,t),
\end{align}
we only need to consider the values of $\eta$ and $\varphi$ within one quadrant, say, $\{(x,y,t)|-L_1\le x\le0,-L_2\le y\le 0,0\le t\le T/4\}$. For \textit{Case I} and \textit{Case II}, \eqref{timesymmetry1}-\eqref{timesymmetry2} imply 
\begin{align}
    \eta(x,y,t) &= \eta(x\pm L_1,y\pm L_2,t),\label{symmetry_shift1}\\
    \varphi(x,y,t) &= \varphi(x\pm L_1,y\pm L_2,t).\label{symmetry_shift2}
\end{align}
Combining these with the fourfold symmetry, we have
\begin{align}
    \eta(-x,-y,t) &= \eta(x\pm L_1,y\pm L_2,t),\label{symmetry_eta}\\
    \varphi(-x,-y,t) &= \varphi(x\pm L_1,y\pm L_2,t).\label{symmetry_phi}
\end{align}
This property is used for \textit{Case I} standing waves to reduce the number of unknowns, as shown in figure \ref{fig:sym}.
For \textit{Case II} and \textit{Case III}, solution have a stronger \textbf{\textit{eightfold symmetry}}, \textit{i.e.} fourfold symmetry plus even symmetry about the two diagonals $x=\pm y$
\begin{align}
    \eta(x,y,t) &= \eta(y,x,t)=  \eta(-y,-x,t),\label{diagonal_eta}\\
    \varphi(x,y,t) &= \varphi(y,x,t)=  \varphi(-y,-x,t).\label{diagonal_phi}
\end{align}
Combining these with \eqref{symmetry_shift1} and \eqref{symmetry_shift2}, we have the following identities for \textit{Case II}
\begin{align}
    \eta(x,y,t) &= \eta(y\pm L,x\mp L,t) = \eta(-y\pm L,-x\mp L,t),\label{symmetry_eta2}\\
    \varphi(x,y,t) &= \varphi(y\pm L,x\mp L,t) = \varphi(-y\pm L,-x\mp L,t),\label{symmetry_phi2}
\end{align}
where $L = L_1 = L_2$. These mean that \textit{Case II} solutions are also symmetric about the four straight lines $x+y = \pm L$, $x-y = \pm L$. We use these symmetries to reduce the computational domain from a quadrant to one of its quarter sub-triangles, as shown in figure \ref{fig:sym2}.
For \textit{Case III}, however, solutions are only symmetric about the two diagonals, as shown in figure \ref{fig:sym3}. 
Moreover, there exist additional spatial symmetries for all three cases at $t=T/4$. Set $t' = 0$ in \eqref{timesymmetry1}-\eqref{timesymmetry4} and use the fourfold symmetry, we have
\begin{itemize}
    \item For \textit{Case I} and \textit{II}
    \begin{align}
        \eta(-x,y,T/4) &= \eta(x\pm L_1,y,T/4) = \eta(x,y\pm L_2,T/4) = \eta(x,-y,T/4),\\
        \varphi(-x,y,T/4) &= -\varphi(x\pm L_1,y,T/4) = -\varphi(x,y\pm L_2,T/4) = \varphi(x,-y,T/4).
    \end{align}
    \item For \textit{Case III}
    \begin{align}
        \eta(-x,-y,T/4) &= \eta(x\pm L_1,y\pm L_2,T/4),\\
        \varphi(-x,-y,T/4) &= -\varphi(x\pm L_1,y\pm L_2,T/4).
    \end{align}
\end{itemize}
These mean that, at $t = T/4$, \textit{Case I} and \textit{Case II} solutions are symmetric about $x = \pm L_1/2$ and $y = \pm L_2/2$ within each quadrant, and \textit{Case III} solutions are symmetric about $x+y = \pm L$ and $x-y = \pm L$, as shown in figure \ref{fig:sym4}.

\subsection{Numerical implementation}
We introduce $I+1,J+1$ and $N+1$ uniform collocation points for $x,y$ and $t$
\begin{align}
    x_i &= \frac{2(i-1)L_1}{I}-L_1,\,\quad i = 1,2,\cdots,I+1,\\
    y_j &= \frac{2(j-1)L_2}{J}-L_2,\quad j = 1,2,\cdots,J+1,\\
    t_n &= \frac{(n-1)T}{N},\qquad\qquad n = 1,2,\cdots,N+1,
\end{align}
and our task is to solve the function values of $\eta$ and $\varphi$ at these points, along with the frequency $\omega$, given a wave amplitude $H$ (this can be the wave amplitude in physical space, the amplitude of a fundamental Fourier mode, or a measure of energy). Based on the spatio-temporal symmetries, we have the following unknowns to solve for the \textit{Case I} standing waves
\begin{itemize}
    \item At $t=0$, $\eta(x_i,y_j,0)$ for $i=1,2\cdots,I/4+1$ and $j=1,2\cdots,J/2+1$;
    \item At $t = t_n$ ($n=2,3,\cdots N/4$), $\eta(x_i,y_j,t_n)$ and $\varphi(x_i,y_j,t_n)$ for $i=1,2\cdots,I/4+1$ and $j=1,2\cdots,J/2+1$; 
    \item At $t=T/4$, $\eta(x_i,y_j,T/4)$ for $i=1,2\cdots,I/4+1$ and $j=1,2\cdots,J/4+1$, $\varphi(x_i,y_j,T/4)$ for $i=1,2\cdots,I/4$ and $j=1,2\cdots,J/4$.
\end{itemize}
\begin{figure}[h!]
    \centering
    \includegraphics[width=0.8\linewidth]{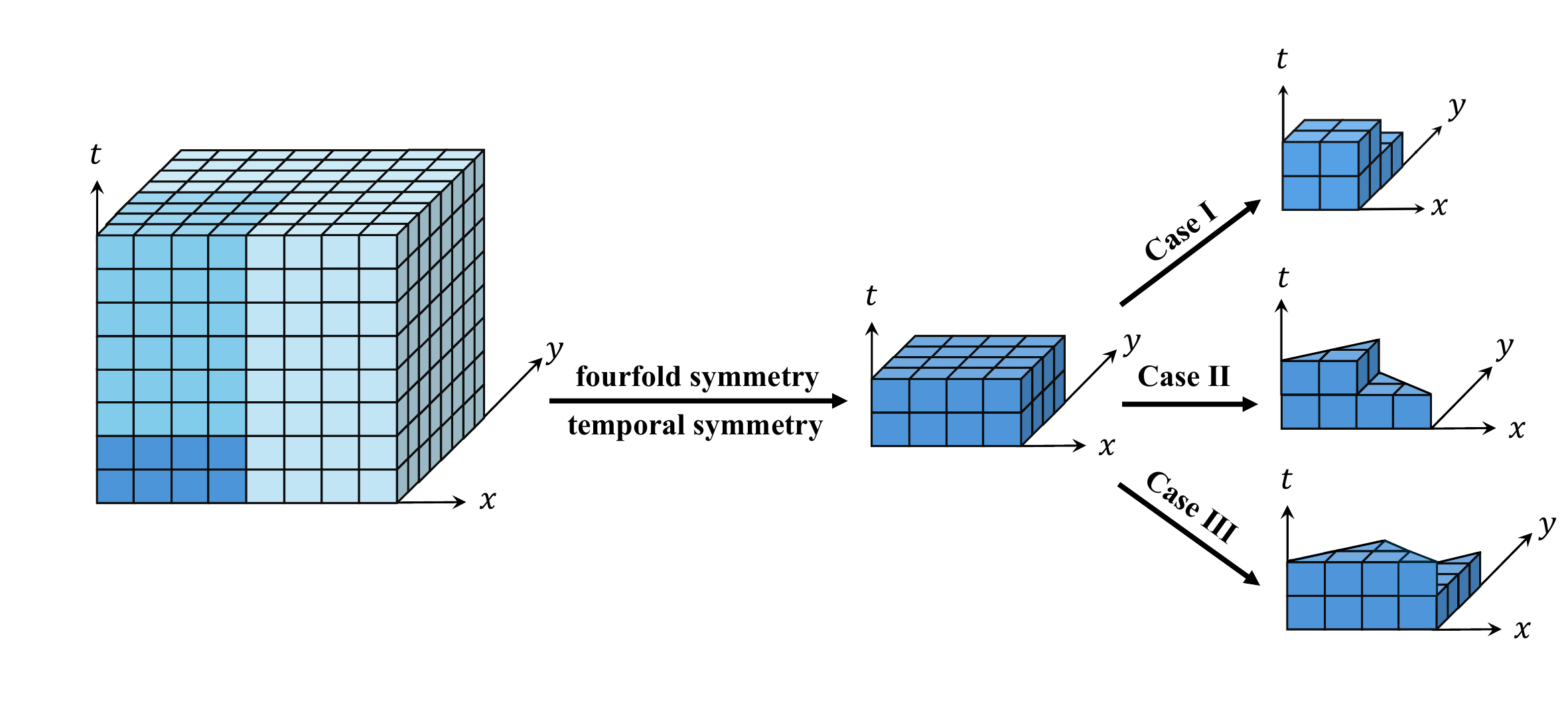}
    \caption{Schematic of the reduced computational domain by using spatio-temporal symmetries.}
    \label{fig:symmetry}
\end{figure}
All together, these give rise to $(I/4+1)(J/2+1)N/2 - J/4 = O(IJN/16)$ unknowns for $\eta$ and $\varphi$. For \textit{Case II} and \textit{Case III}, it is convenient to let $I=J$. The total number of unknowns for $\eta$ and $\varphi$ are reduced to $(I/4+1)^2N/2= O(I^2N/32)$ for \textit{Case II} and $(I/4+1)(I/2+1)N/2 = O(I^2N/16)$ for \textit{Case III}. A schematic is drawn in figure \ref{fig:symmetry} to show the reduced computational domain for all three cases. \eqref{eq1} and \eqref{eq2} give rise to the same number of equations after using the spatio-temporal symmetries. To close the system, we use the following wave-amplitude equation
\begin{align}
    \eta(0,0,0) = H,\label{height equation}
\end{align}
for given values of $H$. 

The system is solved via Newton's method by setting the iterative tolerance to $10^{-10}$. The actual residuals for convergent solutions are usually several orders smaller than this value. All derivatives and operations involving the DNOs are calculated via a pseudo-spectral method and FFT. Once we get a convergent solution, we use continuation method to search for other solutions along the solution branch.

\section{Numerical results}

\subsection{Comparisons with existing results}

We first compare our numerical solutions with some existing results. For two-dimensional gravity-capillary standing waves in finite water depth $h$, Concus \cite{concus1962standing} derived their third-order Stokes expansion and the nonlinear dependence of $\omega$ on $\epsilon$, the amplitude of the fundamental Fourier mode. For $k=1$, this relation reads (using our scalings)
\begin{align}
    \omega \sim \omega_0+\frac{1}{2}\epsilon^2\omega_2 + O(\epsilon^3),\label{concus}
\end{align}
where
\begin{align}
    \omega_0 =\sqrt{2\tanh(h)},\qquad \omega_2 = \frac{-2\omega_0^5-39\omega_0-66\omega_0^{-3}+2592\omega_0^{-7}}{320(1-6\omega_0^{-4})},
\end{align}
and $\epsilon$ represents the amplitude of $\cos(x)\cos(\omega t)$ mode.
\begin{figure}[h!]
    \centering
    \includegraphics[width=0.45\linewidth]{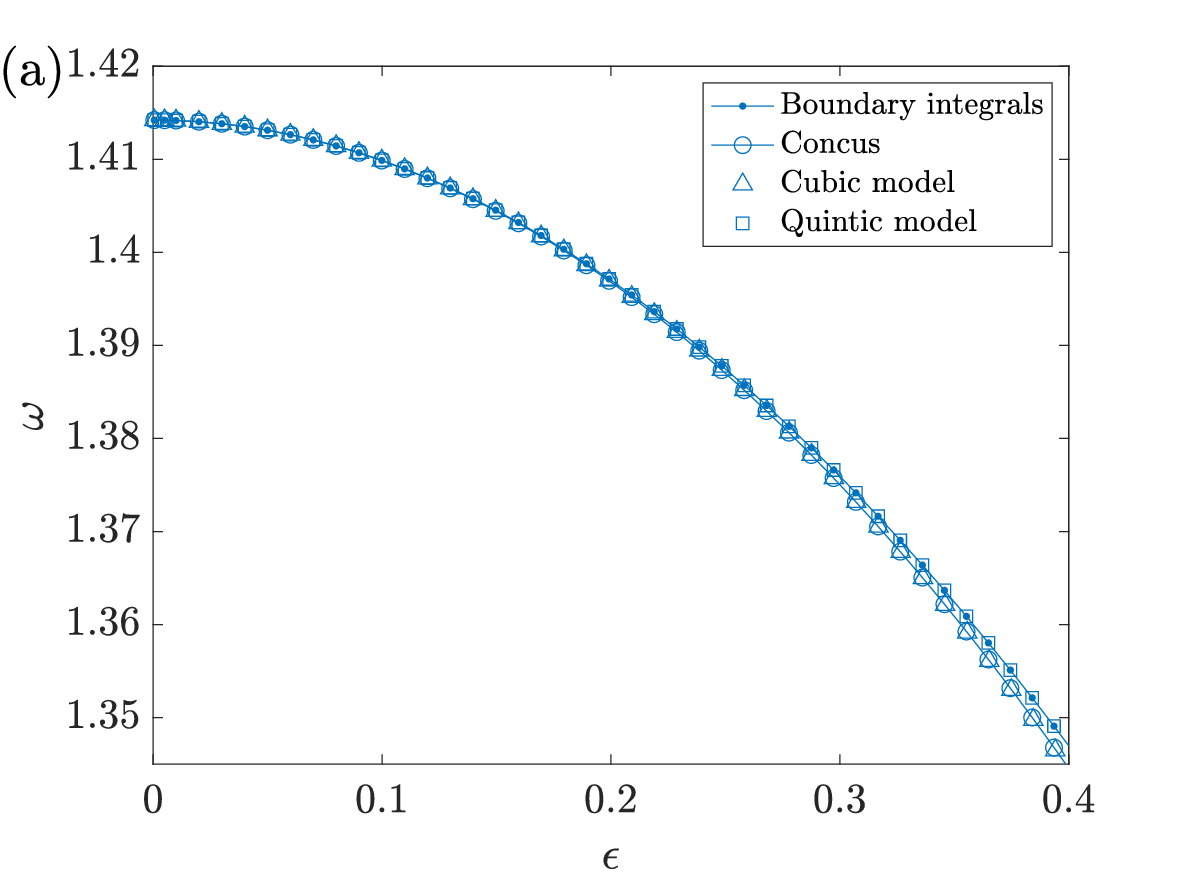}
    \includegraphics[width=0.45\linewidth]{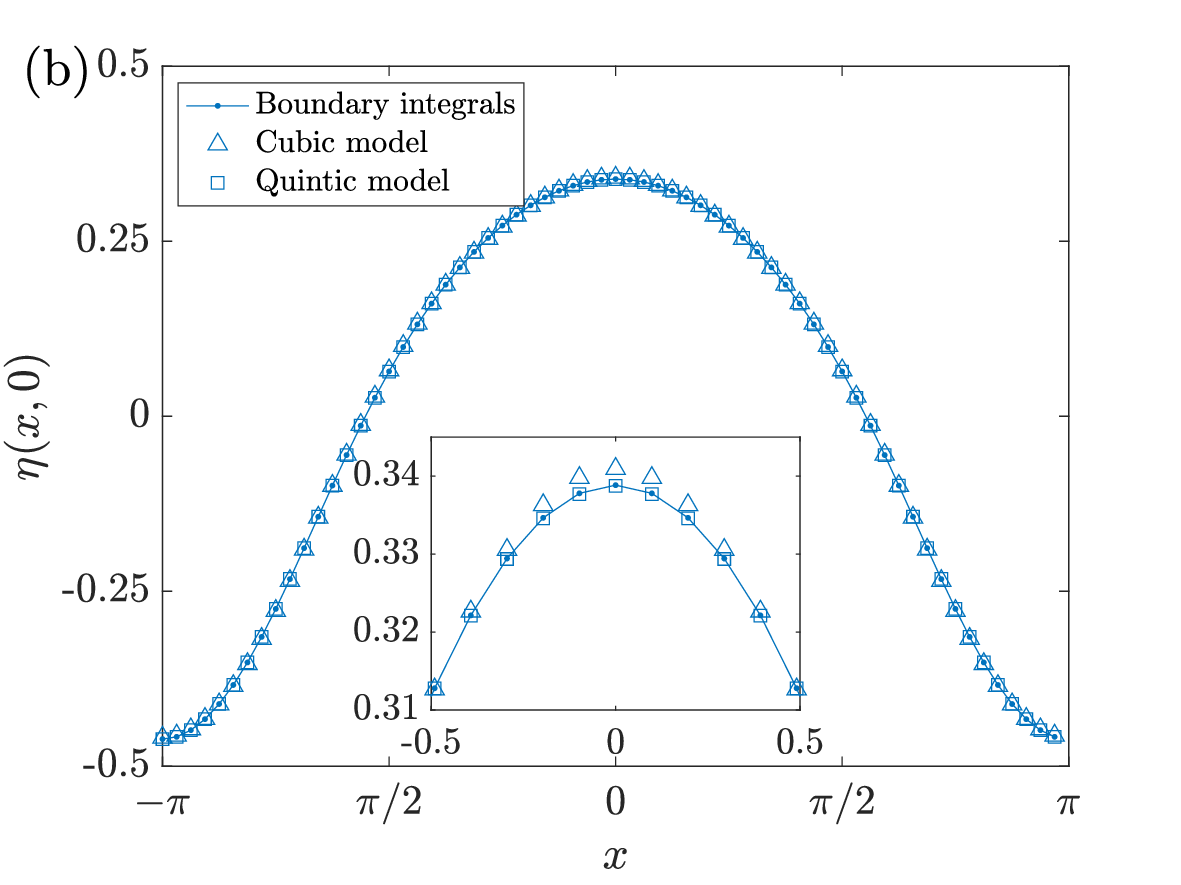}
    \caption{Comparisons between Concus's asymptotic result \eqref{concus}, and the numerical results using  boundary-integral method, the cubic model, and the quintic model for $k=1, h=\infty$. (a) $\omega$ versus $\epsilon$. (b) Initial wave profiles with a crest-to-trough amplitude of $0.8$ ($\epsilon = 0.394$).}
    \label{fig:compare1}
\end{figure}
Figure \ref{fig:compare1} compares Concus's asymptotic result and our numerical results of boundary-integral formulation (see Appendix 2), the cubic model, and the quintic model, for $k = 1, h = \infty$. Panel (a) exhibits the bifurcation curve plotted under $\omega$ and $\epsilon$. All results show perfect consistency and are indistinguishable when $\epsilon<0.2$. As $\epsilon$ gradually increases towards $0.4$, the cubic model agrees well with Concus's analytic result, while the quintic model gives almost identical results to the boundary-integral method, indicated by both the frequency curves and the wave profiles in (b).

Another comparison is made for the \textit{Case II} three-dimensional gravity standing waves in finite-depth water. Verma \& Keller \cite{verma1962three} derived the second-order Stokes-type expansion and the asymptotic expression for the nonlinear frequency, which has the same form as \eqref{concus}. For $k=l=1$, the coefficients are 
\begin{align}
    \omega_0 = \sqrt{\sqrt{2}\tanh(\sqrt{2}h)},\quad \omega_2 = -\frac{(3\omega_0^4-2)^2}{16\omega_0(\tanh(2h)-2\omega_0^2)} + \frac{36\omega_0^{-7}-24\omega_0^{-3}+5\omega_0-23\omega_0^{5}}{64},
\end{align}
and $\epsilon$ represents the amplitude of $\cos(x)\cos(y)\cos(\omega t)$. 
\begin{figure}[h!]
    \centering
    \includegraphics[width=0.45\linewidth]{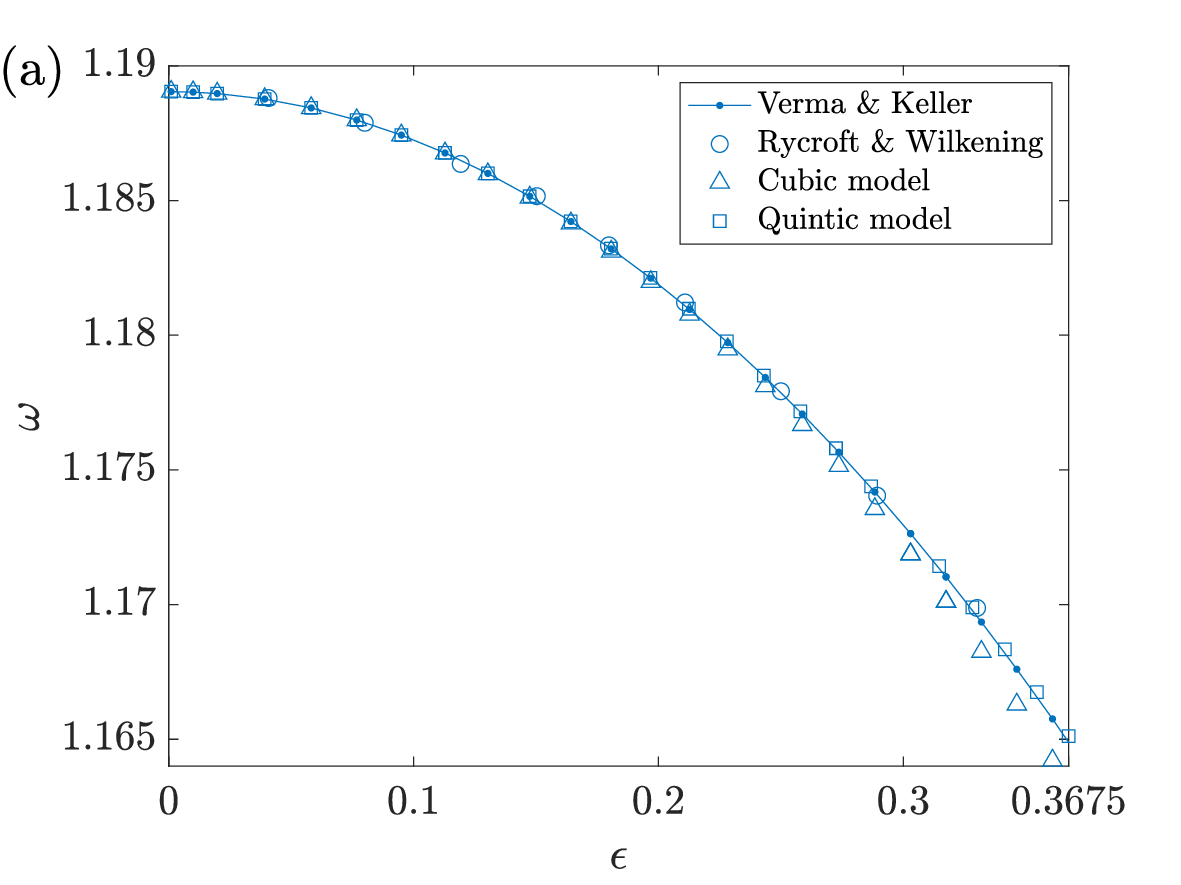}
    \includegraphics[width=0.45\linewidth]{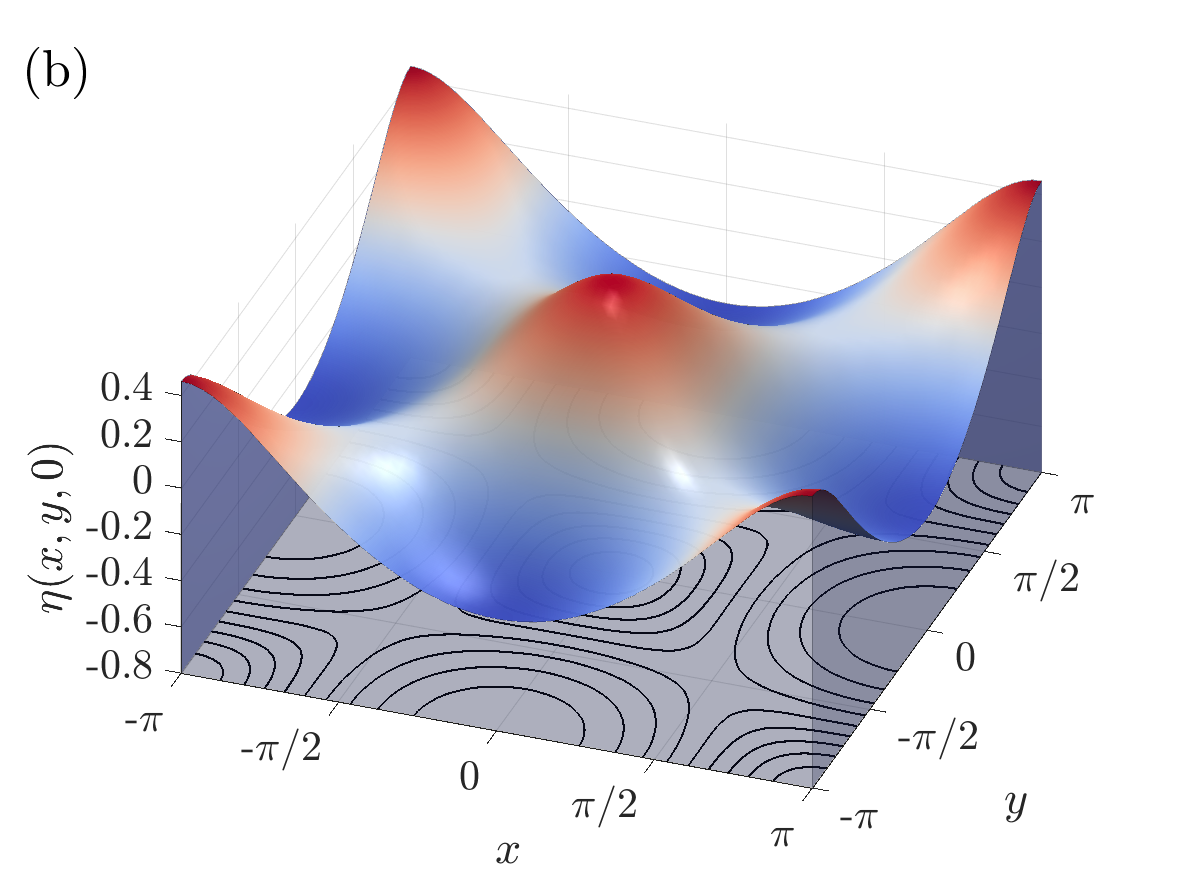}
    \caption{Comparisons between the asymptotic result of Verma \& Keller \cite{verma1962three}, numerical results of Rycroft \& Wilkening \cite{rycroft2013computation}, the cubic model, and the quintic model for \textit{Case II} gravity standing waves with $k=l=1, h=\pi$. (a) $\omega$ versus $\epsilon$. (b) An initial profile obtained from the quintic model with $H=0.46$ ($\epsilon=0.3675$).}
    \label{fig:compare2}
\end{figure}
Figure \ref{fig:compare2} (a) compares the numerical results obtained from the the cubic and the quintic models, asymptotic result of Verma \& Keller, and the numerical result of Rycroft \& Wilkening, for $k = l = 1, h = \pi$. The cubic model slightly deviates other results when $\epsilon>0.2$ but still exhibits a qualitative agreement. In contrast, the quintic model agrees remarkably with the existing asymptotic and numerical results. In panel (b), we exhibit the initial wave profile for $H = 0.46$ ($\epsilon = 0.3675$) obtained from the quintic model. Interested readers may compare it with those numerical solutions shown in \cite{rycroft2013computation} to see the similarity.

In the following sections we shall only exhibit the numerical results obtained from the quintic model owing to its excellent consistency with the original full potential-flow formulation.

\subsection{Standing waves and bifurcations}

\begin{figure}[h!]
    \centering
    \includegraphics[width=0.32\linewidth]{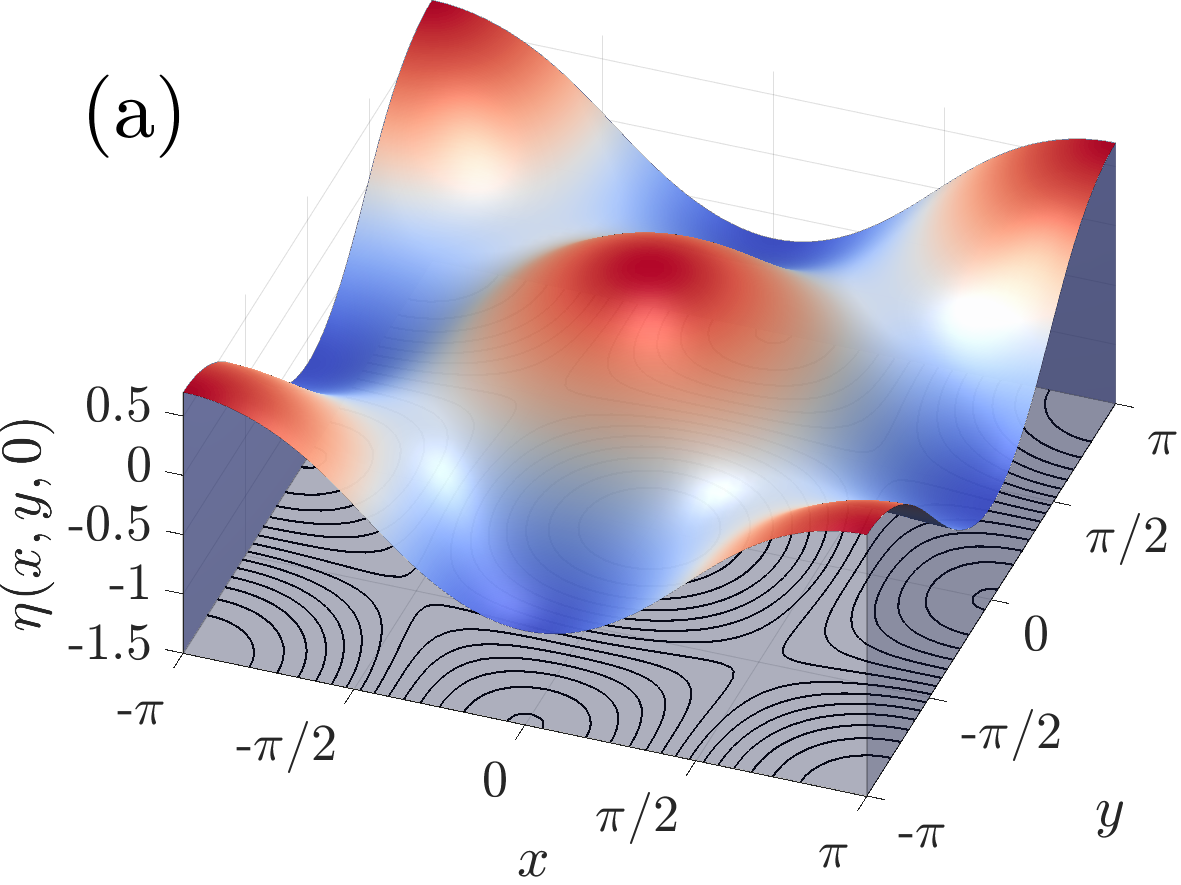}
    \includegraphics[width=0.32\linewidth]{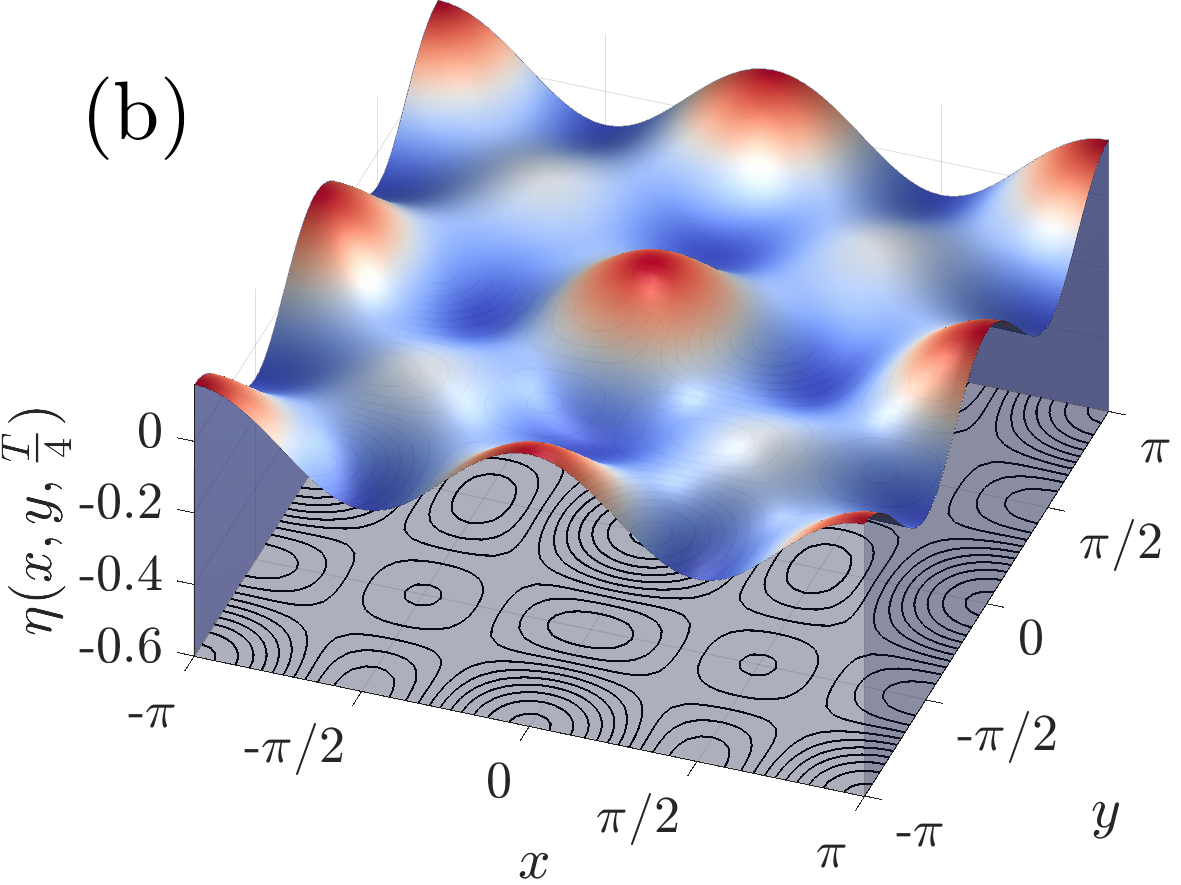}
    \includegraphics[width=0.32\linewidth]{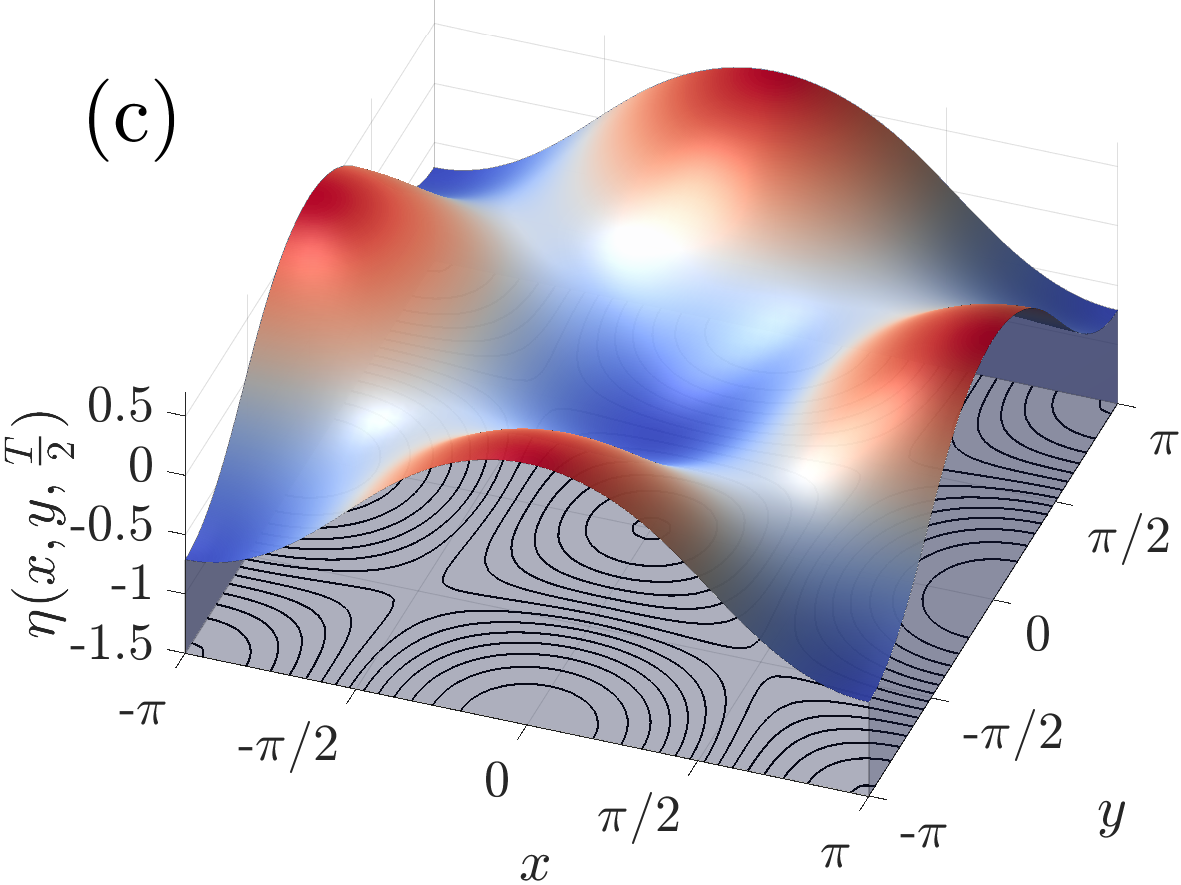}
    \caption{Surface profiles of a \textit{Case II} standing wave for $k = l = 1,H=0.7$ at $t = 0,T/4$ and $T/2$.}
    \label{fig:solution1}
\end{figure}
\begin{figure}[h!]
    \centering
    \includegraphics[width=0.5\linewidth]{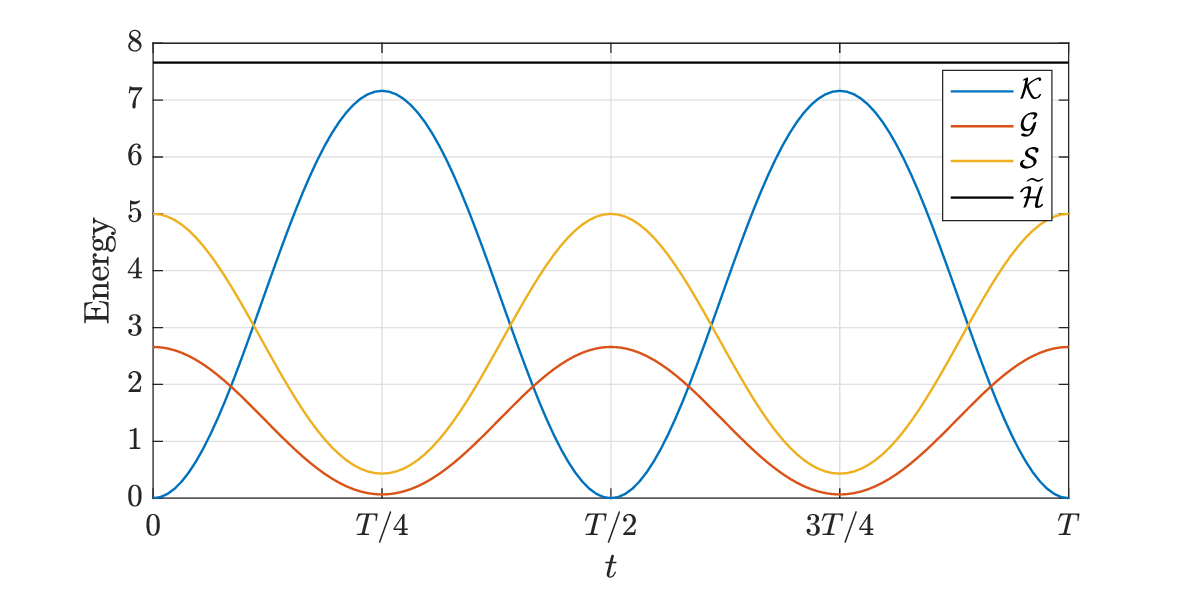}
    \caption{Energy distribution over one temporal period for the standing wave shown in figure \ref{fig:solution1}. $\mathcal K, \mathcal G, \mathcal S$, and $\widetilde{\mathcal H}$ represent kinetic, gravitational, surface-tension, and total energies.}
    \label{fig:solution1.5}
\end{figure}
We first show representative \textit{Case II} standing waves with $k = l = 1$. In contrast to the gravity standing waves featuring pyramidal-shaped crests, gravity-capillary standing waves exhibit rounded crests and troughs. Figure \ref{fig:solution1} displays a solution with $H=0.7,\omega = 1.867$, calculated using $128\times128\times 128$ grid points. Panels (a-c) correspond to the wave profiles at $t=0,T/4$ and $T/2$, respectively. At $t=T/4$, the surface develops multiple crests and troughs with the maximum wave amplitude decreasing to $0.16$. Figure \ref{fig:solution1.5} exhibits the energy distribution among the kinetic, gravitational, and surface-tension parts over one temporal period. The kinetic energy vanishes every half period, while the potential energy reaches its minimum, $0.497$, at $t = T/4,3T/4,\cdots$. The total energy remains at a constant level of $7.661$ perfectly, with the maximum relative error being $7\times 10^{-8}$. 

\begin{figure}[h!]
    \centering
    \includegraphics[width=0.45\linewidth]{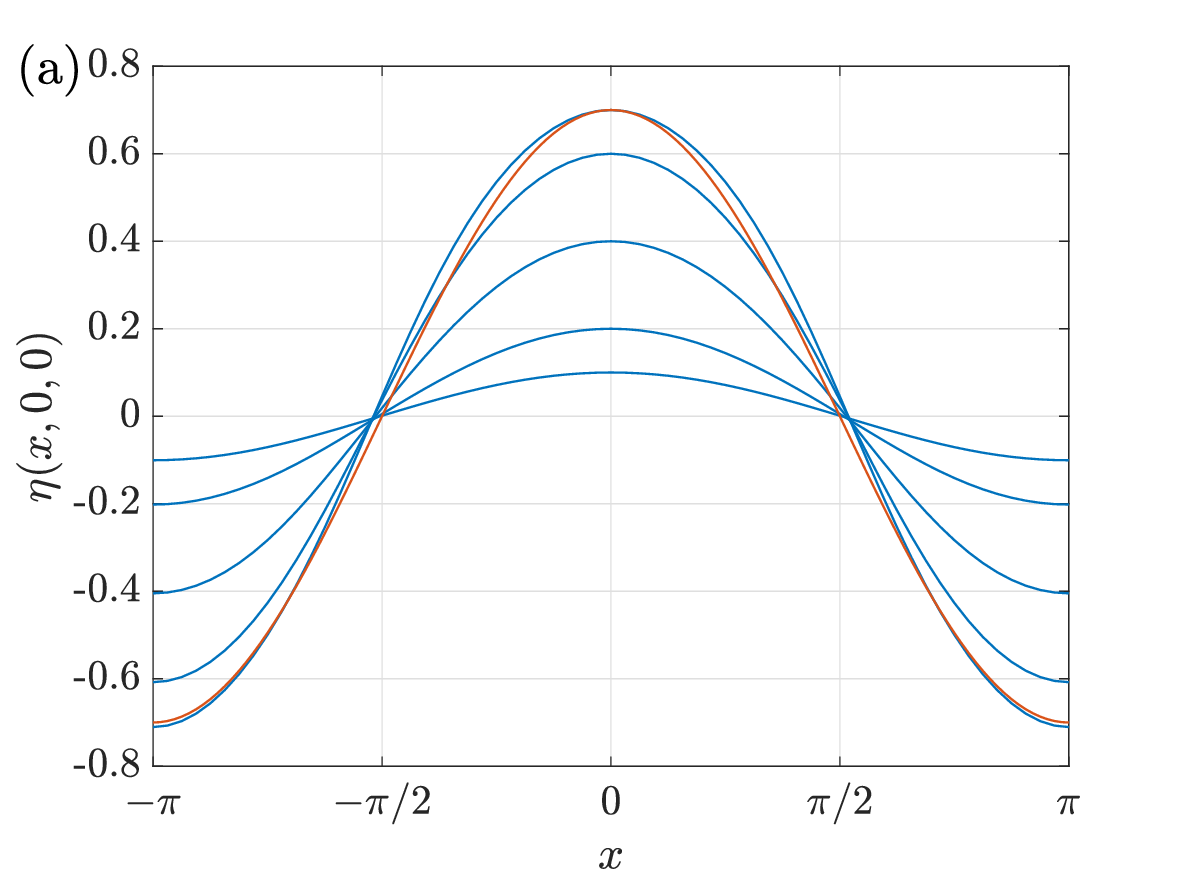}
    \includegraphics[width=0.45\linewidth]{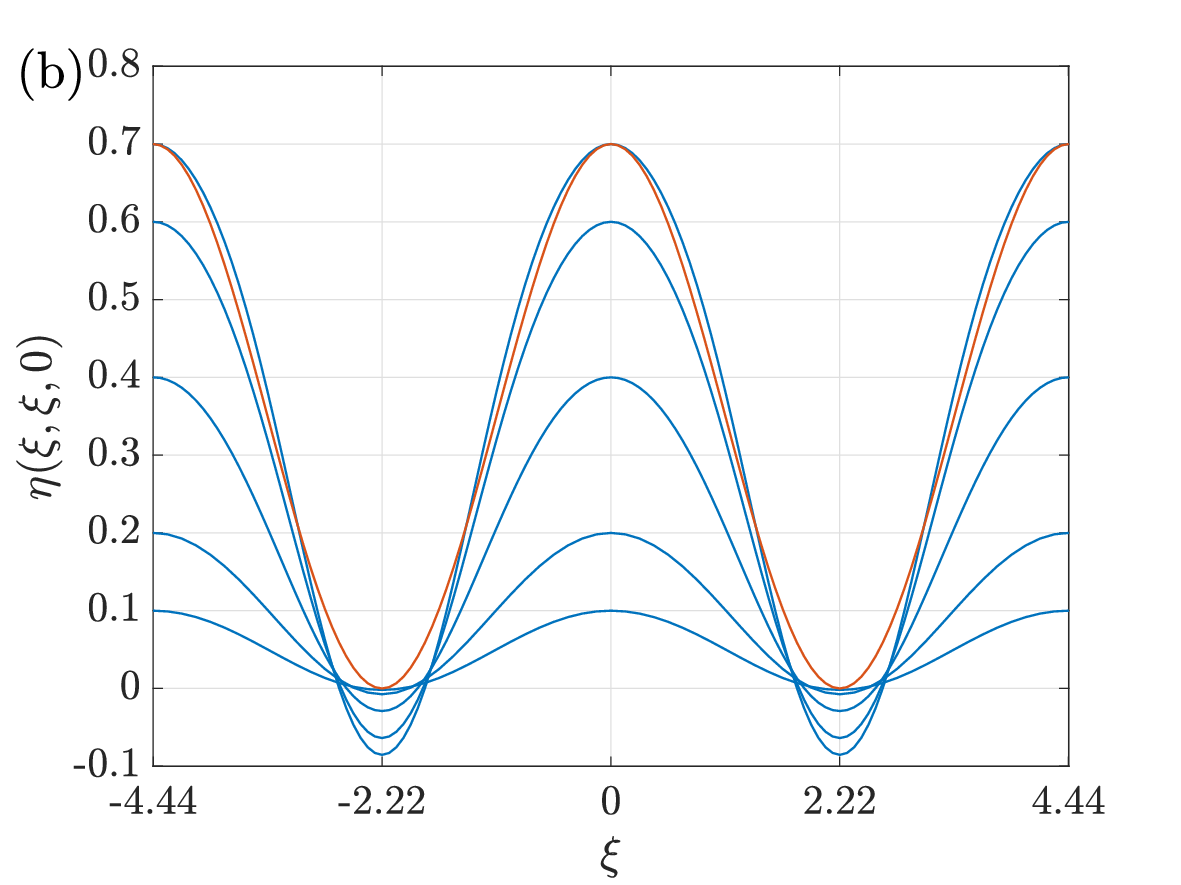}
    \caption{(a) Initial cross-sections of \textit{Case II} standing waves for $H=0.1,0.2,0.4,0.6,0.7$. The red curve corresponds to $0.7\cos(x)\cos(y)$. (b) Initial diagonal cross-sections of the same waves in (a).}
    \label{fig:solution1.5}
\end{figure}
Figure \ref{fig:solution1.5} shows the initial cross-sections of \textit{Case II} standing waves with successively increasing wave amplitudes $H=0.1,0.2,0.4,0.6,0.7$. Panel (a) corresponds to the cross-sections on the plane $y=0$, which are nearly sinusoidal. As shown by the comparison with $0.7\cos(x)\cos(y)$, the crest of the nonlinear solution is slightly broader than the linear solution. This is also shown in panel (b), where we plot the cross-sections along the diagonal $x=y=\xi /\sqrt{2}$. Note that $\xi =\pm\pi/\sqrt{2}\approx \pm 2.22$ are nodes of linear standing waves, but the nonlinear solutions exhibit negative surface elevations at these points. 

\begin{figure}[h!]
    \centering
    \includegraphics[width=0.42\linewidth]{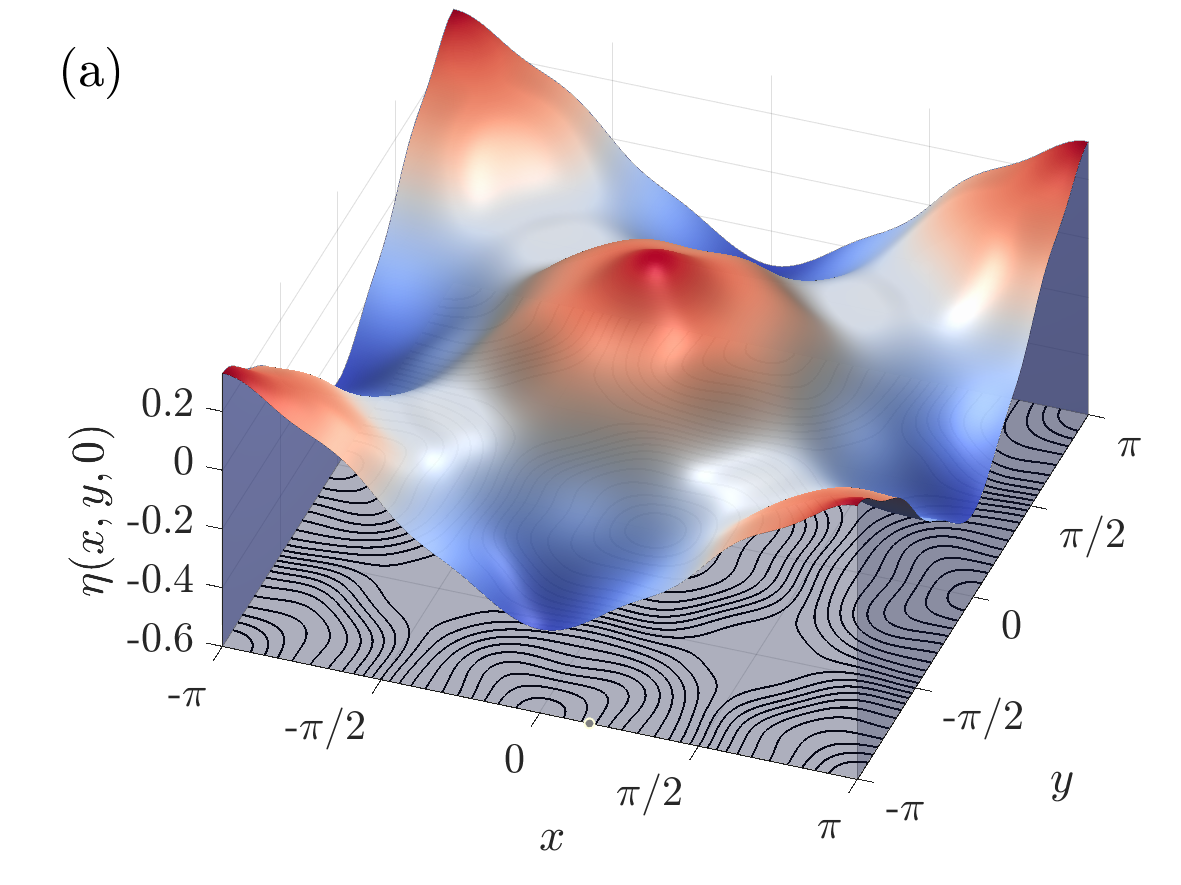}
    \includegraphics[width=0.42\linewidth]{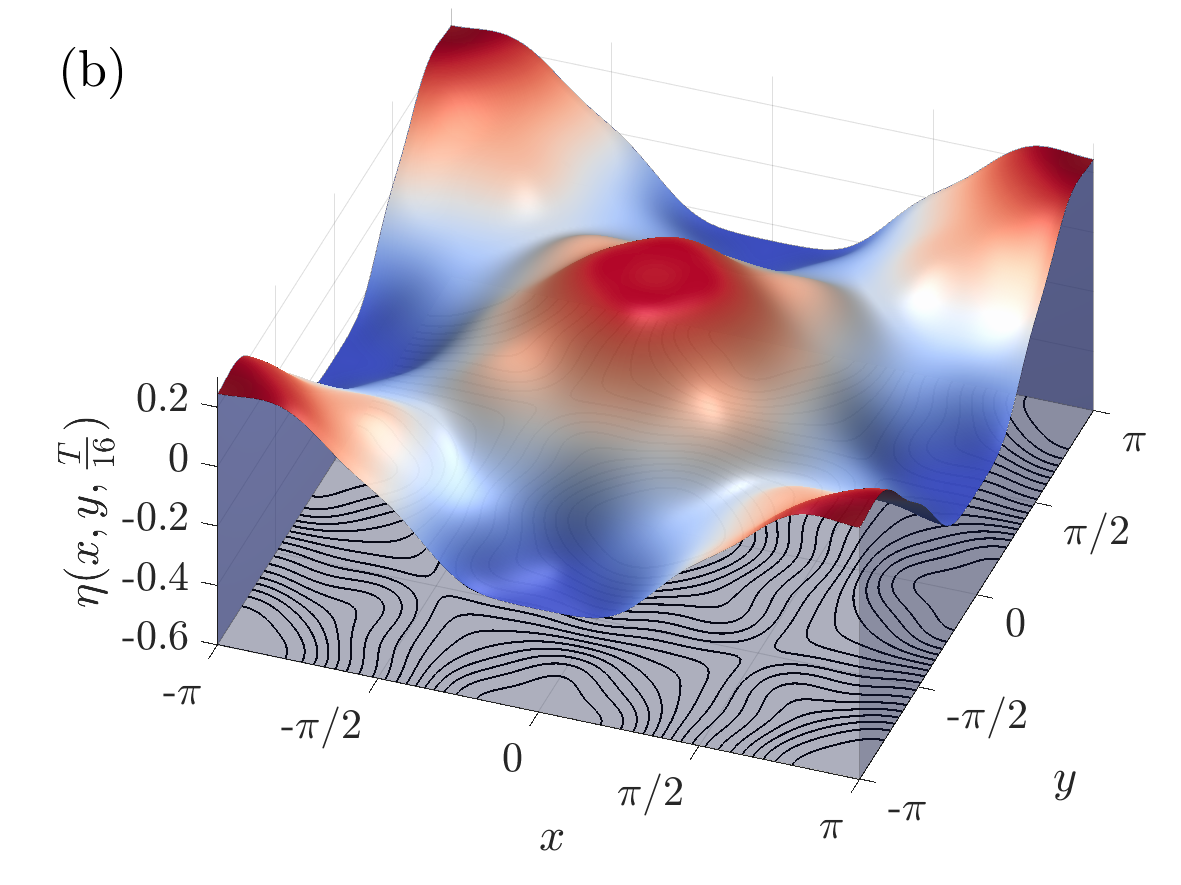}
    \includegraphics[width=0.42\linewidth]{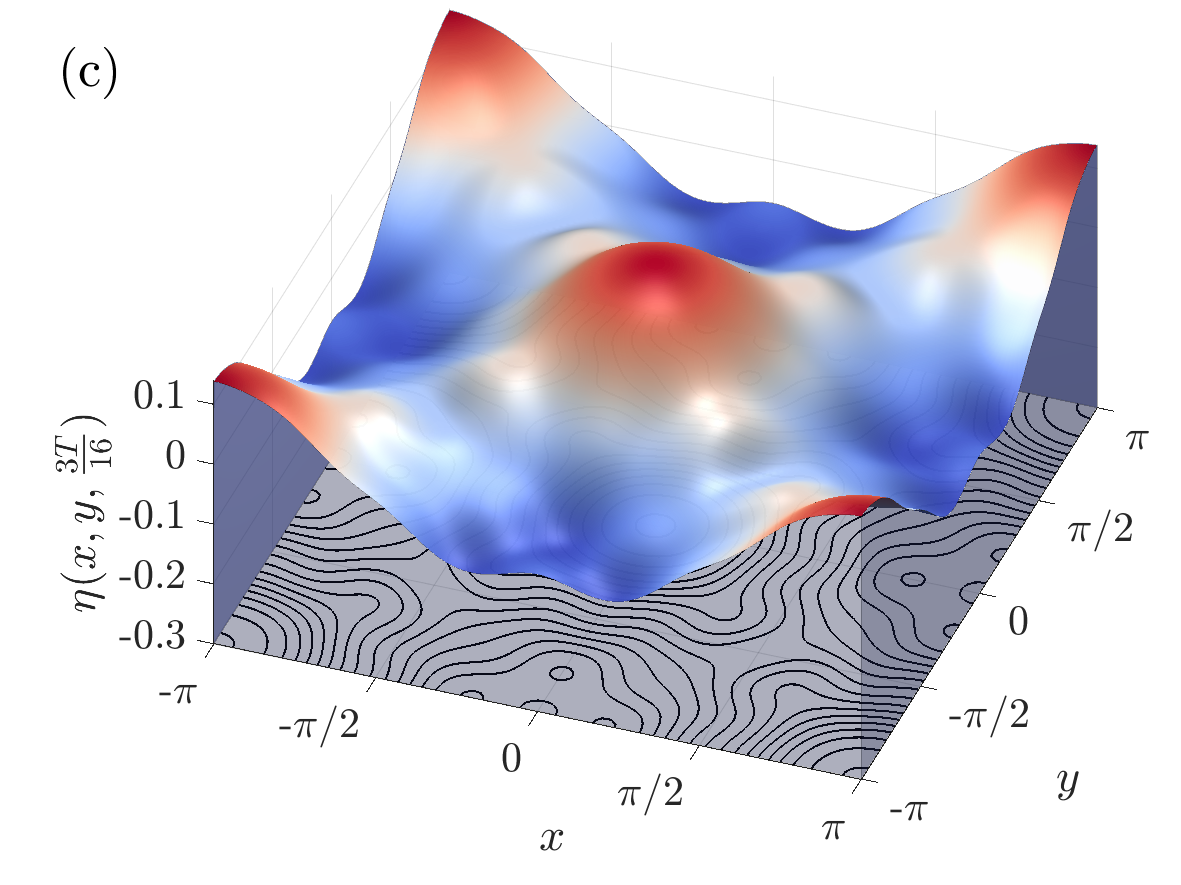}
    \includegraphics[width=0.42\linewidth]{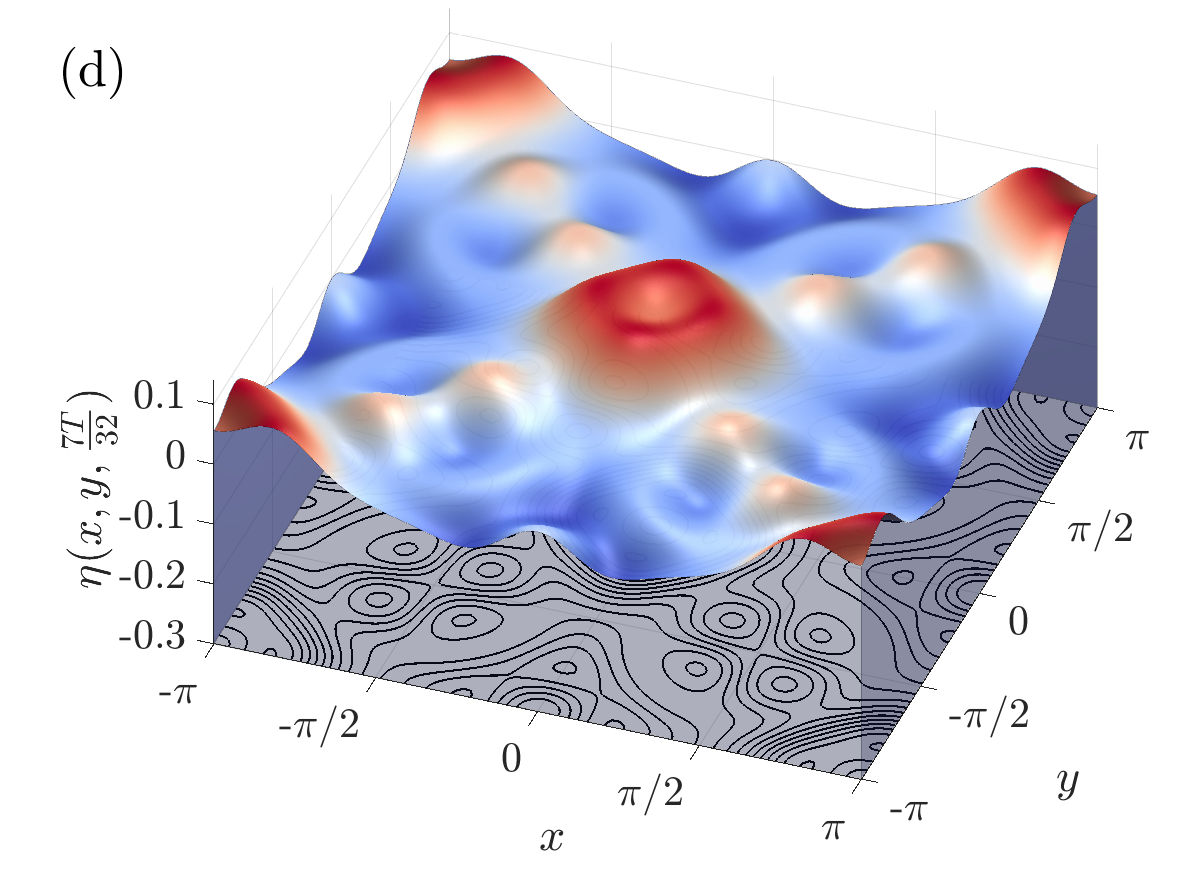}
    \includegraphics[width=0.42\linewidth]{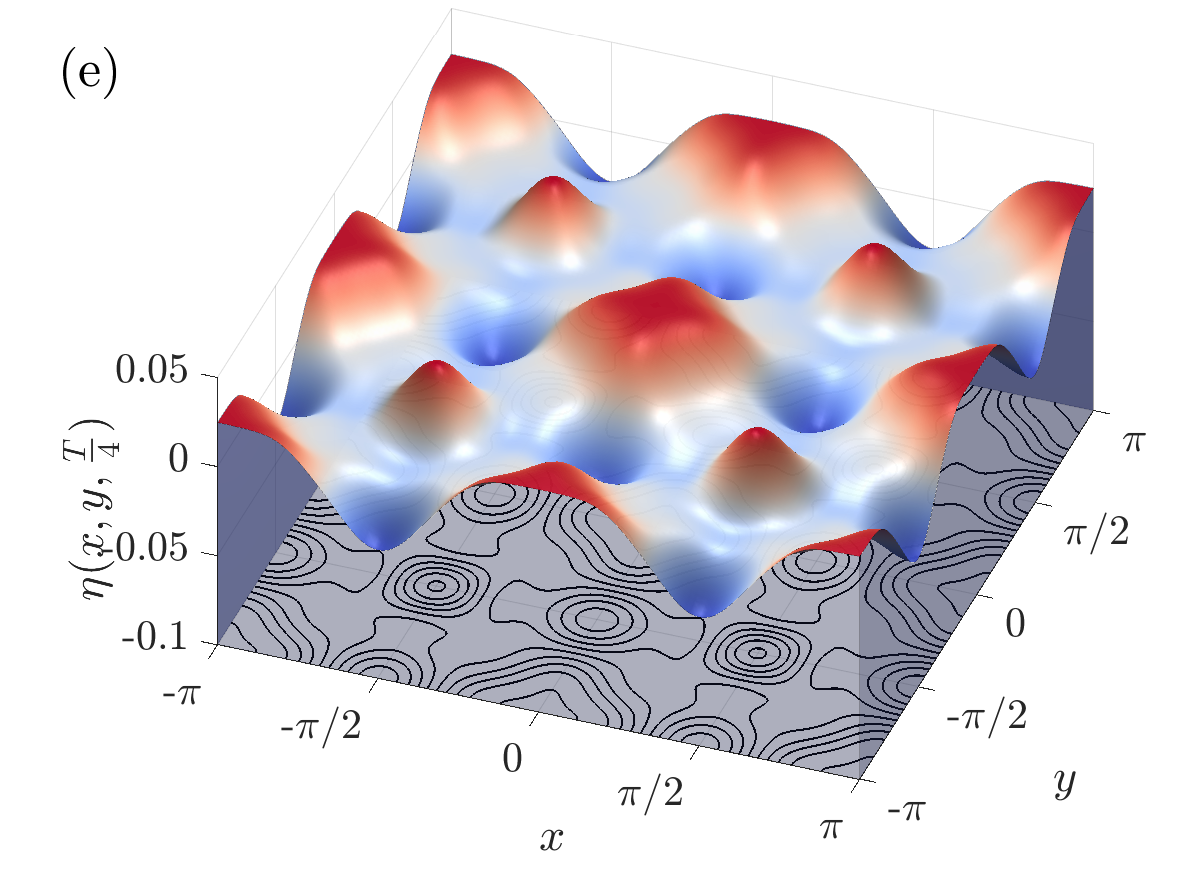}
    \includegraphics[width=0.42\linewidth]{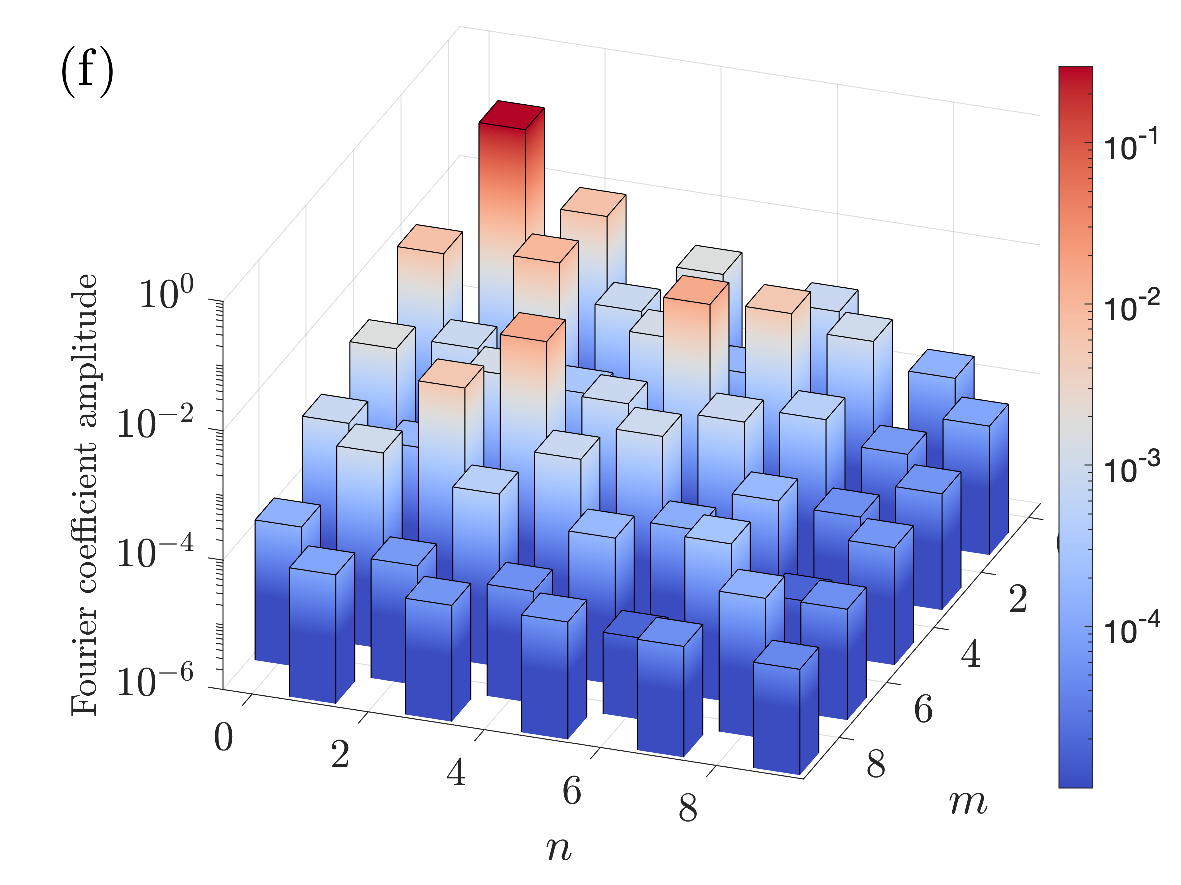}
    \caption{(a-e) Surface profiles of a resonant \textit{Case II} standing wave with $k = l = 1, H=0.33$ at $t = 0,T/16,3T/16,7T/32$ and $T/4$. (f) Amplitudes of the first $100$ Fourier coefficients for $\eta(x,y,0)$.}
    \label{fig:solution2}
\end{figure}

A common feature for two-dimensional standing waves is the existence of various nonlinear resonant solutions \cite{smith1999branching,wilkening2011breakdown,wilkening2012overdetermined}. We observe a similar phenomenon in three dimensions. Figure \ref{fig:solution2} shows a representative \textit{Case II} resonant standing wave with $k=l=1,H=0.33,\omega = 2.0192$, calculated using $128\times 128\times 128$ grid points. In contrast to the smooth surface profiles shown in figure \ref{fig:solution1}, the resonant standing wave exhibits complex surface wrinkles. A local hump and dip alternately emerge on top of the crest at a faster oscillating frequency relative to the fundamental Fourier mode, as shown in panels (a-c). The surface develops a volcano-shaped appearance with four saddles along the diagonals at $t=7T/32$, as shown in panel (d). At $t=T/4$, the wave amplitude drops to $0.0247$ approximately, and the surface profile shows a similar pattern to that in figure \ref{fig:solution1} (b) but has much flatter crests and four sharper spikes at $(\pm \pi/2,\pm\pi/2)$. Panel (f) shows the Fourier amplitudes associated with $\cos(nx)\cos(my)$ mode for $\eta(x,y,0)$. Note that all Fourier modes with odd-valued sum $n+m$ vanish, in consequence of \eqref{timesymmetry1} and \eqref{timesymmetry2}. Owing to the eightfold symmetry, the spectrum is invariant under the interchange of $n$ and $m$. As can be clearly seen, the three dominant Fourier components are $\cos(x)\cos(y)$, $\cos(3x)\cos(5y)$, and $\cos(5x)\cos(3y)$. According to the dispersion relation \eqref{dispersion}, the two latter modes have the same linear frequency $\sqrt{35\sqrt{34}} \approx 7.075 \omega$, thus yielding a resonance of the seventh temporal harmonic. The eighth temporal harmonic, dominated by the $\cos(2x)\cos(6y)$ and $\cos(6x)\cos(2y)$ modes, is also nearly resonant, as indicated by linear frequency $\sqrt{41\sqrt{40}} \approx 7.975 \omega$.

\begin{figure}[h!]
    \centering
    \includegraphics[width=0.45\linewidth]{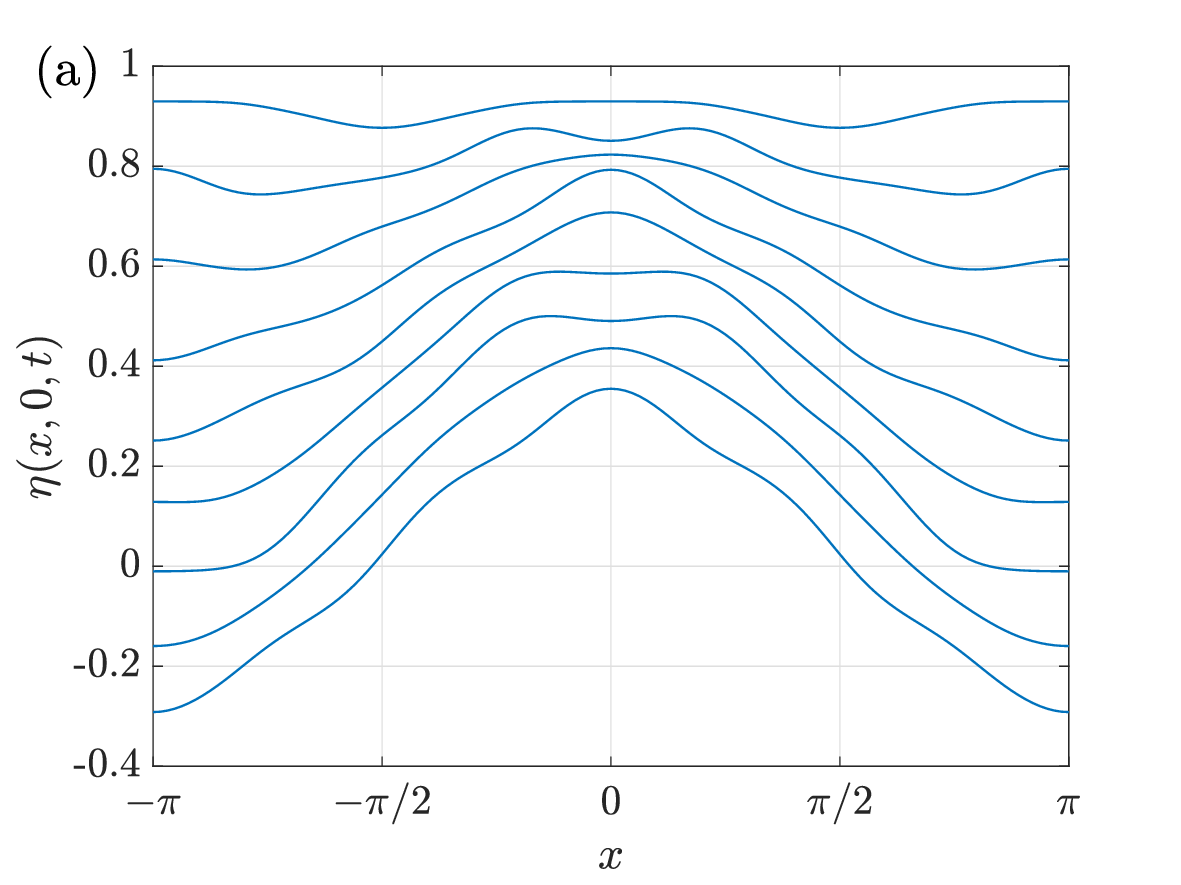}
    \includegraphics[width=0.45\linewidth]{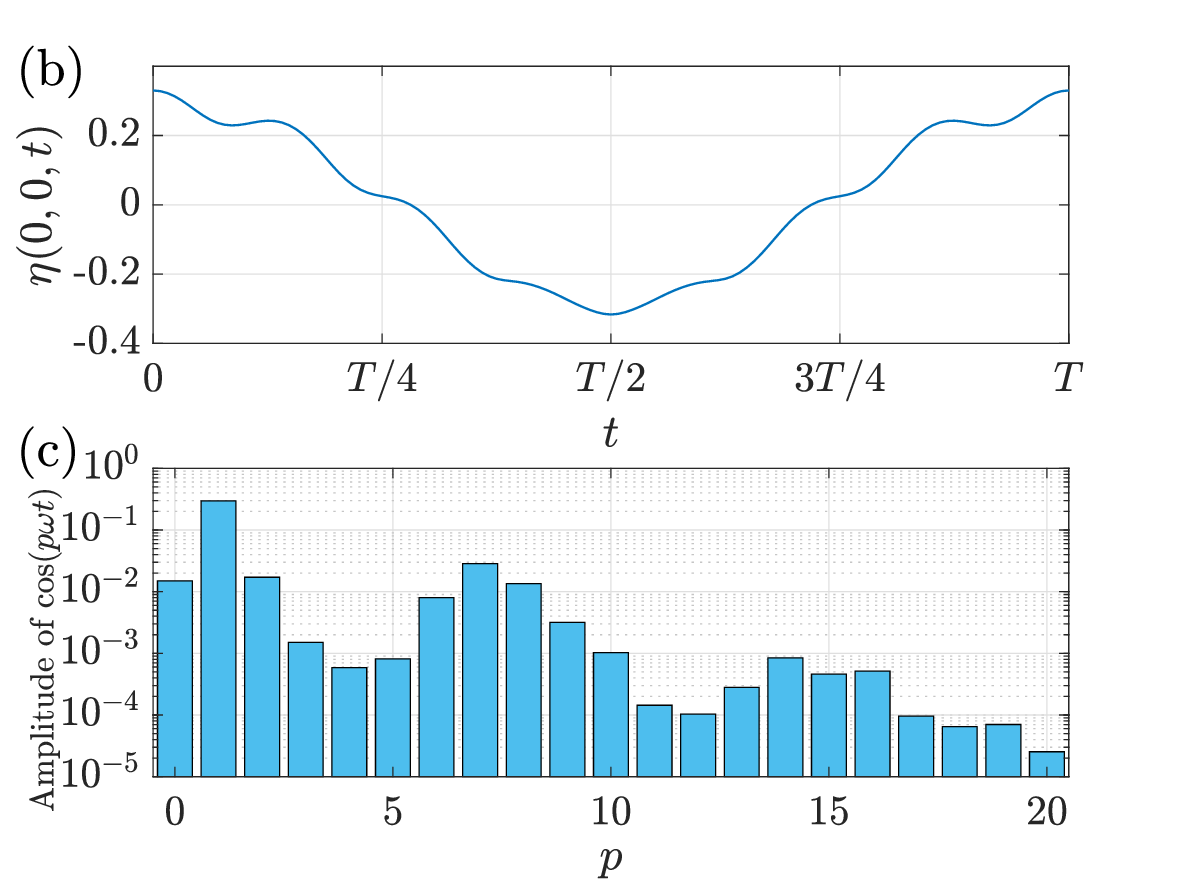}
    \caption{(a) Cross-sections of the standing wave shown in figure \ref{fig:solution2} at $t = 0,T/32,T/16,\cdots,T/4$ from bottom to top. The profiles are vertically shifted for the clarity. (b) $\eta(0,0,t)$ over one temporal period. (c) Amplitudes of the temporal Fourier modes $\cos(p\omega t)$ from $p = 0$ to $20$.}
    \label{fig:solution2.5}
\end{figure}
\begin{figure}[h!]
    \centering
    \includegraphics[width=0.45\linewidth]{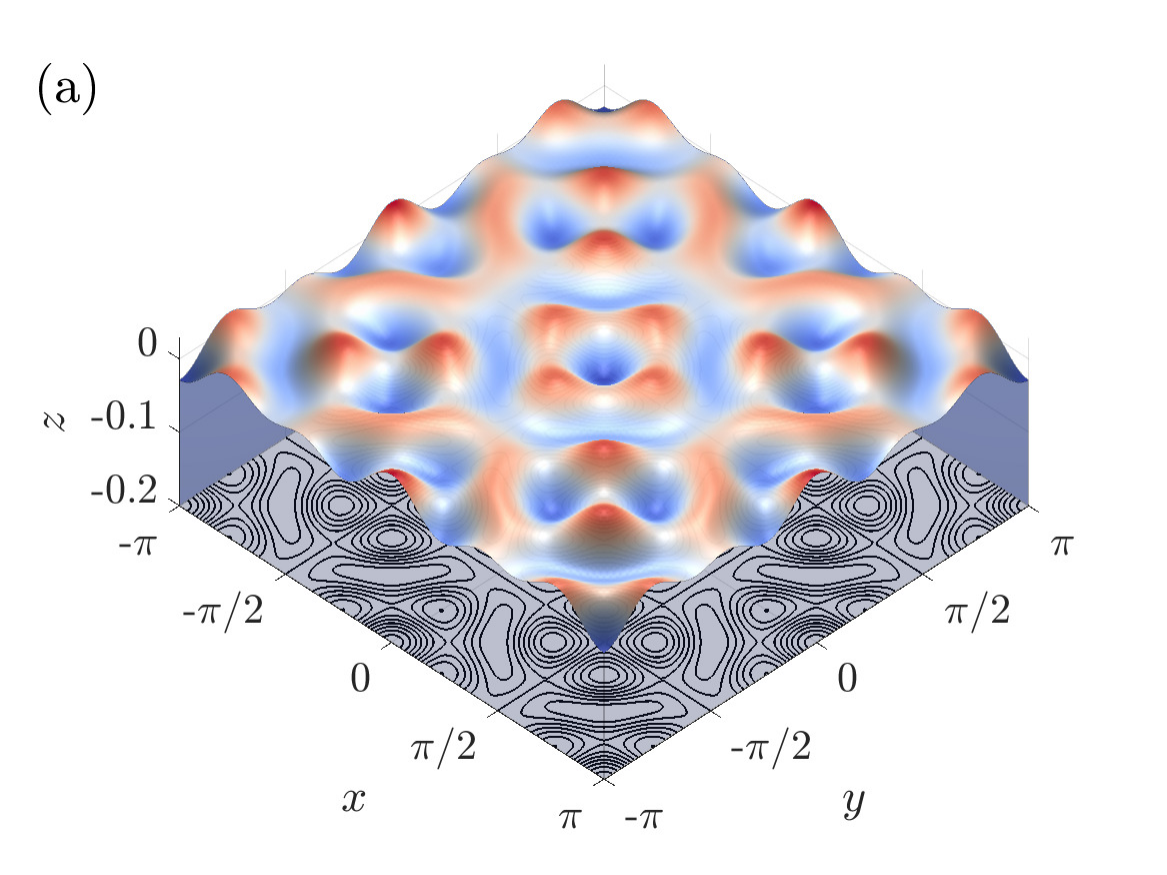}
    \includegraphics[width=0.45\linewidth]{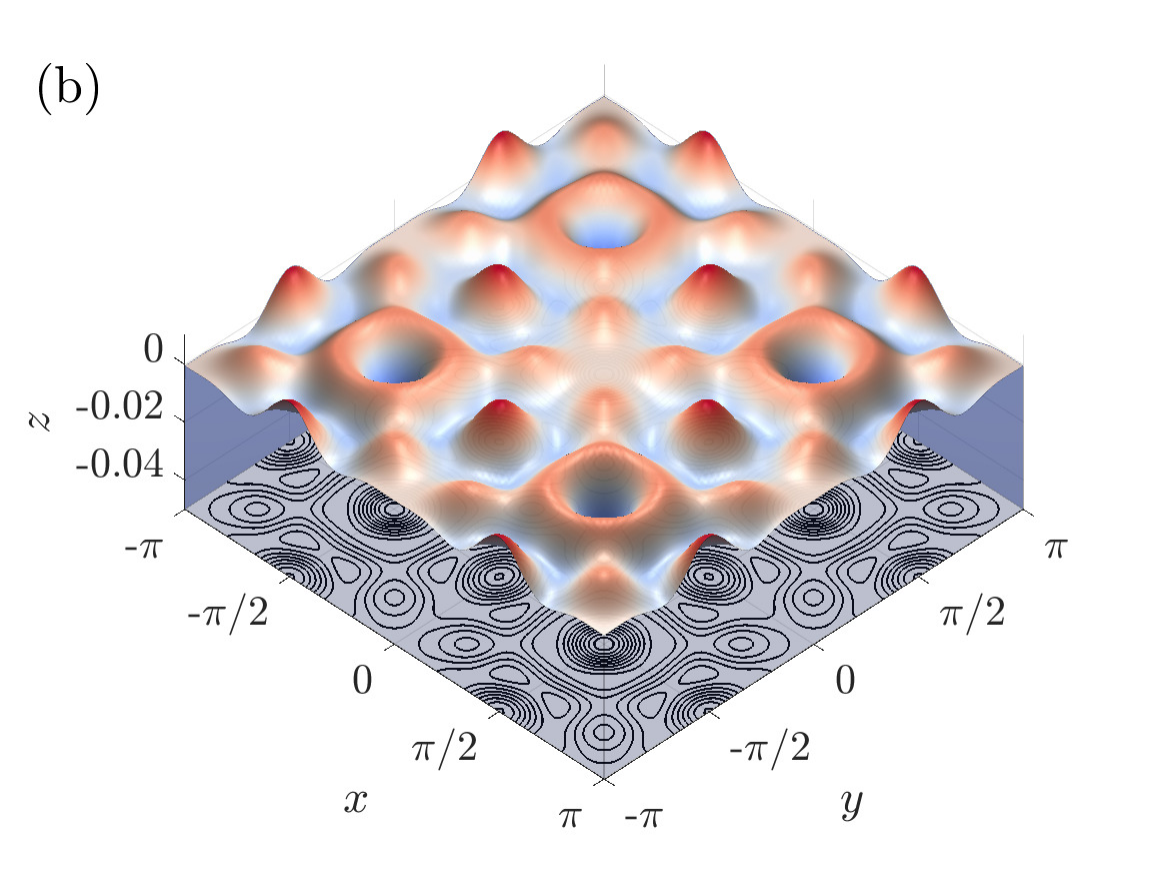}
    \caption{ (a) The coefficient of $\cos(7\omega t)$. (b) The coefficient of $\cos(8\omega t)$.}
    \label{fig:solution3}
\end{figure}
The same resonant wave is shown in figure \ref{fig:solution2.5}. Panel (a) presents the cross-sections of the wave profile on the plane $y=0$ over a quarter temporal period, exhibiting five shorter waves clearly. Panel (b) shows the time history of $\eta(0,0,t)$ over a temporal period, which indicates seven minor oscillations. To measure the amplitude of each temporal harmonic, we take Fourier transform of $\eta(x,y,t)$ with respect to $t$ and display the infinity norm associated with $\cos(p\omega t)$. As can be clearly seen, the five dominant components correspond to $p = 1, 7, 2, 0$, and $8$. Figure \ref{fig:solution3} shows the coefficients (as functions of $x$ and $y$) of the seventh and eighth temporal harmonic, which are dominated by $\cos(3x)\cos(5y)+\cos(5x)\cos(3y)$ and $\cos(2x)\cos(6y)+\cos(6x)\cos(2y)$, respectively.

\begin{figure}[h!]
    \centering
    \includegraphics[width=0.8\linewidth]{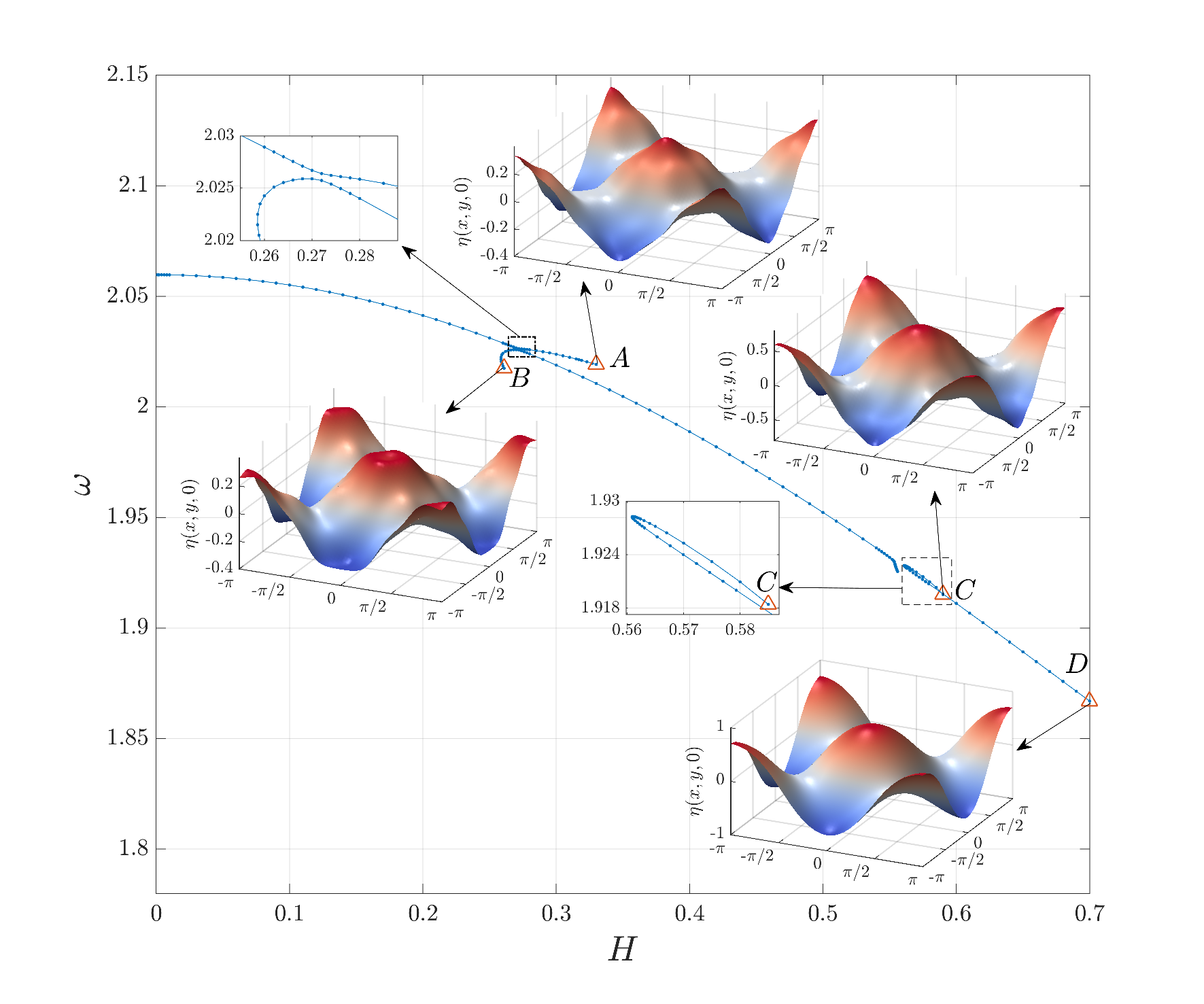}
    \caption{Bifurcations of \textit{Case II} standing waves for $k = l = 1$ and representative initial wave profiles.}
    \label{fig:solution4}
\end{figure}
Figure \ref{fig:solution4} shows the bifurcation of \textit{Case II} standing waves for $k = l = 1$, which grows from the infinitesimal linear solution with the frequency $\omega = 2.0598$. Following the primary branch where no resonances occur, solutions are characterized by smooth surfaces with rounded crests and troughs. A large-amplitude solution, which is labelled by the triangle $D$, is identical to the standing wave shown in figure \ref{fig:solution1}. The nonlinear frequency $\omega$ monotonically decreases with $H$, until $\omega\approx 2.026$ when the bifurcation breaks into two disjoint branches. The local structure is shown in the small box nearby. Moving along each branch in a specific direction, the resonance between the fundamental and seventh temporal harmonics gradually becomes significant. Two representative resonant solutions are labelled by $A$ and $B$. The latter one, which features a local hump on top of the crest, is the same solution shown in figure \ref{fig:solution2}. In contrast, solution $A$ exhibits a local dip on the crest, revealing that the two branches of resonant solutions are out of opposite phases in the seventh temporal harmonic. When $\omega\approx 1.93$, we observe a second break-up. A resonant solution with $H = 0.59$ is labelled by $C$ and shown. This break-up is caused by resonances among multiple modes, including the seventh, eighth, and ninth temporal harmonics. While there exist more break-ups following the bifurcation, we do not present them because of the similar structure.

\begin{figure}[h!]
    \centering
    \includegraphics[width=0.42\linewidth]{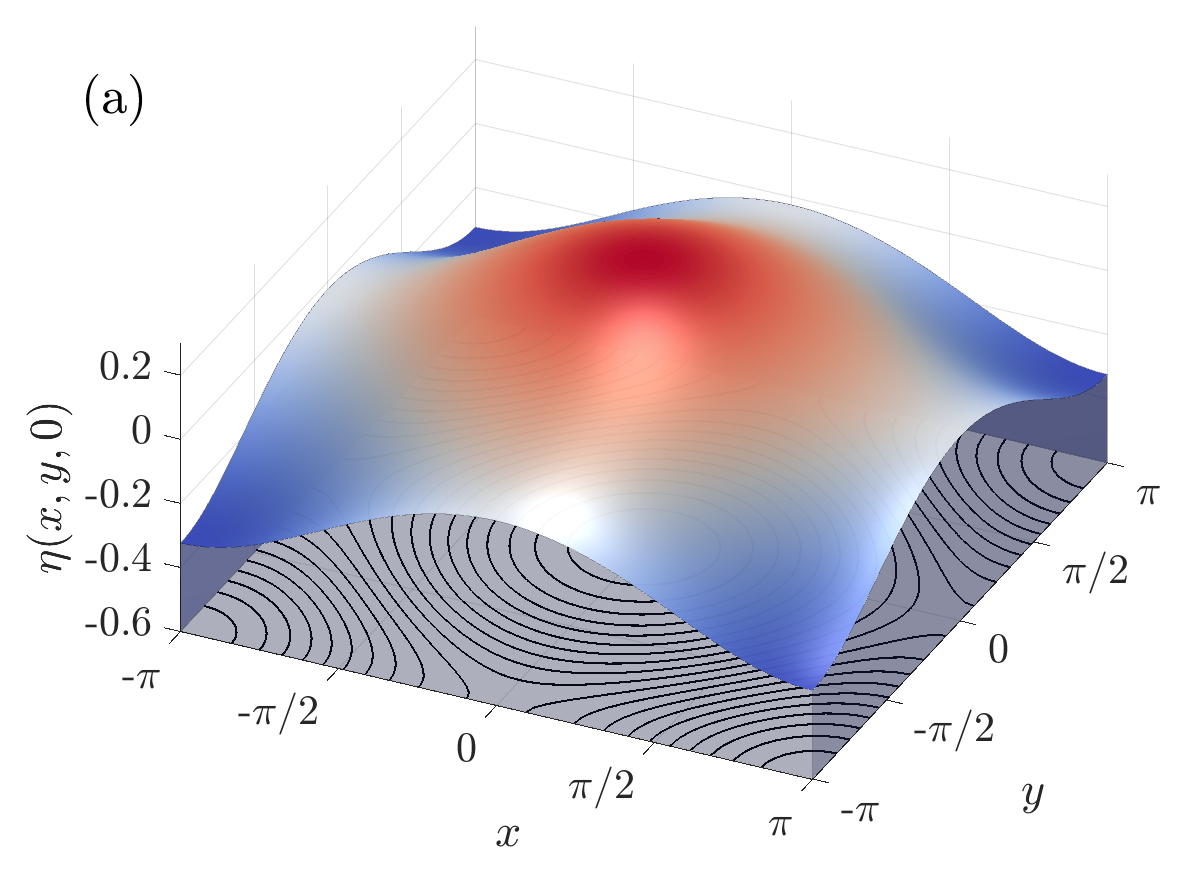}
    \includegraphics[width=0.42\linewidth]{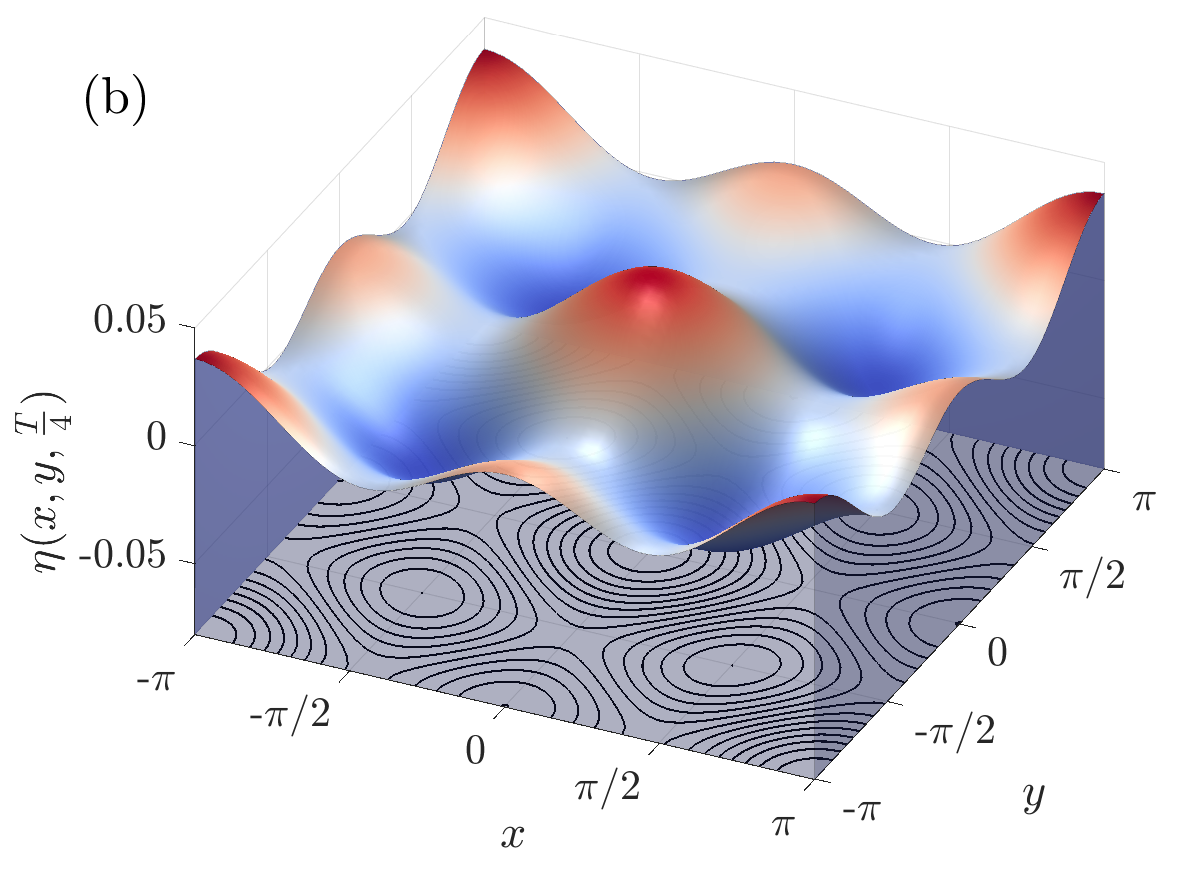}
    \includegraphics[width=0.42\linewidth]{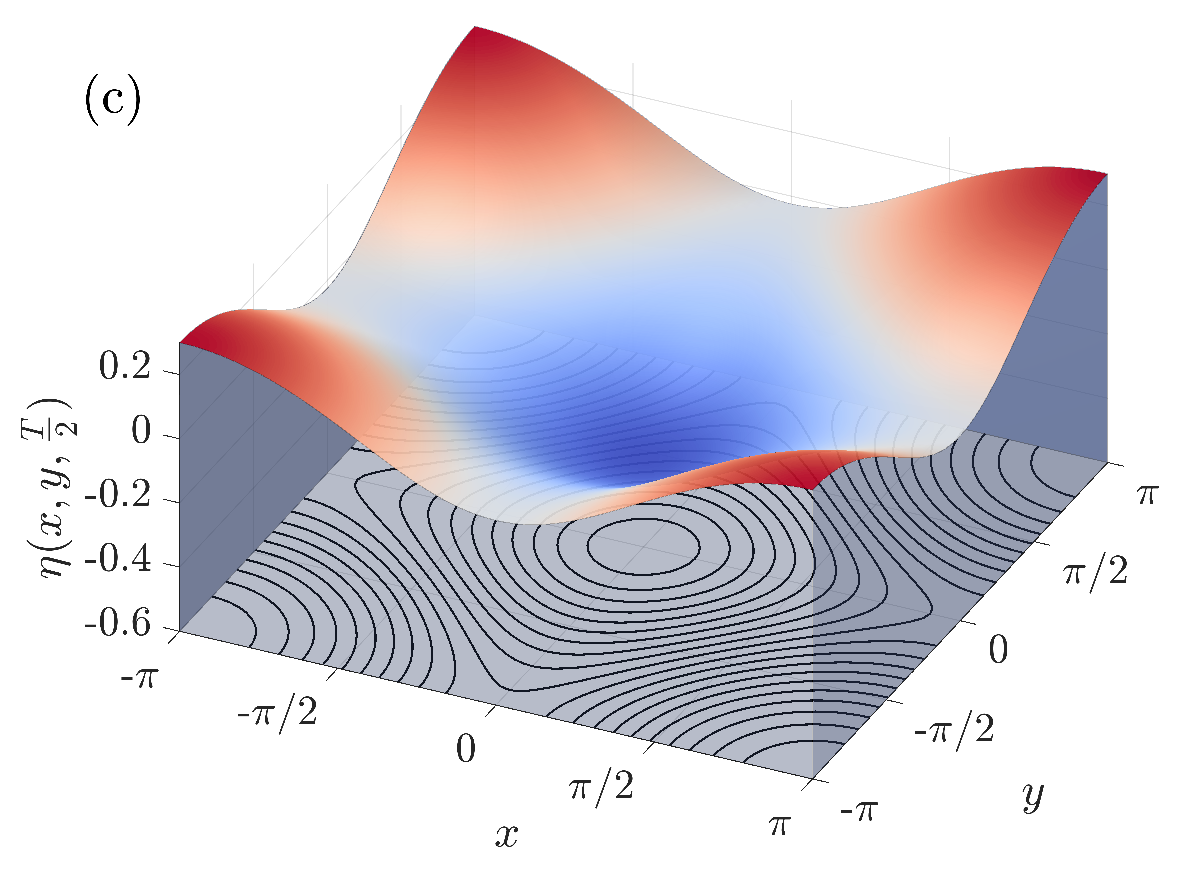}
    \includegraphics[width=0.42\linewidth]{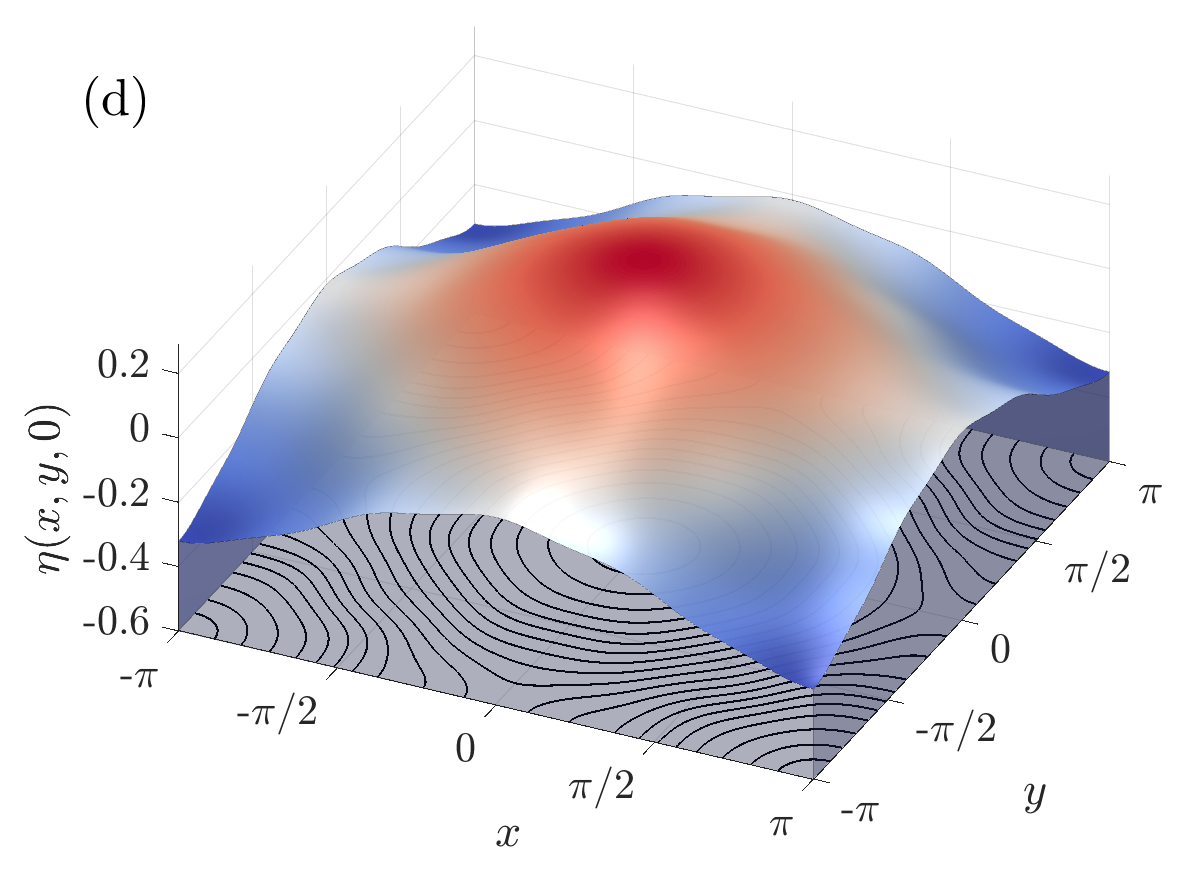}
    \caption{Surface profiles of \textit{Case III} standing waves for $k = 1,l = 0$. (a-c) A non-resonant solution with $H=0.3$ at $t = 0,T/4$, and $T/2$. (d) A resonant solution with $H=0.29$ at $t = 0$.}
    \label{fig:solution5}
\end{figure}
Figure \ref{fig:solution5} shows typical \textit{Case III} standing waves for $k = 1, l = 0$. The surface profiles of a non-resonant solution with $H = 0.3,\omega = 1.3979$ are shown in panels (a-c), corresponding to $t = 0,T/4$, and $T/2$. Note that this solution is essentially the same as the $\textit{Case II}$ standing wave with $k=l=\sqrt{2}/2, H=0.3$ after a rotation of the $(x,y)$-plane by $\pi/4$. This is also reflected by the fact that the shift connecting $\eta(x,y,0)$ and $\eta(x,y,T/2)$ is along the diagonal direction, which becomes the $x$- or $y$-direction after the rotation. Panel (d) exhibits a resonant standing wave with $H=0.29,\omega = 1.3994$. The resonance is primarily between the fundamental and ninth temporal harmonics, which is dominated the $\cos(2x)\cos(5y)+\cos(5x)\cos(2y)$ modes. 

\begin{figure}[h!]
    \centering
    \includegraphics[width=0.8\linewidth]{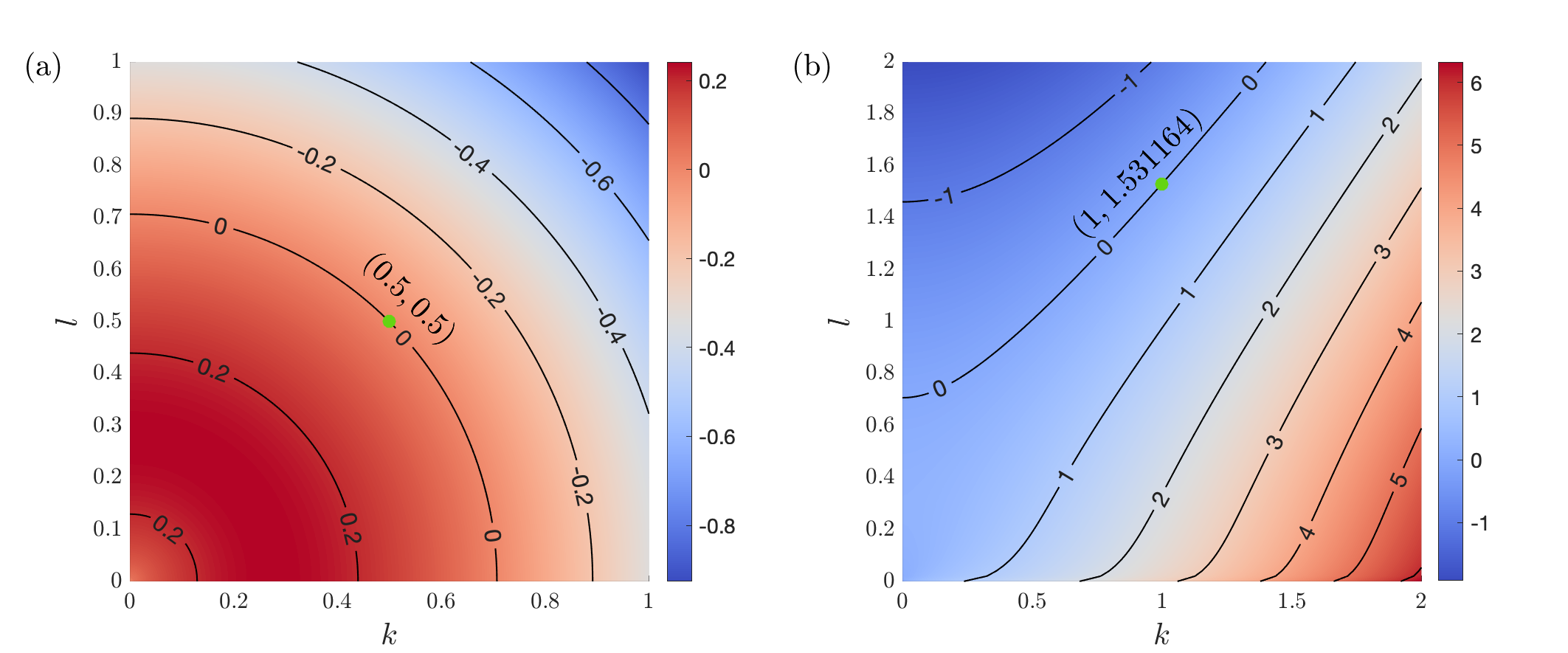}
    \caption{Values and typical level sets of the left-hand side of \eqref{resonanc2} for (a) $\alpha = \beta = 1$, and for (b) $\alpha = -1, \beta = 1$. The green dots denote particular values of $\kappa$ supporting resonant standing waves.}
    \label{fig:resonance}
\end{figure}

In previous studies of gravity-capillary waves, it is known that the three-wave resonance accounts for the classical Wilton ripples, travelling waves where the fundamental and second-order Fourier modes move at the same speed (other types of Wilton ripples exist but correspond to high-order resonances). To further investigate resonant standing waves, we consider the same mechanism here. Given a triad wavenumbers $\boldsymbol{\kappa}$, $\widetilde{\boldsymbol{\kappa}}$, and $\breve{\boldsymbol{\kappa}} = \boldsymbol{\kappa}+\widetilde{\boldsymbol{\kappa}}$, the condition of three-wave resonance reads
\begin{align}
    \omega(\breve{\boldsymbol{\kappa}}) - \omega(\boldsymbol{\kappa}) - \omega(\widetilde{\boldsymbol{\kappa}}) = 0,\label{resonanc2}
\end{align}
where $\omega(\boldsymbol{\kappa})$ is evaluated from \eqref{dispersion} with $\kappa =|\boldsymbol{\kappa}|$. Here we let $\boldsymbol{\kappa} = (k,l)$ and $\widetilde{\boldsymbol{\kappa}} = (\alpha k, \beta l)$, with $\alpha$ and $\beta$ being rational numbers. Figure \ref{fig:resonance} (a) shows the left-hand side of \eqref{resonanc2} for $\alpha = \beta = 1$. The zero level set is identical to a circle of radius $\sqrt{2}/2$, and each point on it corresponds to a collinear resonant triad. To construct \textit{Case II} standing waves, we take
\begin{align}
\begin{cases}
    \boldsymbol{\kappa} = \widetilde{\boldsymbol{\kappa}} = (0.5,0.5), \quad \breve{\boldsymbol{\kappa}} = (1,1),\\
    \omega(\boldsymbol{\kappa}) = \omega(\widetilde{\boldsymbol{\kappa}}) = \dfrac{\sqrt{3\sqrt{2}}}{2}\approx 1.0299, \quad \omega(\breve{\boldsymbol{\kappa}}) = \sqrt{3\sqrt{2}} \approx 2.0598,
\end{cases}
\end{align}
along with their transverse counterparts. These give rise to resonant standing waves featuring square-shaped patterns. Other possible wavenumbers can be found by calculating the intersections of the resonant circle and the straight line $l = rk$, where $r$ is a rational number. To find non-collinear resonant triads, we let $\alpha = -1, \beta = 1$. The left-hand side of \eqref{resonanc2} is shown in figure \ref{fig:resonance} (b). In particular, we find the following resonant triad numerically (accurate up to the sixth decimal place)
\begin{align}
\begin{cases}
    \boldsymbol{\kappa} = (1,1.531164),\quad \widetilde{\boldsymbol{\kappa}} = (-1,1.531164), \quad \breve{\boldsymbol{\kappa}} = (0, 3.062328),\\
    \omega(\boldsymbol{\kappa}) =\, \omega(\widetilde{\boldsymbol{\kappa}}) = 2.818705, \quad \omega(\breve{\boldsymbol{\kappa}}) = 5.637410.
\end{cases}
\end{align}

\begin{figure}[h!]
    \centering
    \includegraphics[width=0.4\linewidth]{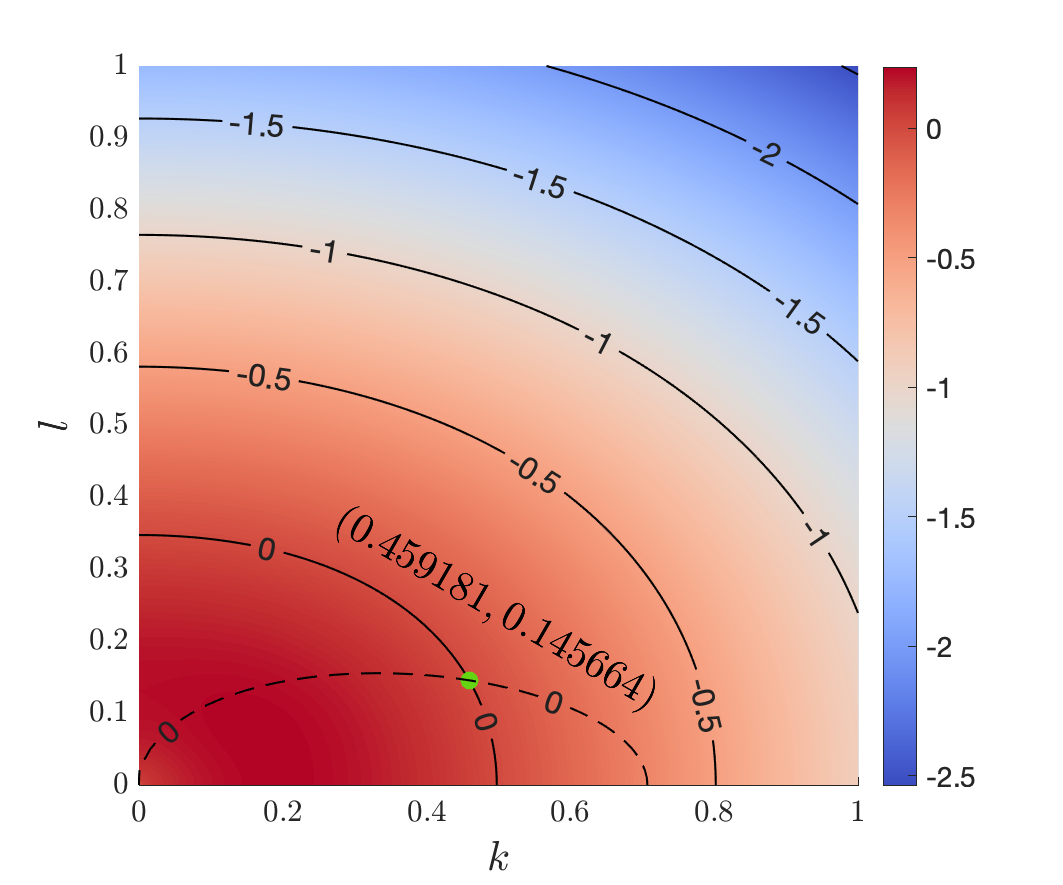}
    \caption{Values and typical level sets of the left-hand side of \eqref{resonanc2} for $\alpha = 2, \beta = 4$. The dashed line represents the zero level set of the the left-hand side of \eqref{resonance_frequency} for $\gamma = 2$. The green dot represents a particular value of $\kappa$ supporting resonant standing waves.}
    \label{fig:resonance2}
\end{figure}
In general, \eqref{resonanc2} alone is not sufficient to support resonant standing waves. The other necessary condition is
\begin{align}
    \omega(\widetilde{\boldsymbol{\kappa}}) - \gamma\omega(\boldsymbol{\kappa}) =0, \quad \gamma\in \mathbb Q,\label{resonance_frequency}
\end{align}
which means that the frequencies of the resonant triads must be rationally related. The previous two cases correspond to $ \gamma = 1$. Figure \ref{fig:resonance2} shows the left-hand side of \eqref{resonanc2} for $\alpha = 2, \beta = 4$. The dashed line, which represents the zero level set of the left hand-side of \eqref{resonance_frequency} with $\gamma = 2$, intersects the resonant curve at $k = 0.459181, l = 0.145664$ (accurate up to the sixth decimal place). Thus a particular choice supporting resonant standing waves is
\begin{align}
\begin{cases}
    \boldsymbol{\kappa} = (0.459181,0.145664),\quad \widetilde{\boldsymbol{\kappa}} = (0.918362,0.582656), \quad \breve{\boldsymbol{\kappa}} = (1.377543, 0.728320),\\
    \omega(\boldsymbol{\kappa}) = 0.770405,\quad \omega(\widetilde{\boldsymbol{\kappa}}) = 1.540811, \quad \omega(\breve{\boldsymbol{\kappa}}) = 2.311216.\label{triads}
\end{cases}
\end{align}

\begin{figure}[h!]
    \centering
    \includegraphics[width=0.42\linewidth]{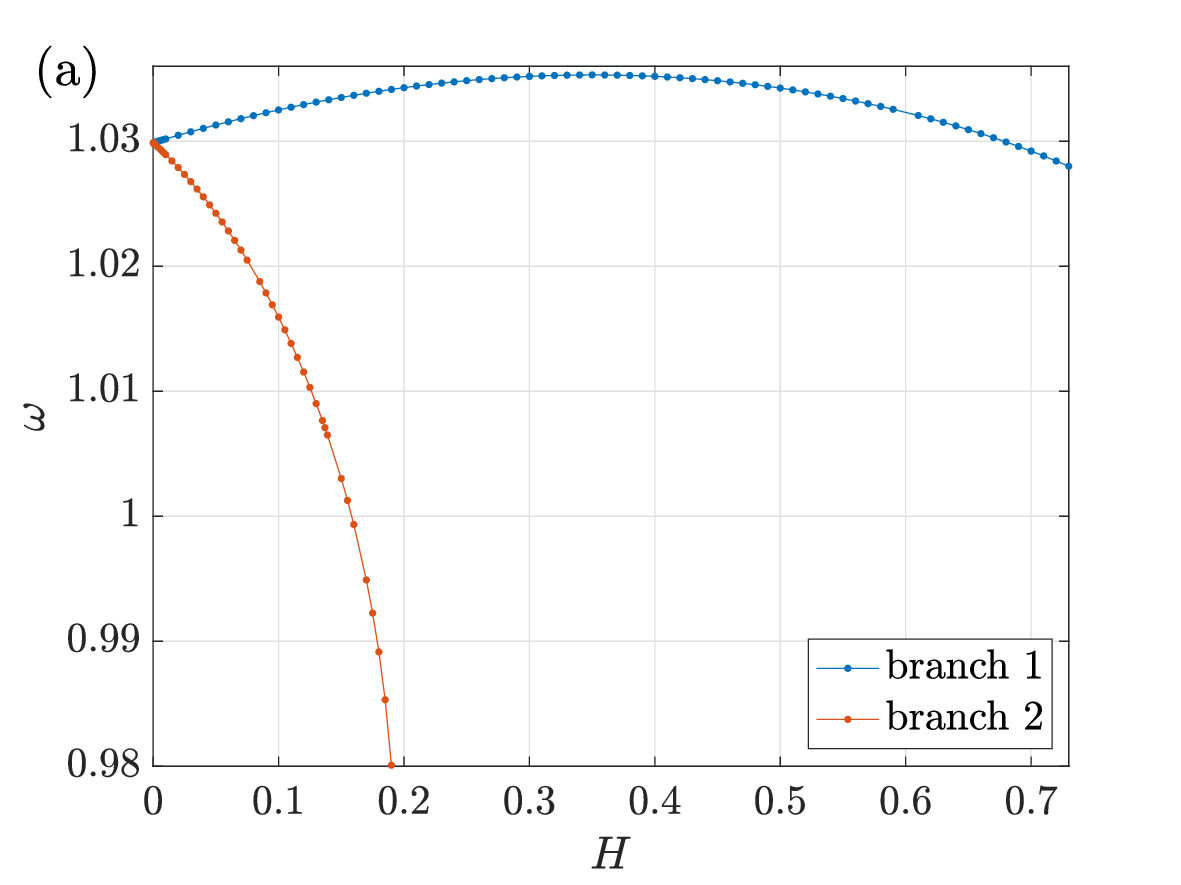}
    \includegraphics[width=0.42\linewidth]{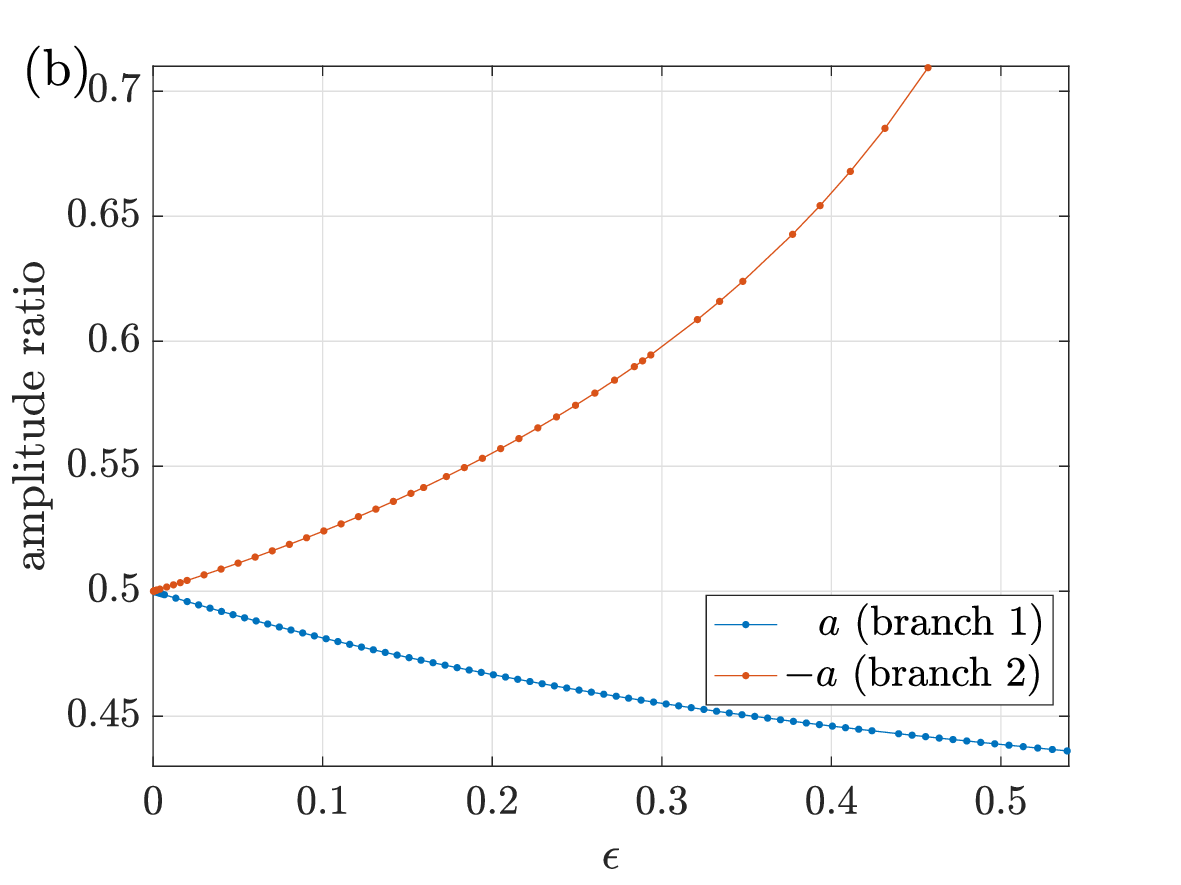}
    \caption{Two solution branches of \textit{Case II} standing waves for $k = l = 0.5$. (a) $\omega$ versus $H$. (b) Amplitude ratio versus $\epsilon$.}
    \label{fig:bif_0.5_0.5}
\end{figure}
\begin{figure}[h!]
    \centering
    \includegraphics[width=0.3\linewidth]{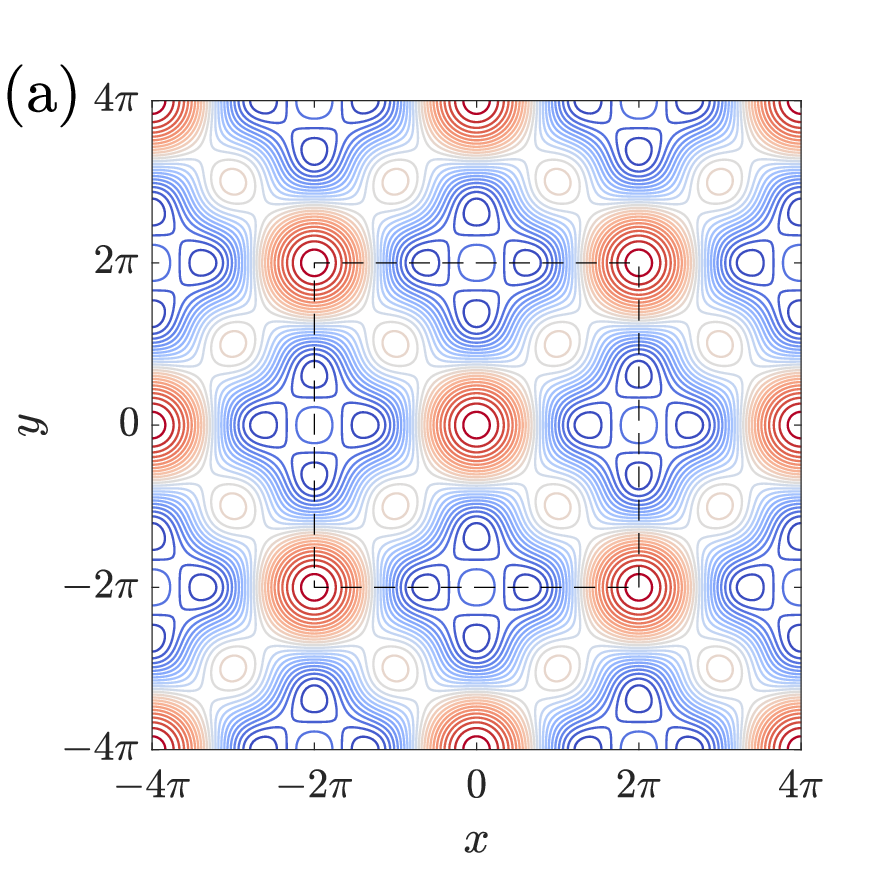}
    \includegraphics[width=0.3\linewidth]{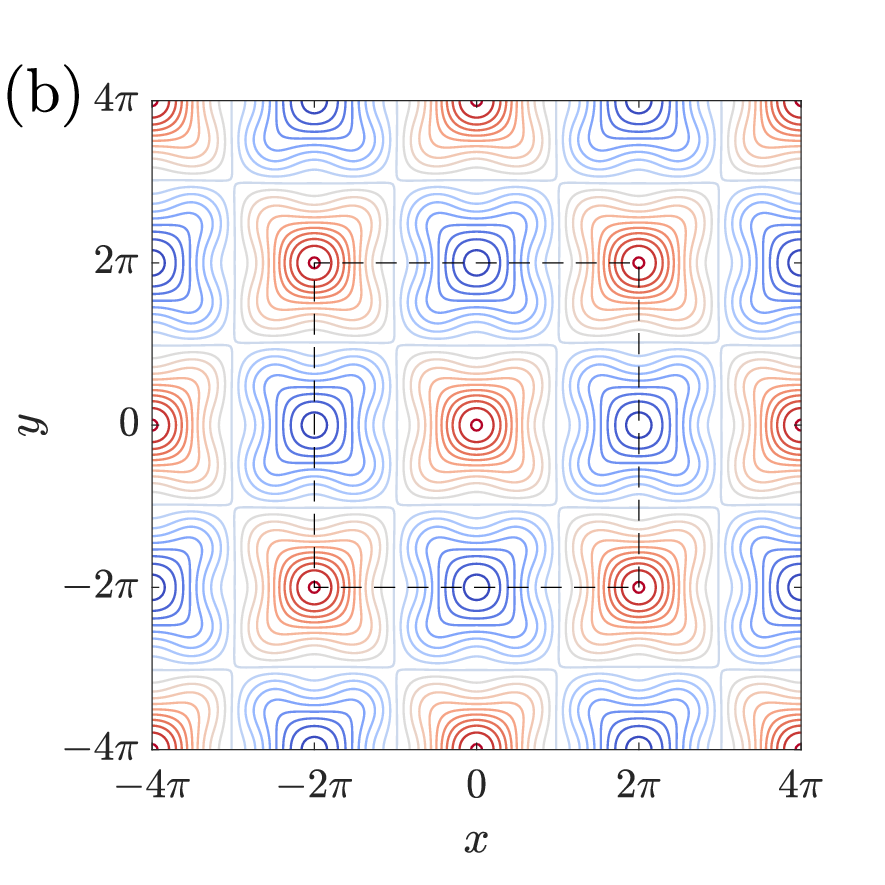}
    \includegraphics[width=0.3\linewidth]{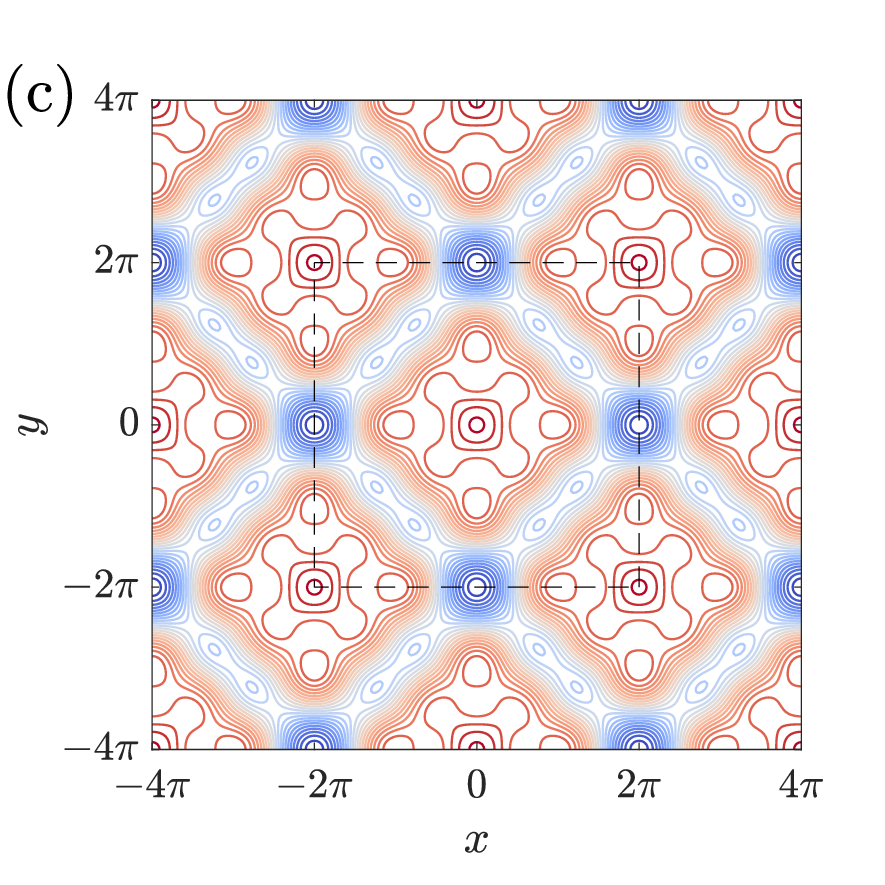}
    \includegraphics[width=0.32\linewidth]{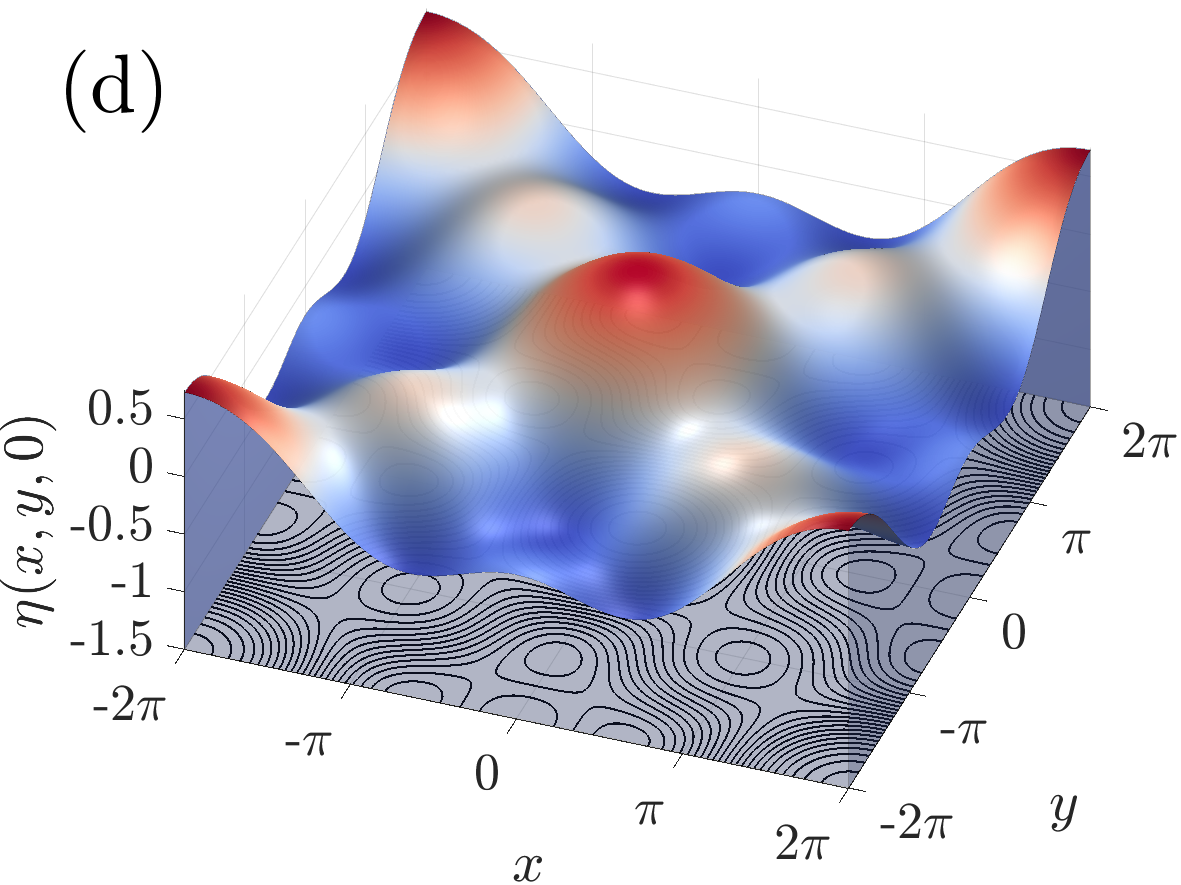}
    \includegraphics[width=0.32\linewidth]{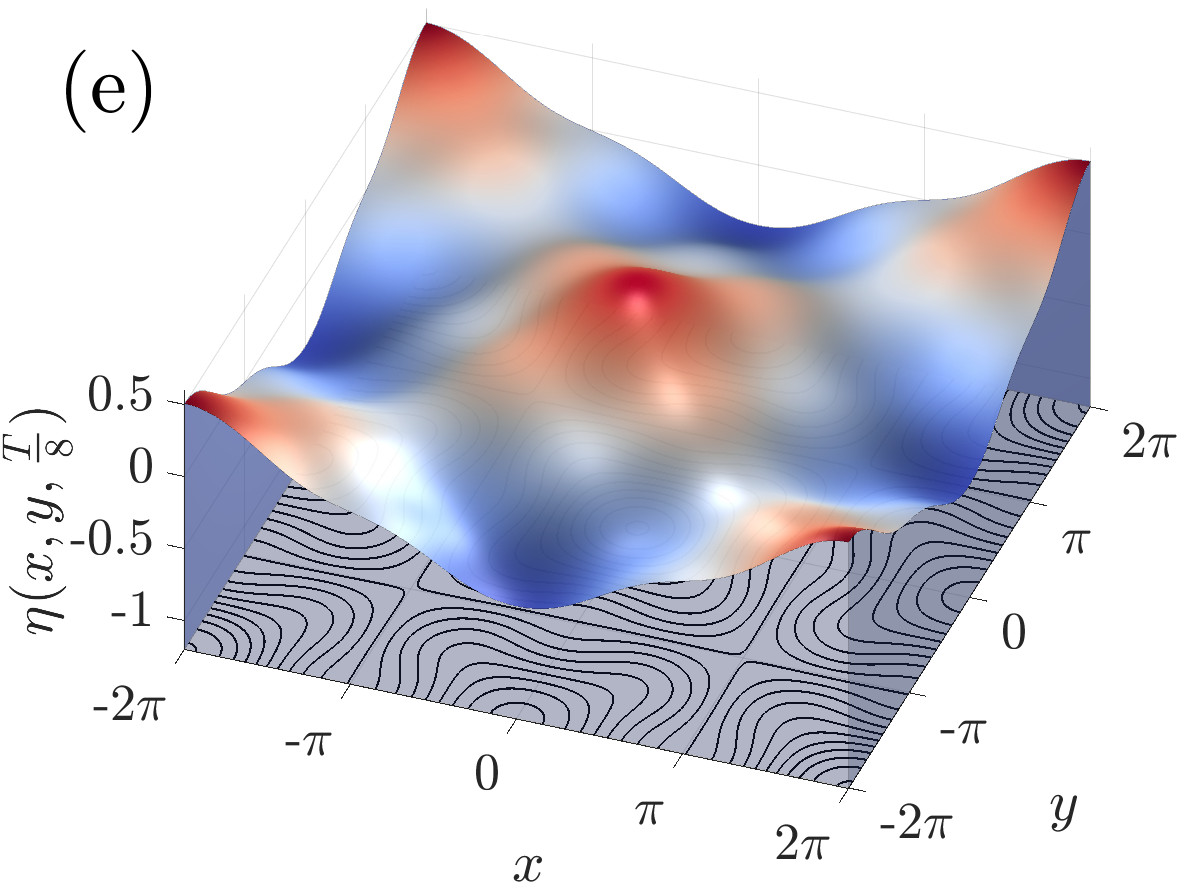}
    \includegraphics[width=0.32\linewidth]{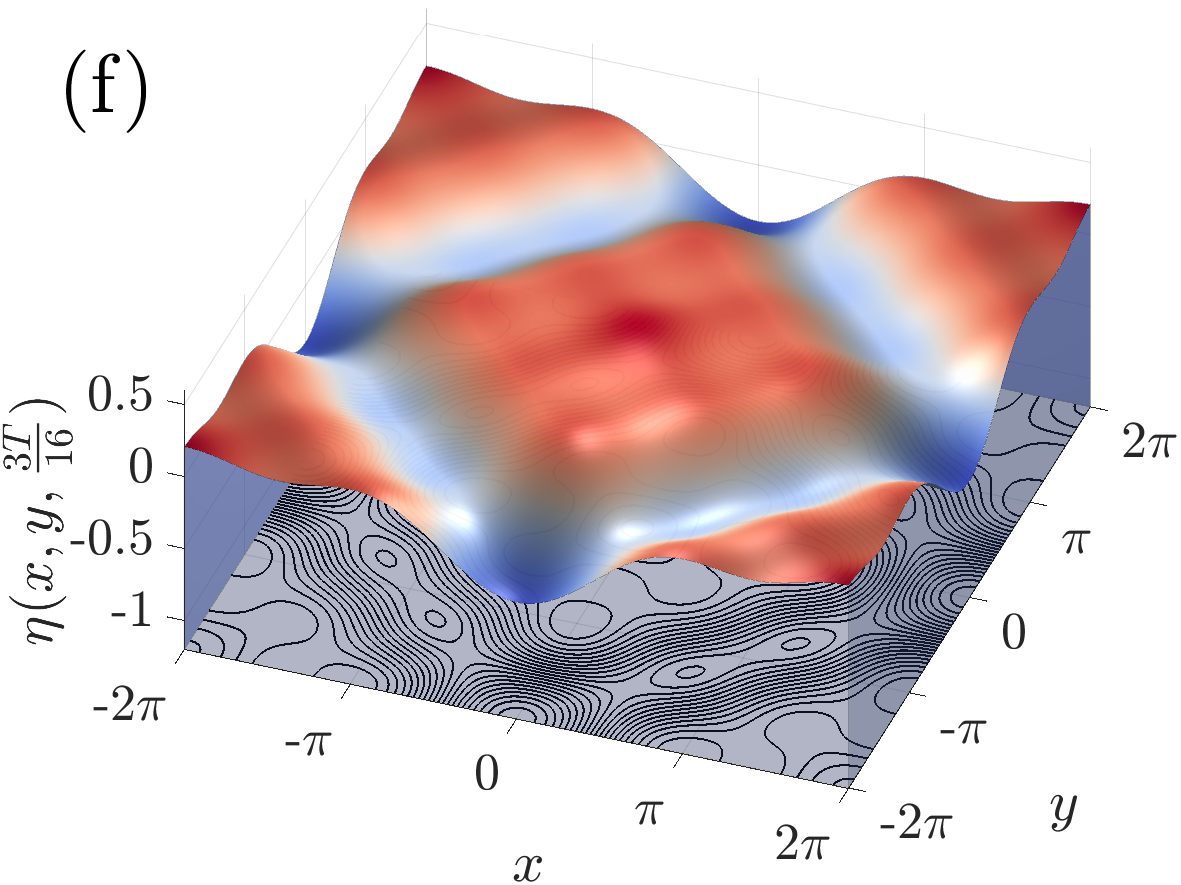}
    \caption{A \textit{Case II} standing wave with $k = l = 0.5, H=0.73$ on branch $1$. (a-c) show the top views of the solution at $t=0,T/8$ and $3T/16$. A periodic cell is surrounded by the dashed lines. The corresponding wave profiles are shown in (d) and (e).}
    \label{fig:solution5.5}
\end{figure}
\begin{figure}[h!]
    \centering
    \includegraphics[width=0.3\linewidth]{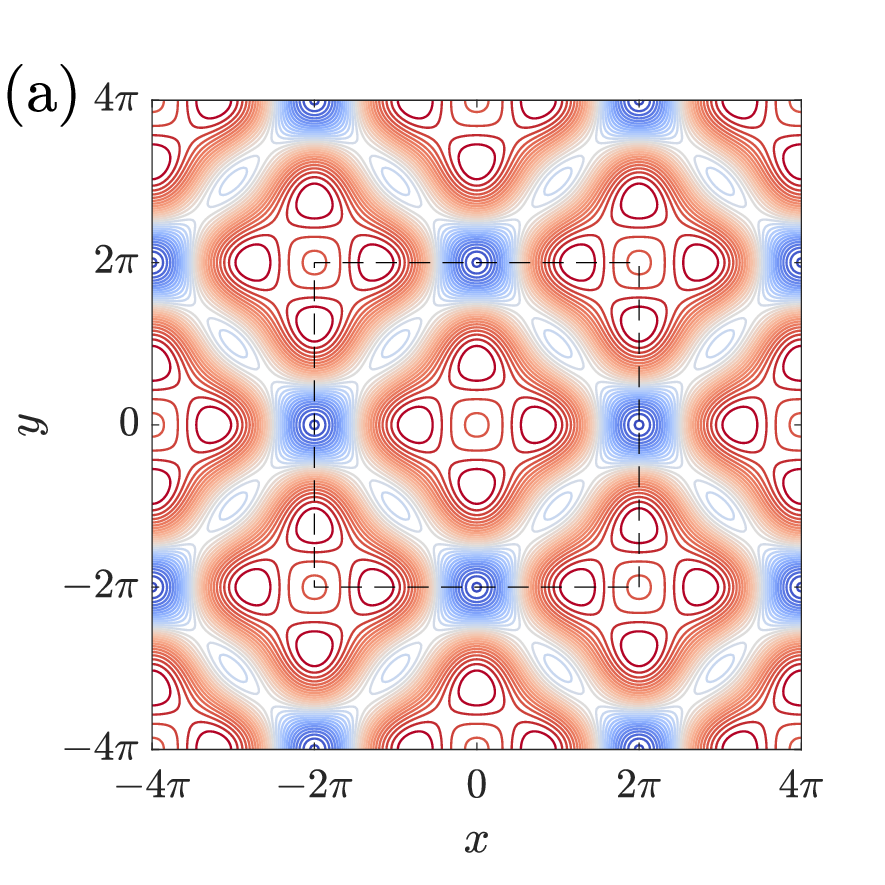}
    \includegraphics[width=0.3\linewidth]{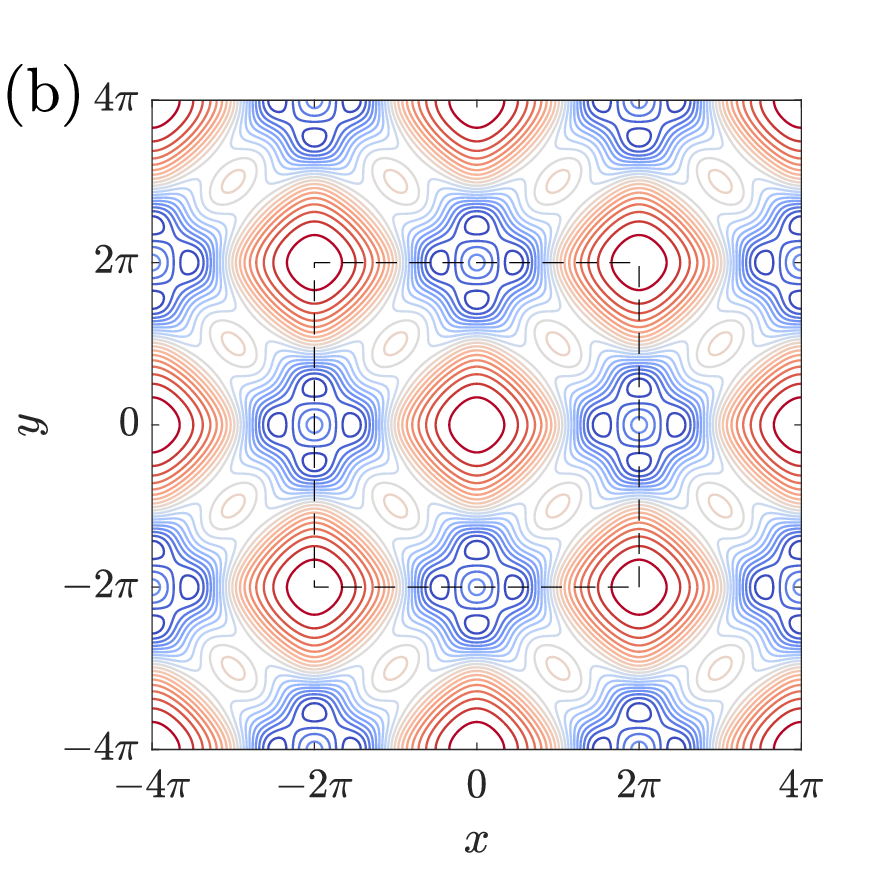}
    \includegraphics[width=0.3\linewidth]{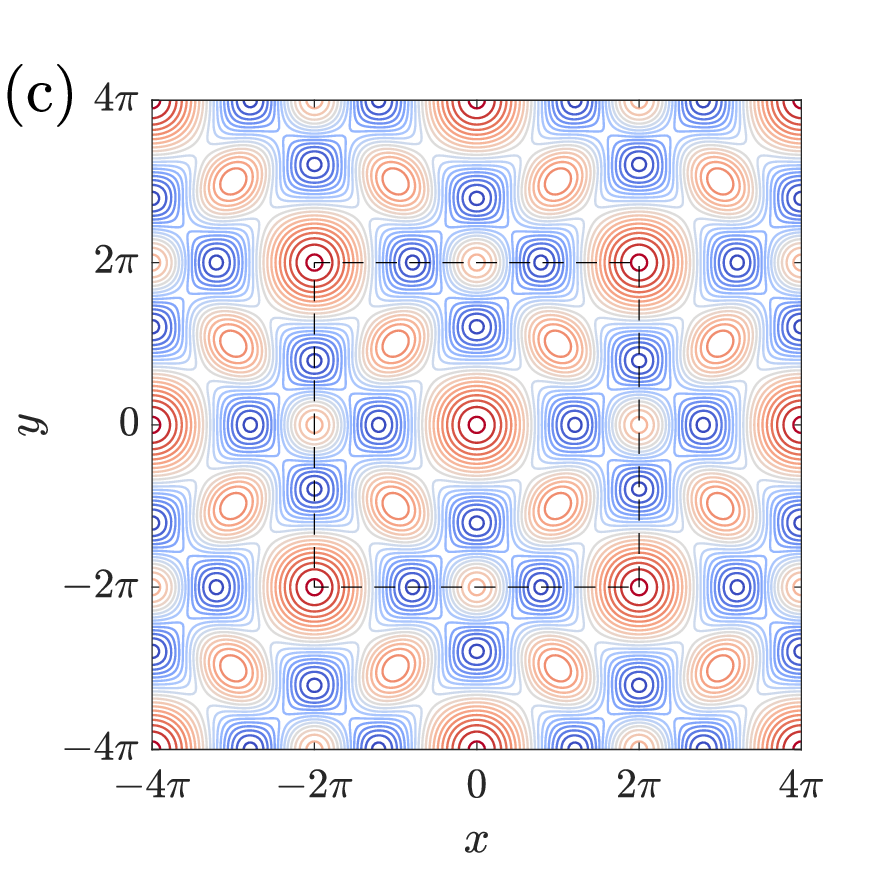}
    \includegraphics[width=0.32\linewidth]{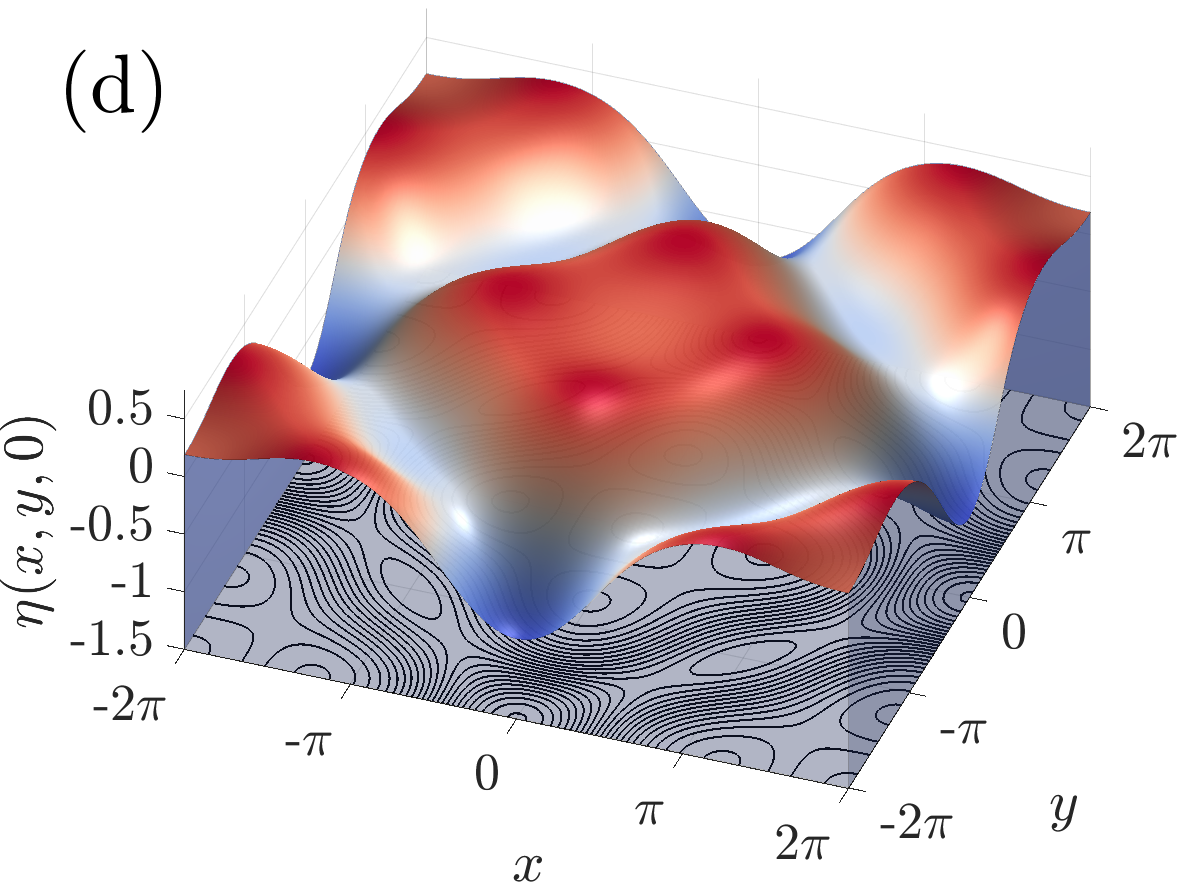}
    \includegraphics[width=0.32\linewidth]{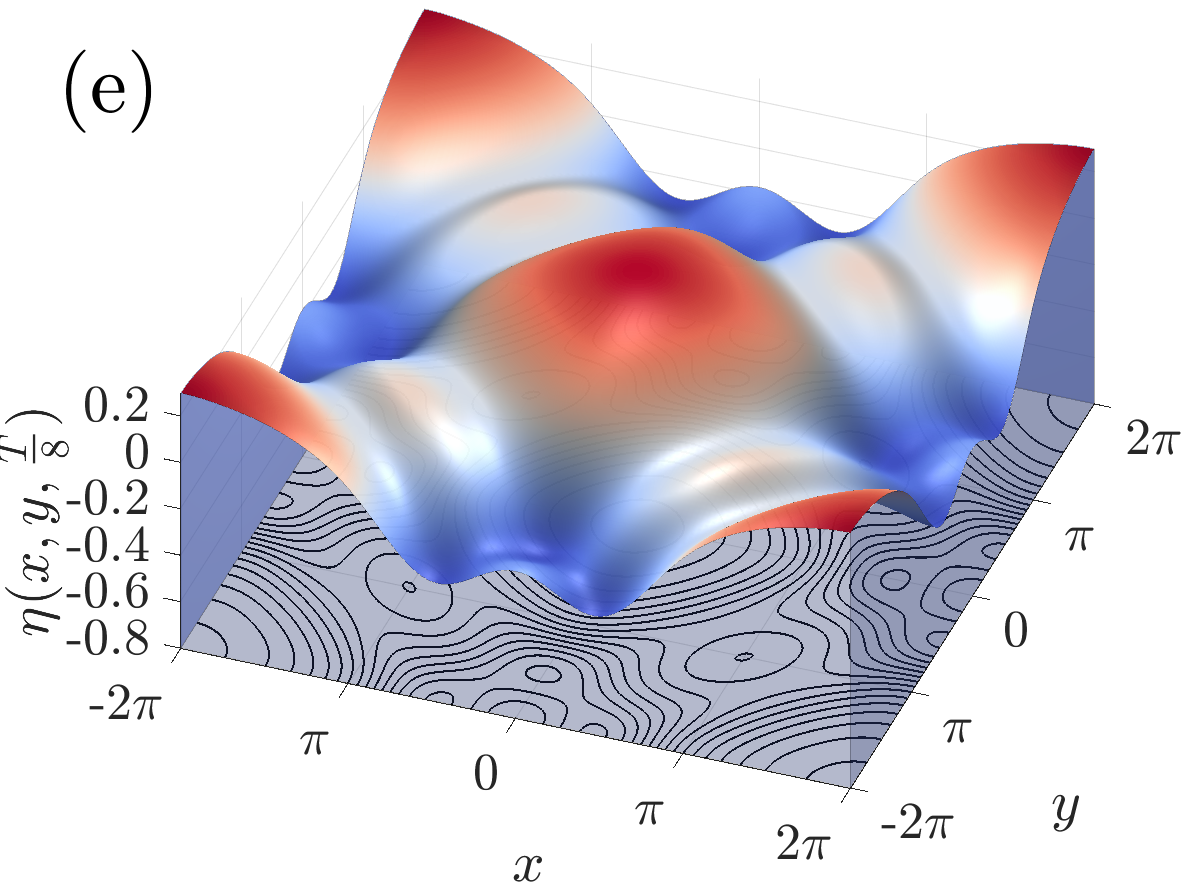}
    \includegraphics[width=0.32\linewidth]{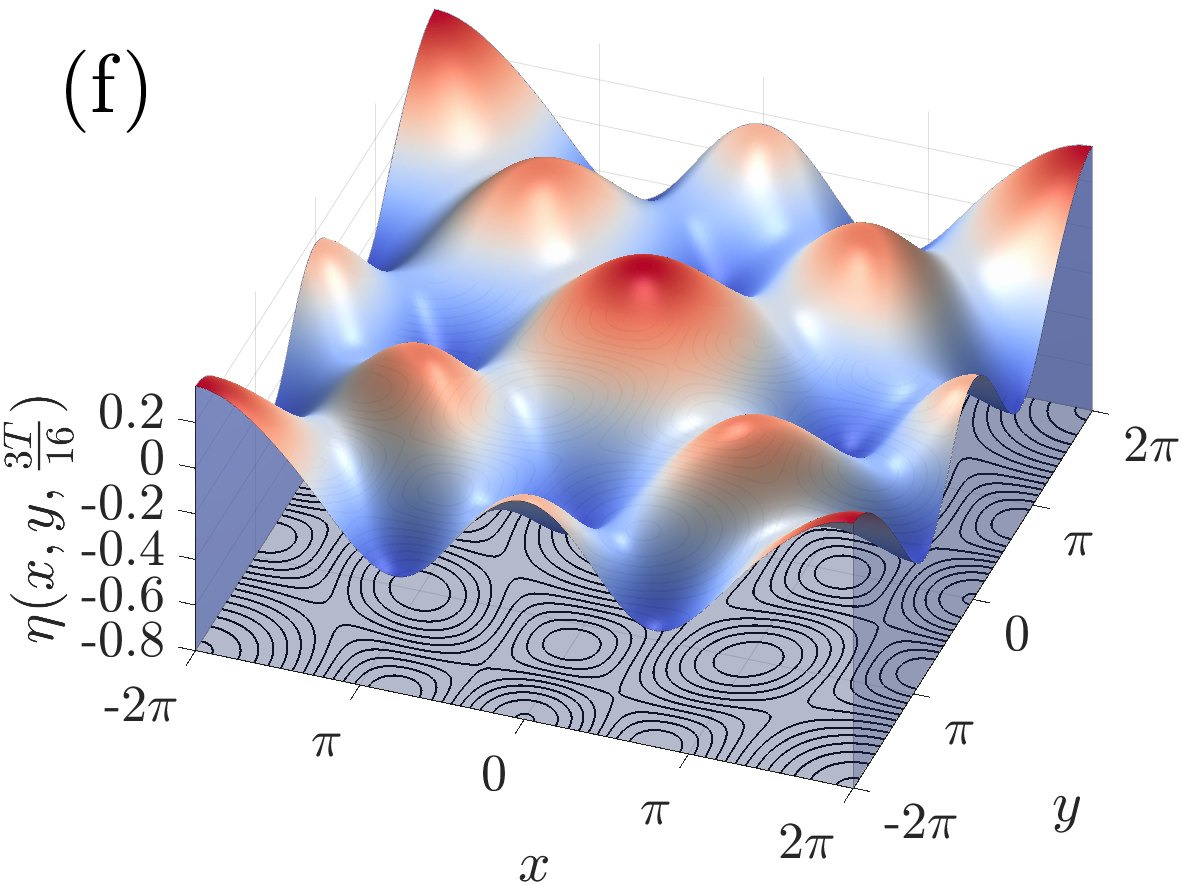}
    \caption{A \textit{Case II} standing wave with $k = l = 0.5, H=0.19$ on branch $2$. (a-c) show the top views of the solution at $t=0,T/8$ and $3T/16$. A periodic cell is surrounded by the dashed lines. The corresponding wave profiles are shown in (d) and (e).}
    \label{fig:solution5.55}
\end{figure}
\begin{figure}[h!]
    \centering
    \includegraphics[width=0.45\linewidth]{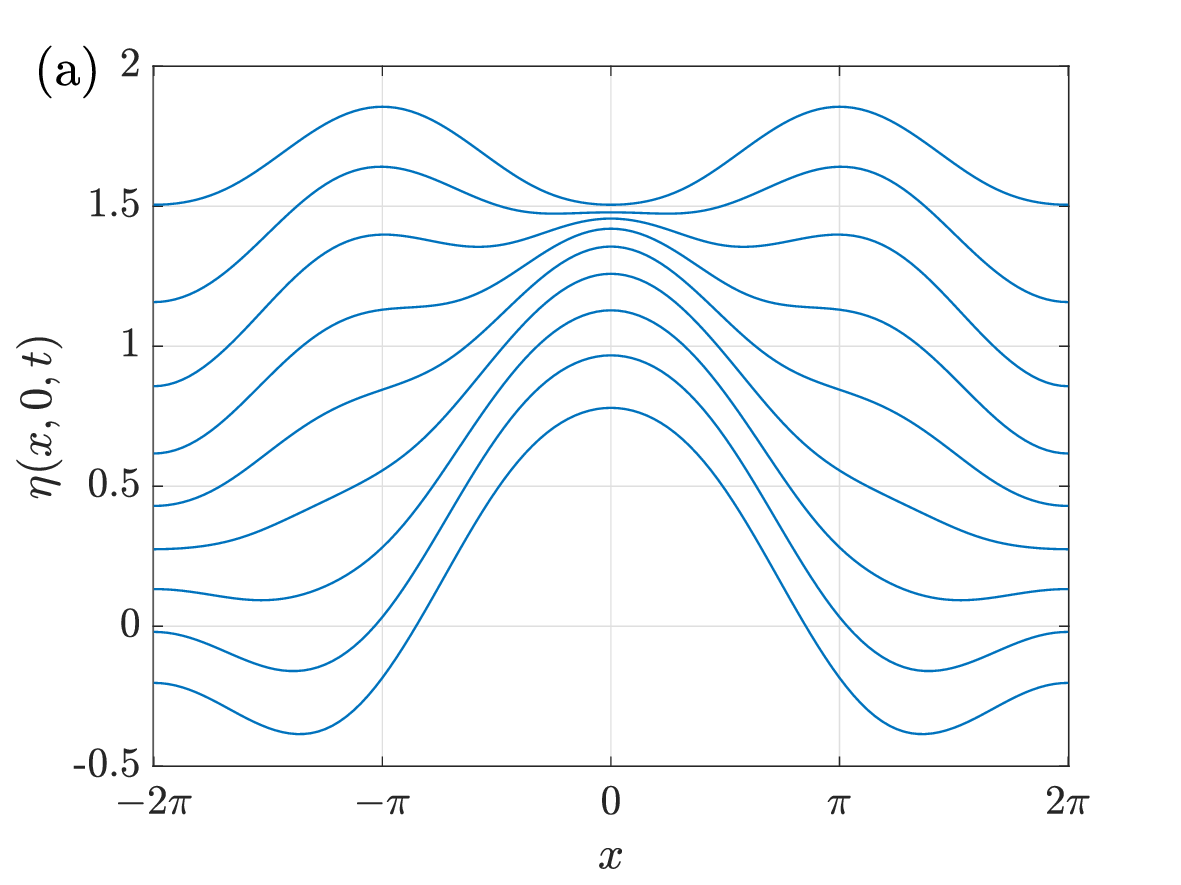}
    \includegraphics[width=0.45\linewidth]{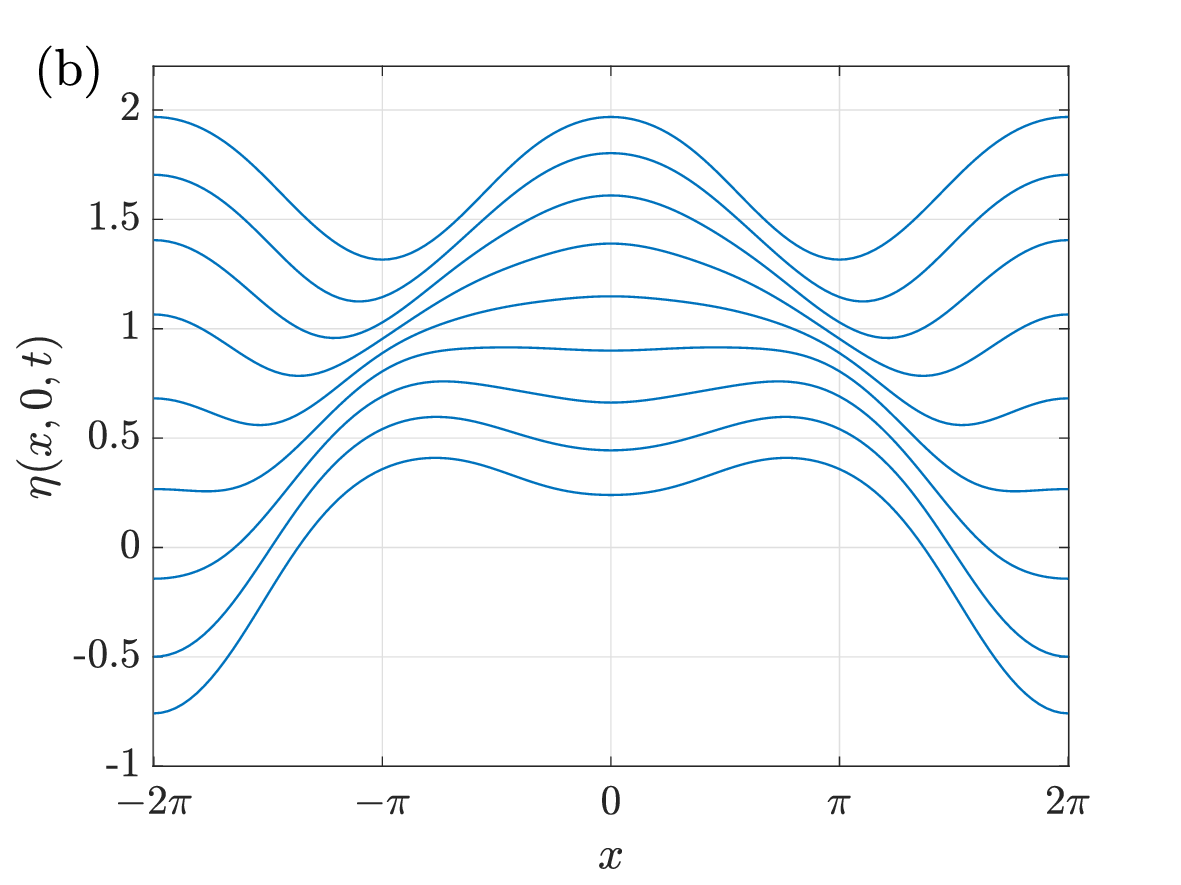}
    \caption{Cross-sections on the plane $y=0$. (a) The branch-$1$ solution with $H=0.73$. (b) The branch-$2$ solution with $H=0.19$. From bottom to top, $t = 0, T/32,T/16,\cdots$,$T/4$.}
    \label{fig:solution5.6}
\end{figure}
We first examine the standing waves with $k=l=0.5$. Figure \ref{fig:bif_0.5_0.5} shows two fundamental solution branches (labelled by $1$ and $2$) bifurcating from infinitesimal linear solutions at the frequency $\omega=1.0299$. For branch $2$, $\omega$ monotonically decreases with $H$. For branch $1$, $\omega$ initially grows for $H < 0.35$, and subsequently decreases with $H$. At small amplitude, $\eta$ can be written as
\begin{align}
    \eta \sim \epsilon \cos(0.5 x)\cos(0.5 y)\cos(\omega t) + a\epsilon \cos(x)\cos(y)\cos(2\omega t) + O(\epsilon^2),\label{stokes1}
\end{align}
where $a$ measures the relative importance of the two resonant modes, thus influencing the wave morphology. Panel (b) shows that $a$ takes opposite signs on the two branches and tends to $\pm 0.5$ as $\epsilon\rightarrow 0$. Figure \ref{fig:solution5.5} displays a representative standing wave with $H=0.73 (\epsilon=0.4361),\omega = 1.02800$ on branch $1$. Panels (a-c) show the top views of the solution at $t = 0,T/8$ and $3T/16$, exhibiting waffle-shaped patterns clearly. The corresponding wave profiles within a periodic cell (the region surrounded by the dashed lines) are shown in panels (d-f). The solution is calculated using $128\times128\times128$ grid points. Similarly, figure \ref{fig:solution5.55} shows the top views and wave profiles of a branch-$2$ standing wave with $H=0.135 (\epsilon=0.7095),\omega = 0.98008$ at $t = 0,T/8$ and $3T/16$. Figure \ref{fig:solution5.6} (a) and (b) exhibit the cross-sections of the two standing waves over a quarter temporal period, resembling the classical Wilton ripples when the fundamental and second-order Fourier components are resonant.

\begin{figure}[h!]
    \centering
    \includegraphics[width=0.42\linewidth]{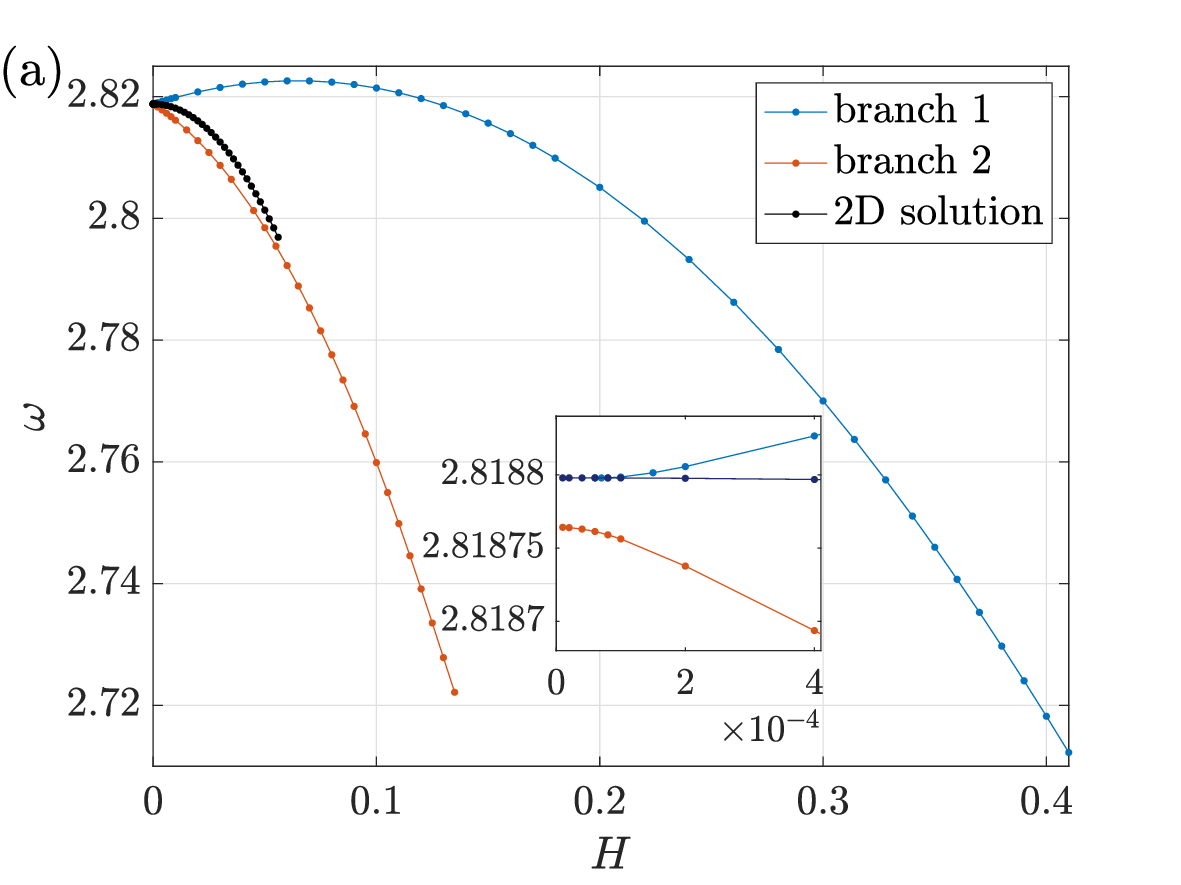}
    \includegraphics[width=0.42\linewidth]{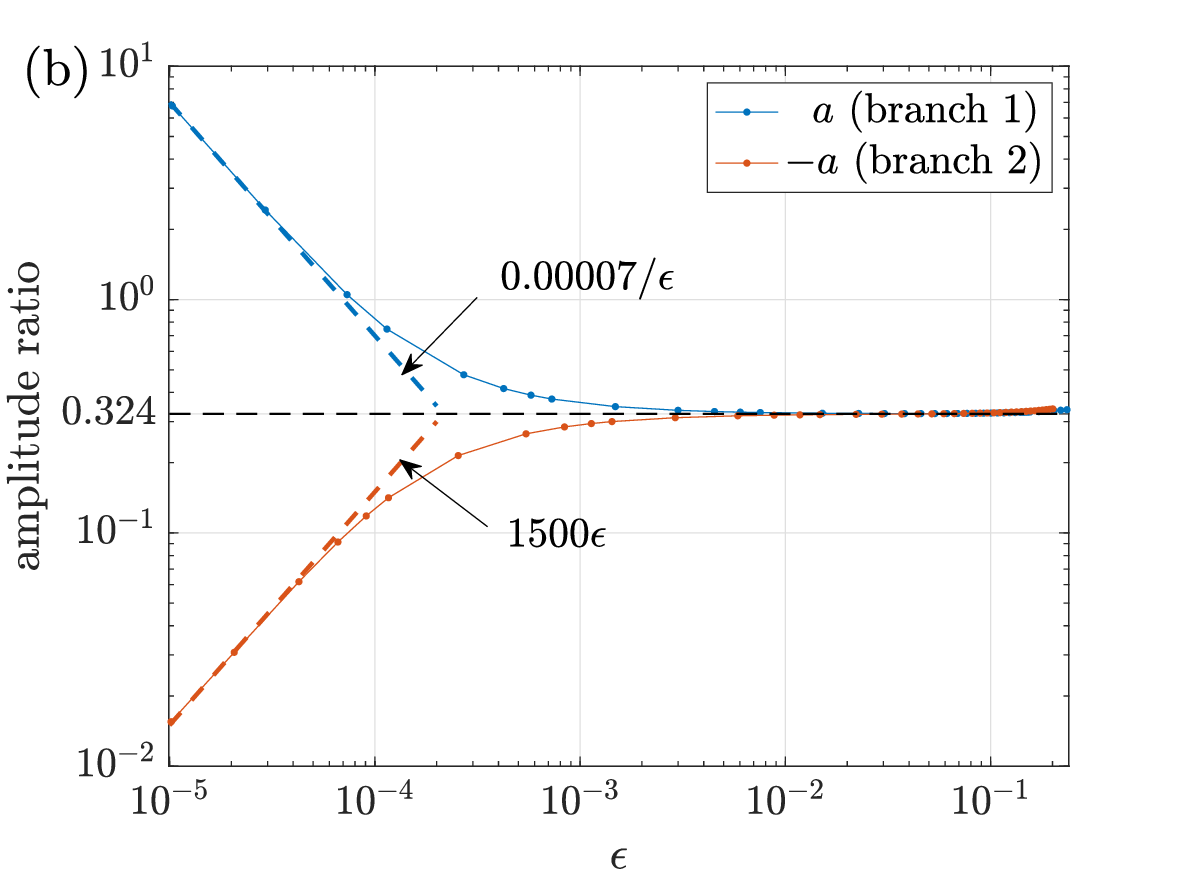}
    \caption{Bifurcations of \textit{Case I} standing waves for $k = 1,l = 1.5312$. (a) $\omega$ versus $H$. The blue and red curves correspond to three-dimensional solution branches, and the black curve represents the two-dimensional solution branch for $k = 0,l = 3.0624$. (b) Amplitude ratio versus $\epsilon$. The blue and red dashed lines represent two fitted analytic relations for small $\epsilon$.}
    \label{fig:bif2}
\end{figure}

Next we examine the standing waves generated by non-collinear resonant triads. Figure \ref{fig:bif2} shows the two solution branches (also labelled by $1$ and $2$) for the \textit{Case I} standing waves with $k = 1,l = 1.5312$. Branch $1$ bifurcates from a two-dimensional standing wave with $k=0,l=3.0624$ at $H = 7\times 10^{-5}, \omega = 2.81880$, while branch $2$ bifurcates from an infinitesimal three-dimensional standing wave at the linear frequency $\omega = 2.81876$. The small frequency gap between the two solution branches at $H=0$ is a consequence of the slight deviation at the fourth decimal place of $l$ from the exact value supporting resonance. The two frequency curves are similar to those shown in figure \ref{fig:bif_0.5_0.5} (a): one branch exhibits a monotonic decrease of frequency with wave amplitude, while the other one displays a slight frequency upshift followed by a subsequent downshift. Following \eqref{stokes1}, we write $\eta$ as
\begin{align}
    \eta \sim \epsilon \cos(x)\cos(1.5312 y)\cos(\omega t) + a\epsilon \cos(3.0624 y)\cos(2\omega t) + O(\epsilon^2).\label{stokes2}
\end{align}
Panel (b) shows that $a$ takes opposite signs on the two solution branches and the two amplitude-ratio curves are approximately symmetric in the $\log$-scale. For $\epsilon>10^{-3}$, $|a|$ stabilises to a nearly constant level of $\pm0.324$. Because of the different bifurcation mechanisms of the two branches, their amplitude ratios exhibit totally different behaviours when $\epsilon<10^{-4}$: for branch $2$, $a \sim 7\times  10^{-5}/\epsilon$, while for branch $1$, $a \sim-0.324^2 \epsilon/(7\times 10^{-5})\approx -1500\epsilon$, which is implied by the symmetry of the two curves.

\begin{figure}[h!]
    \centering
    \includegraphics[width=0.42\linewidth]{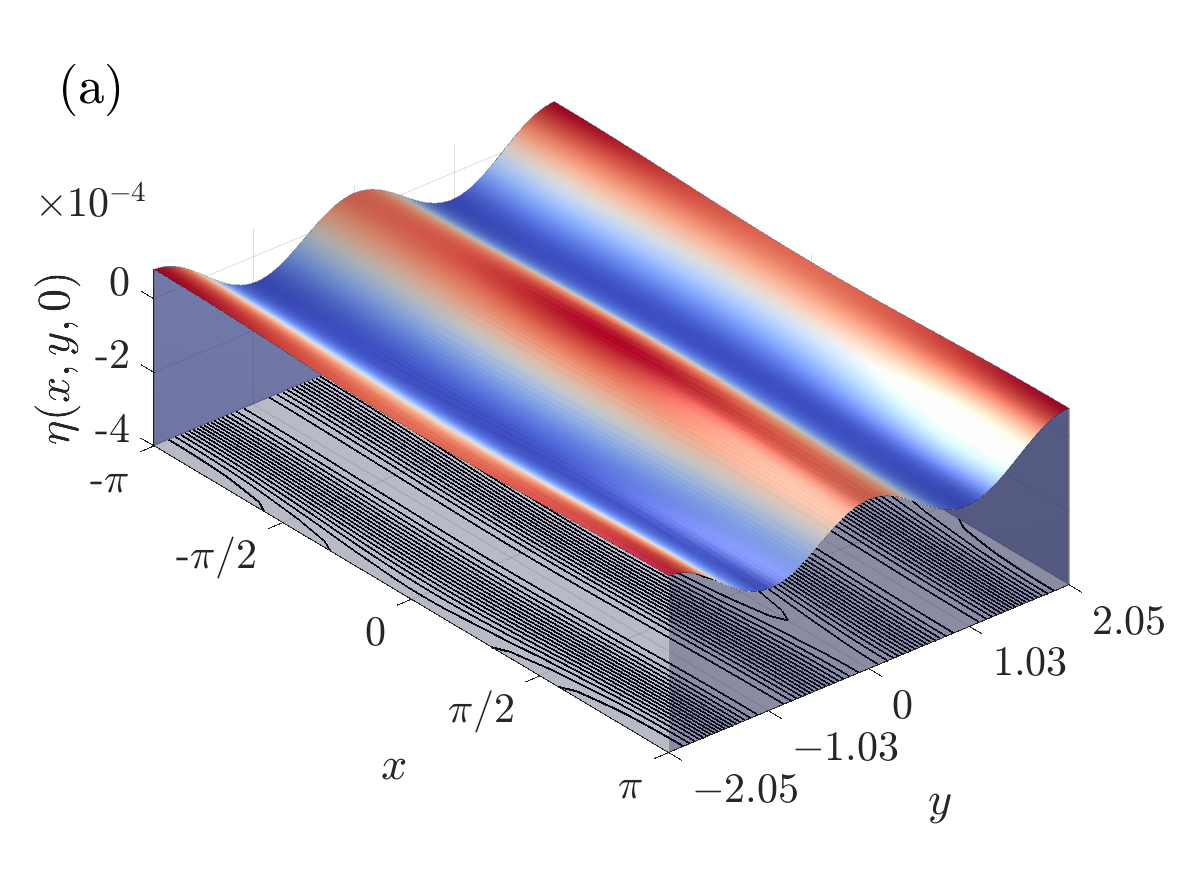}
    \includegraphics[width=0.42\linewidth]{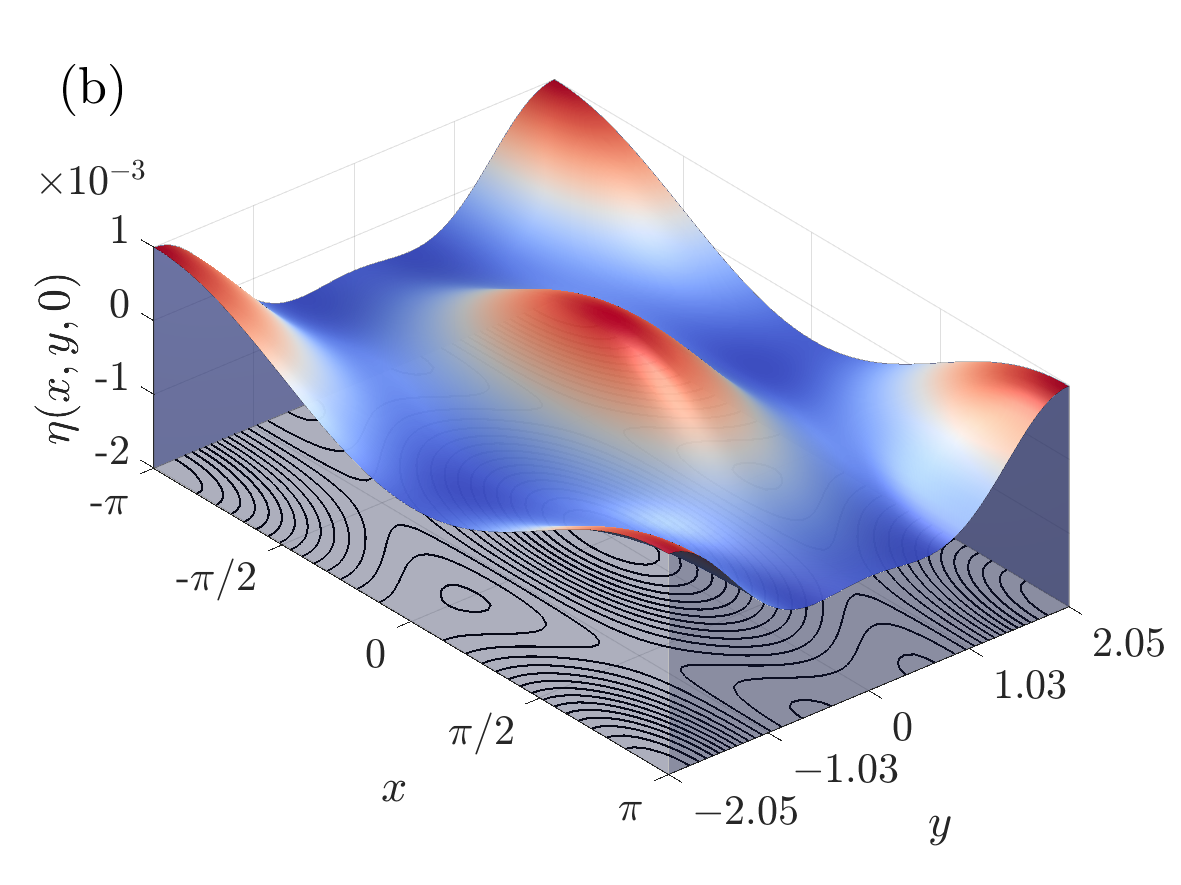}
    \caption{Profiles of small-amplitude \textit{Case I} standing waves for $k = 1$ and $l = 1.5312$ on branch $1$. (a) $H = 8\times10^{-5}$. (b) $H = 10^{-3}$.}
    \label{fig:solution6}
\end{figure}
Figure \ref{fig:solution6} shows the initial profiles of two small-amplitude branch-$1$ solutions with $H=8\times 10^{-5}$ in (a), and with $H=0.001$ in (b). The former is close to the bifurcation point, thus featuring a nearly two-dimensional shape, while the latter develops a three-dimensional form consisting of both resonant modes. 
\begin{figure}[h!]
    \centering
    \includegraphics[width=0.3\linewidth]{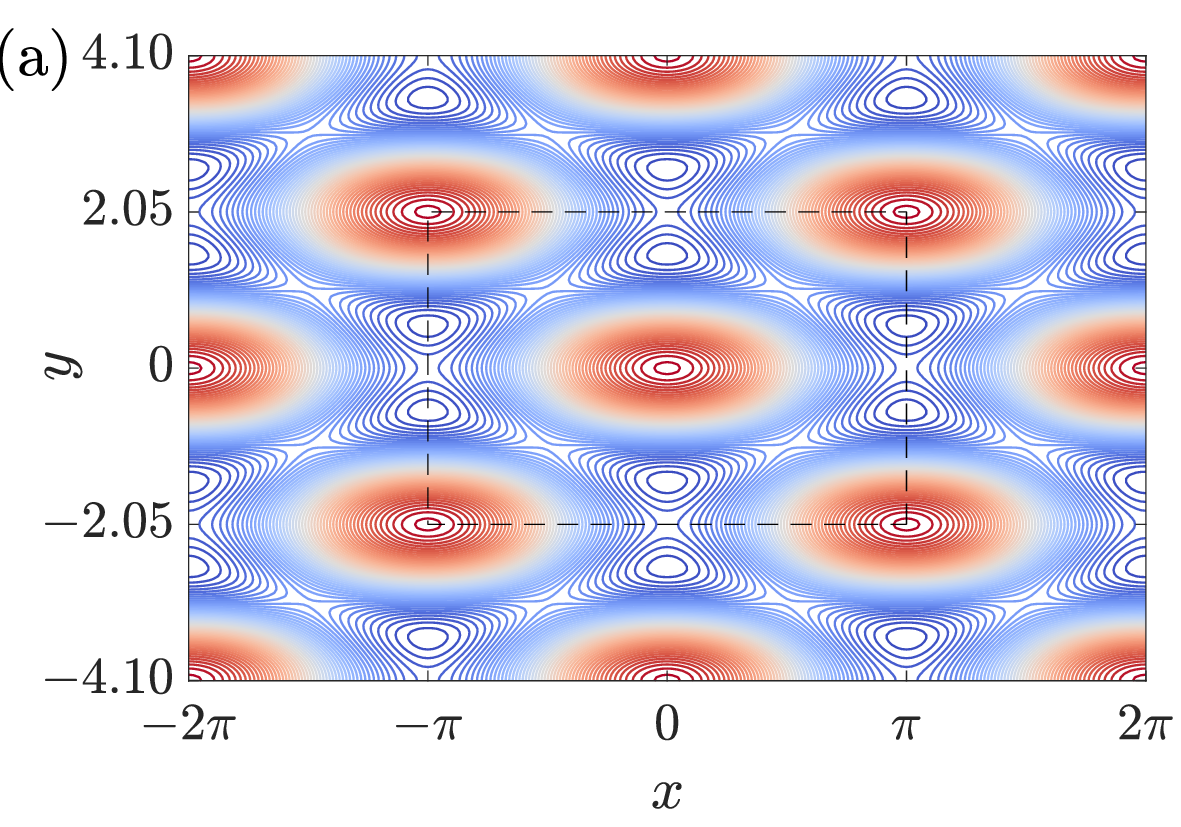}
    \includegraphics[width=0.3\linewidth]{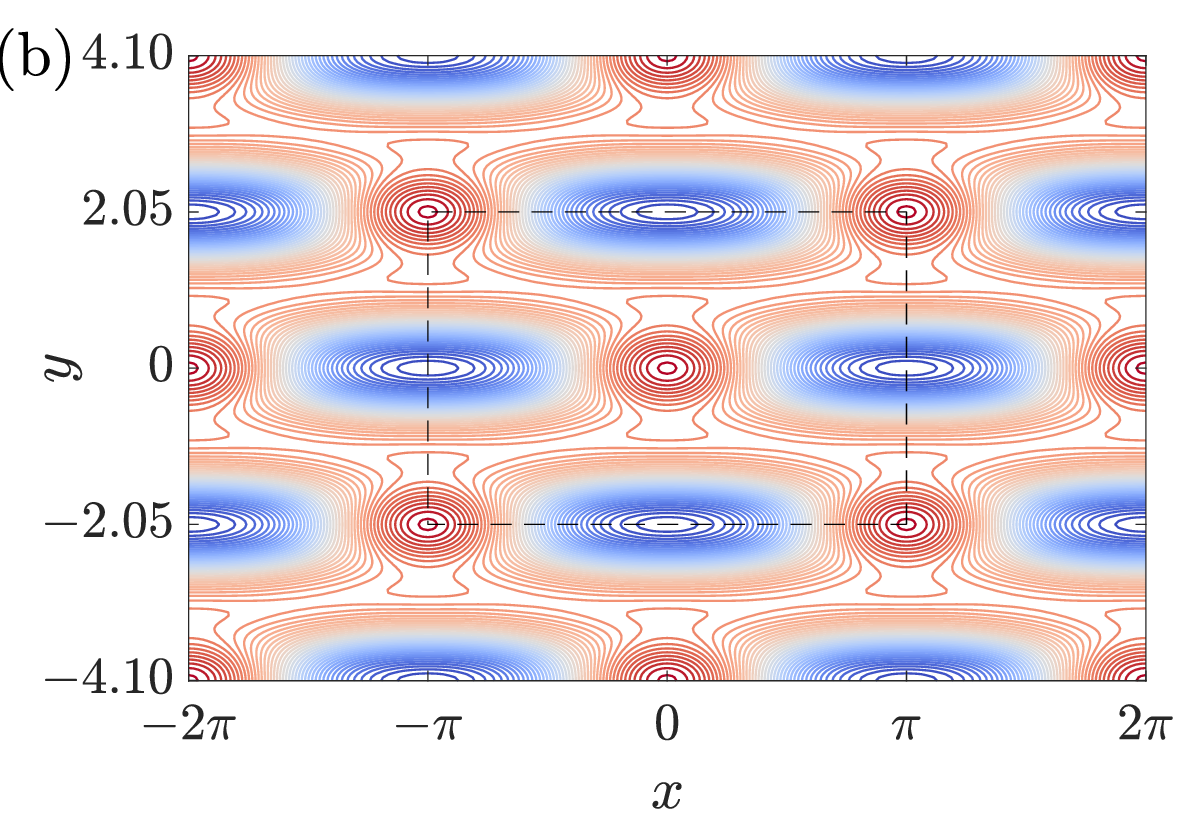}
    \includegraphics[width=0.3\linewidth]{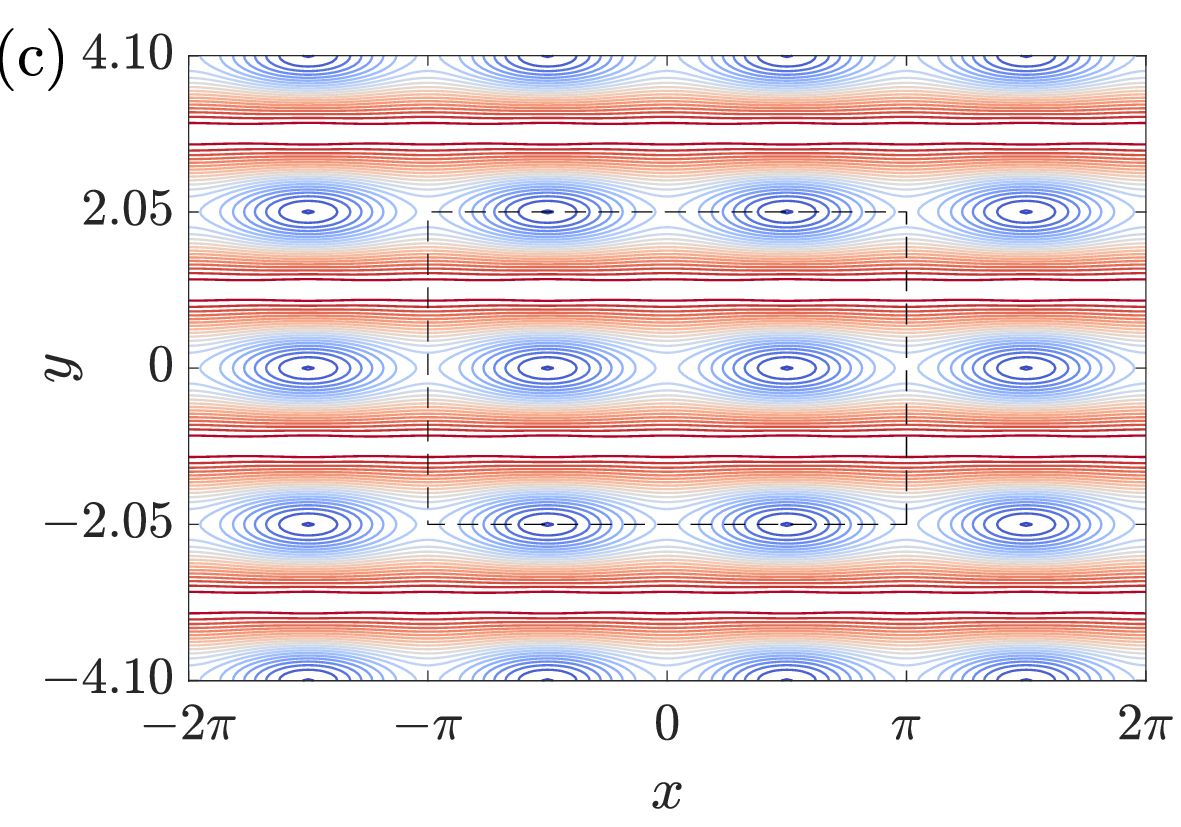}

    \includegraphics[width=0.32\linewidth]{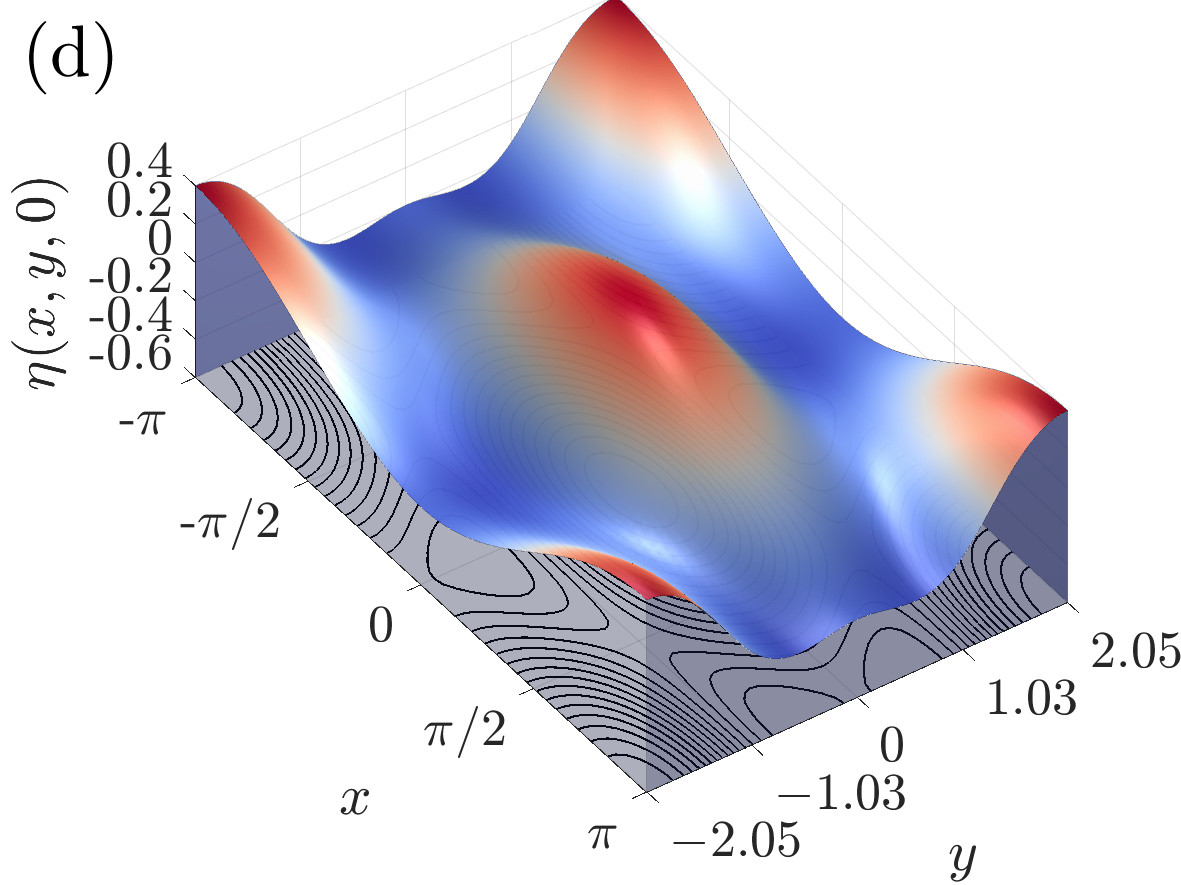}
    \includegraphics[width=0.32\linewidth]{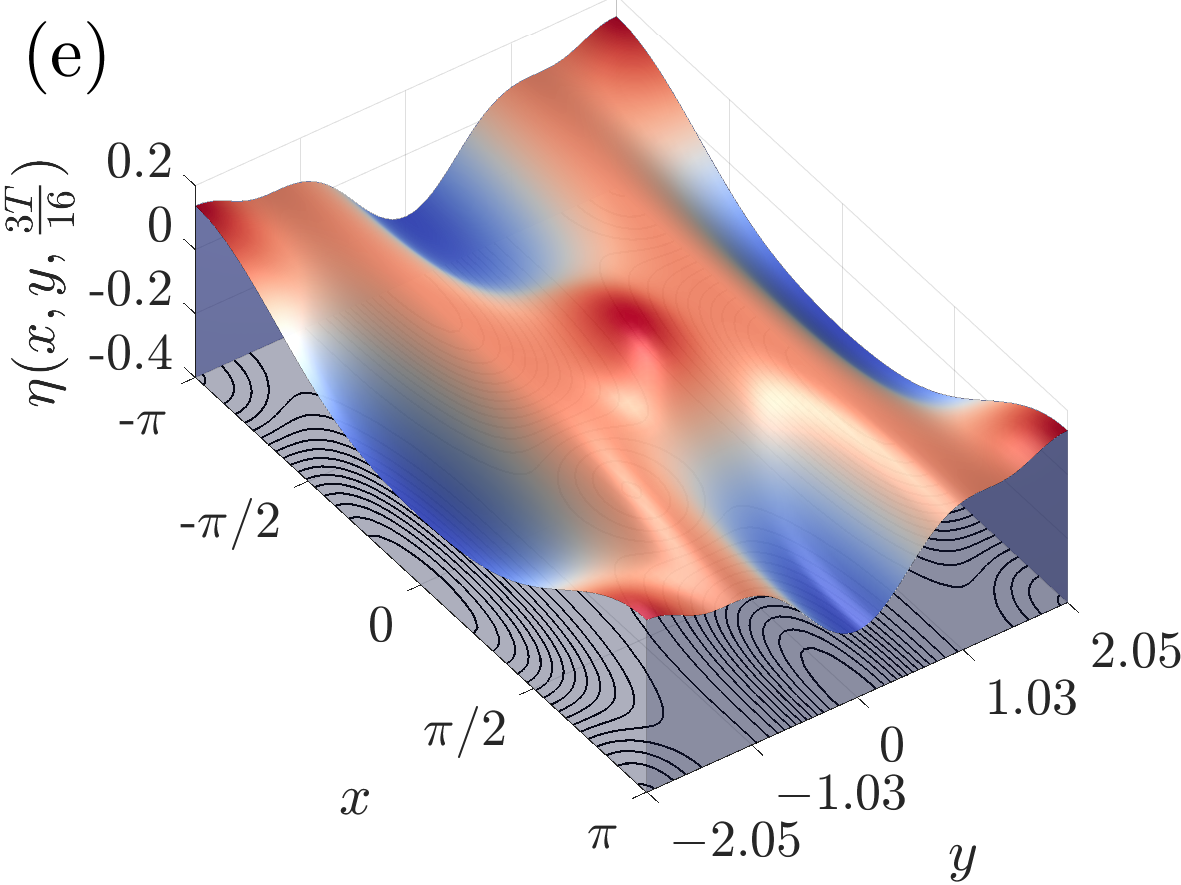}
    \includegraphics[width=0.32\linewidth]{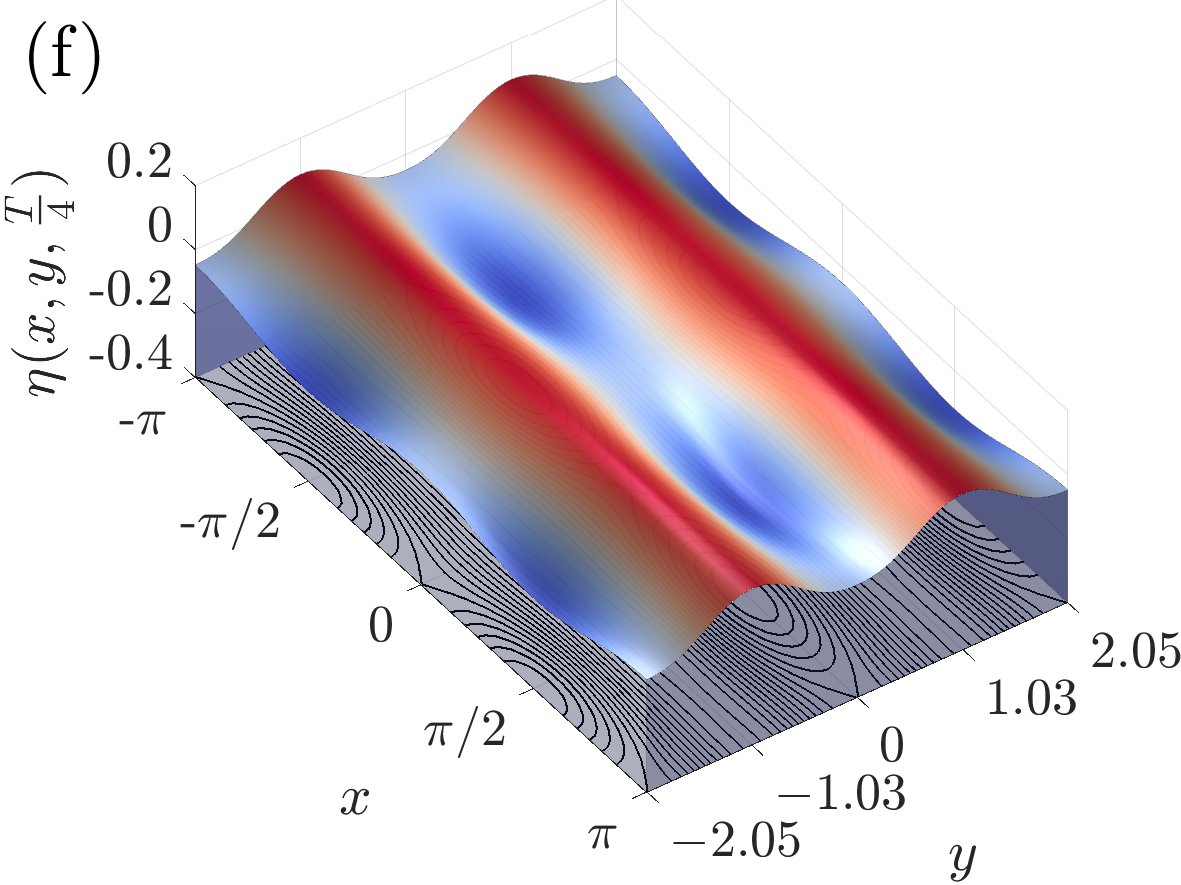}
    \caption{A standing wave with $k = 1,l = 1.5312, H=0.4$ on branch $1$. (a-c) show the top views at $t=0,3T/16$ and $T/4$. The corresponding profiles within the periodic cell surrounded by the dashed lines are shown in (d) and (e).}
    \label{fig:solution6.5}
\end{figure}
\begin{figure}[h!]
    \centering
    \includegraphics[width=0.3\linewidth]{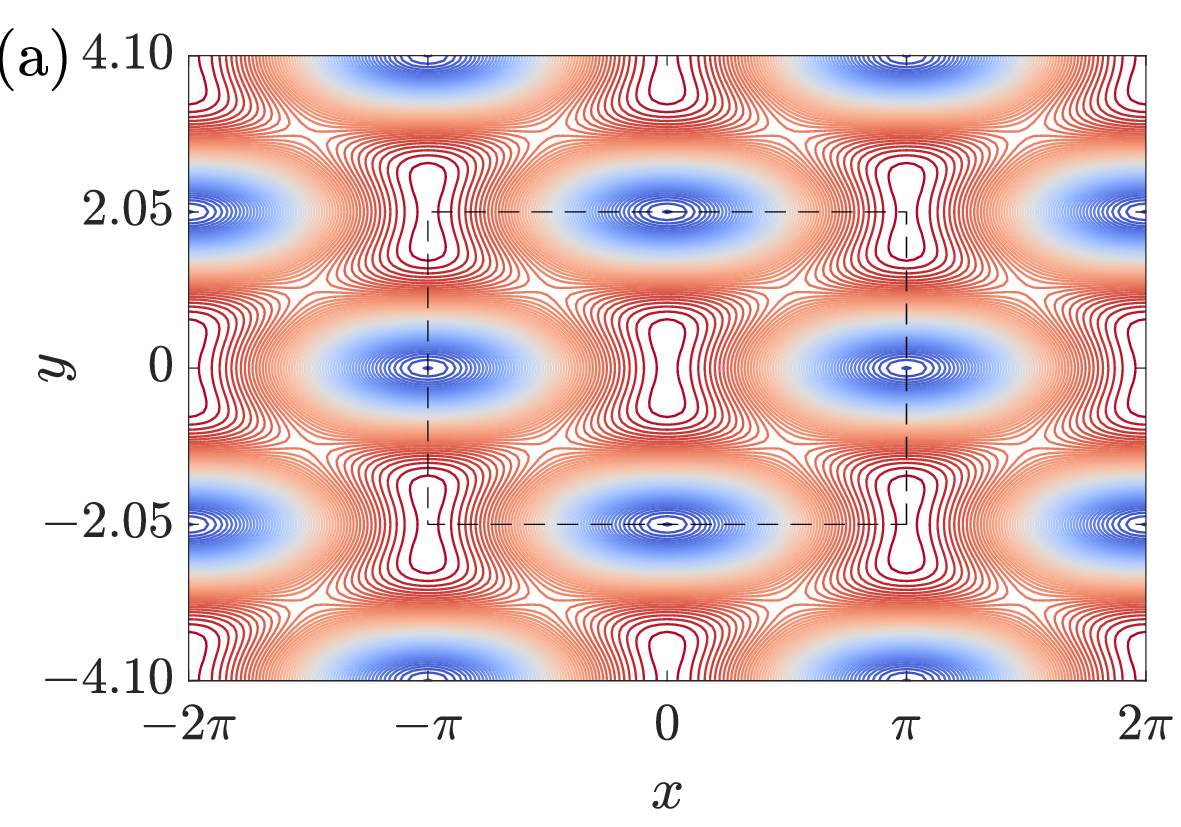}
    \includegraphics[width=0.3\linewidth]{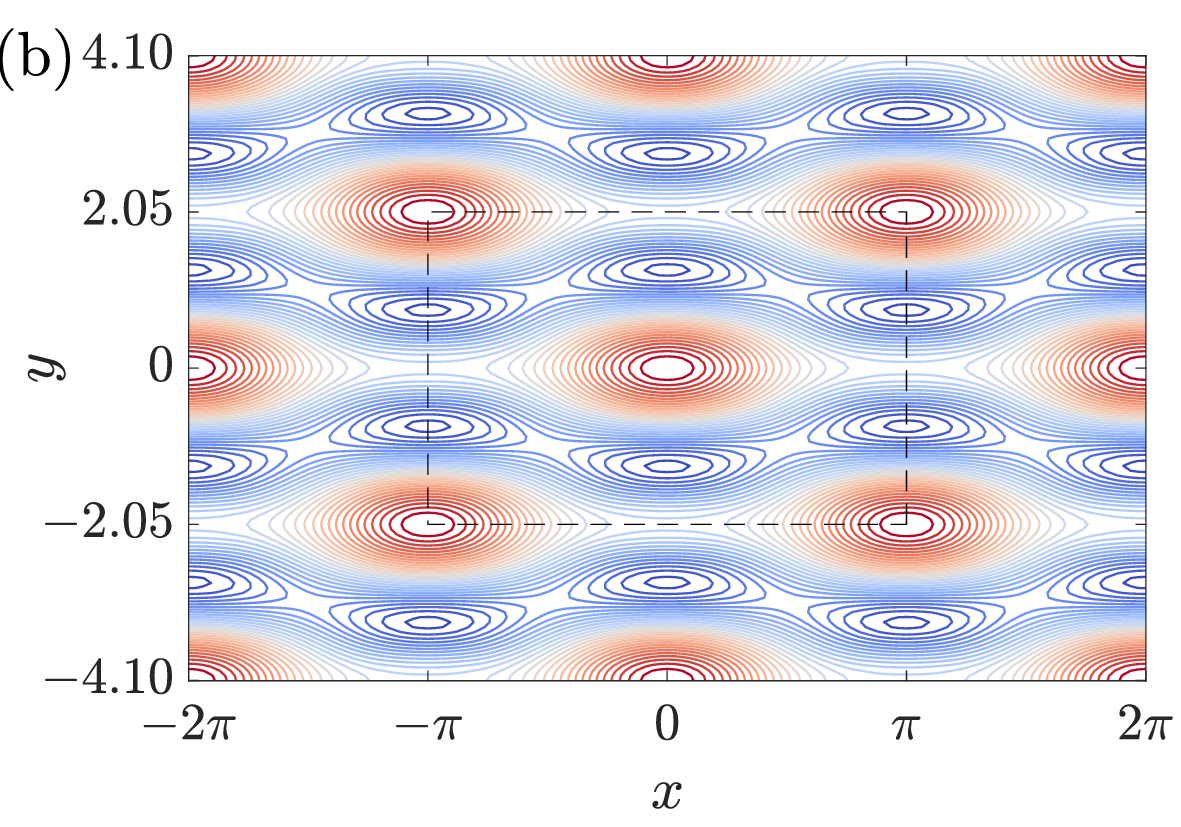}
    \includegraphics[width=0.3\linewidth]{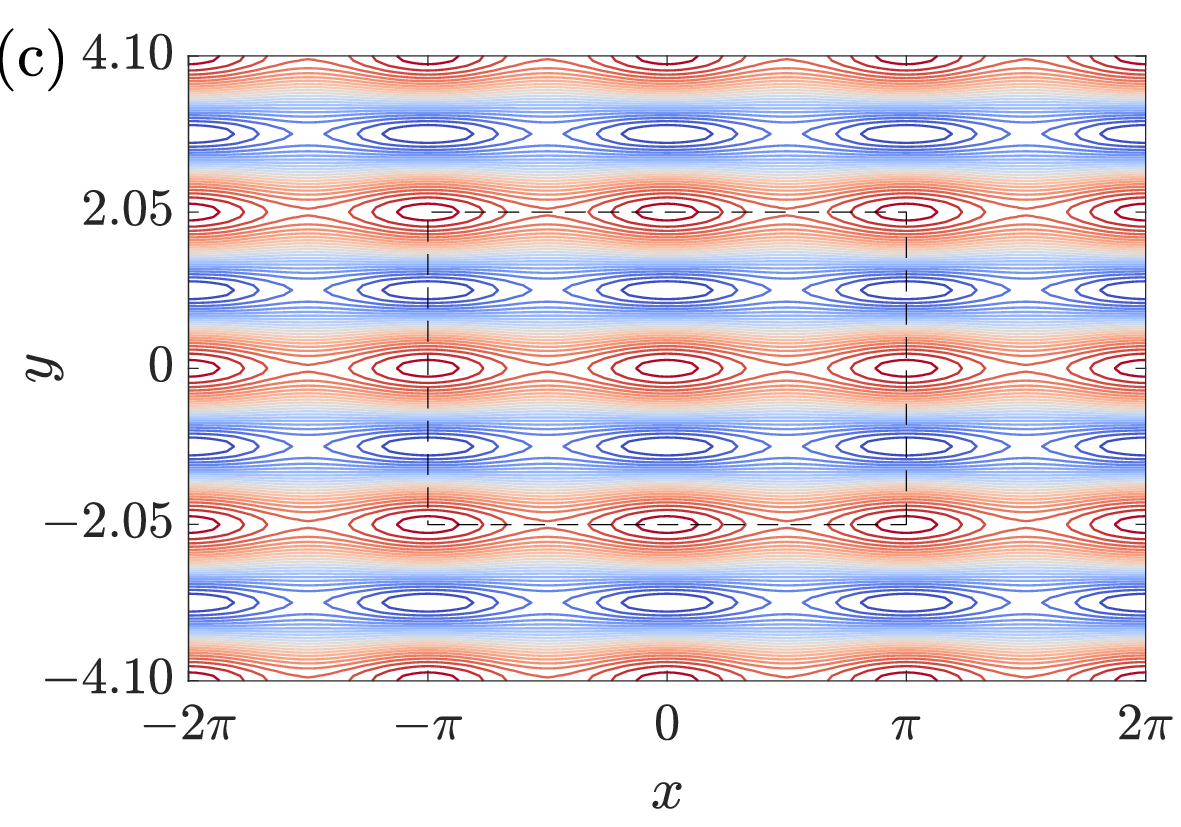}

    \includegraphics[width=0.32\linewidth]{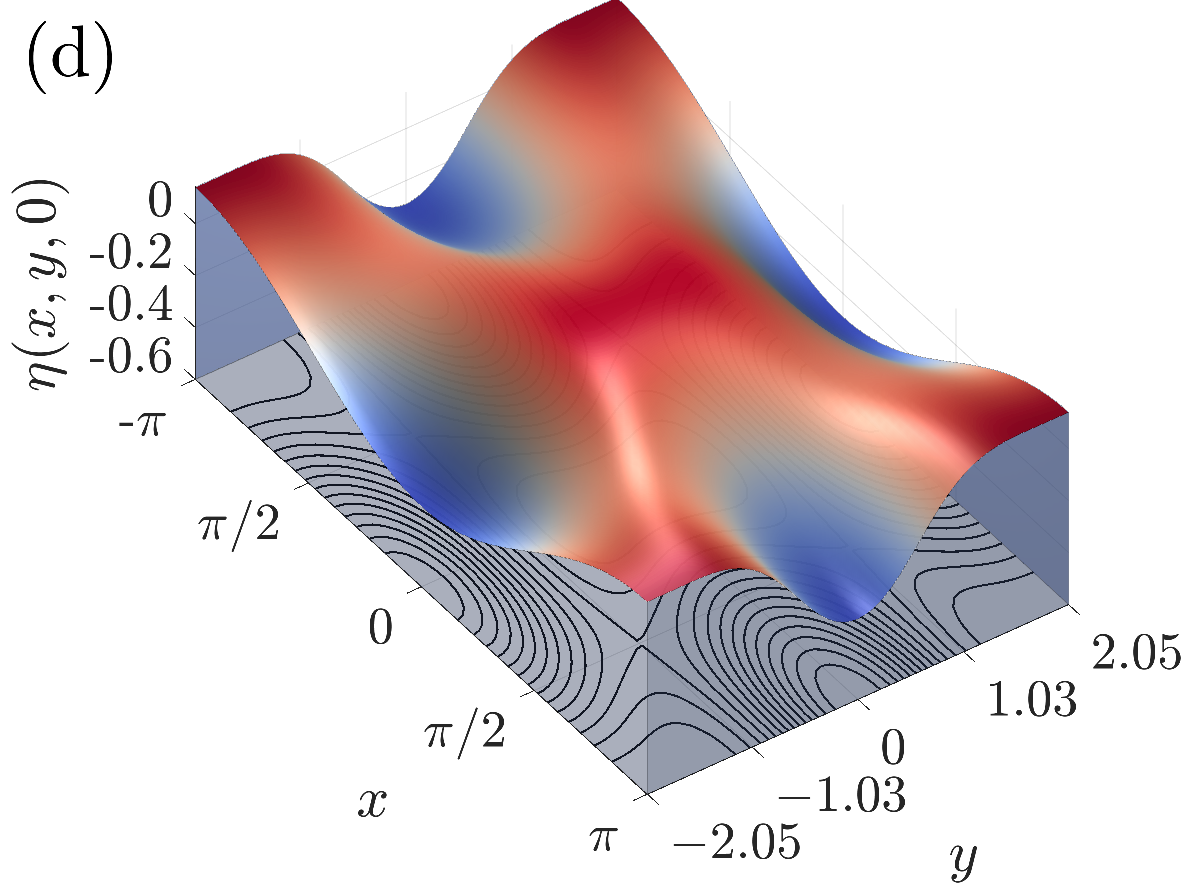}
    \includegraphics[width=0.32\linewidth]{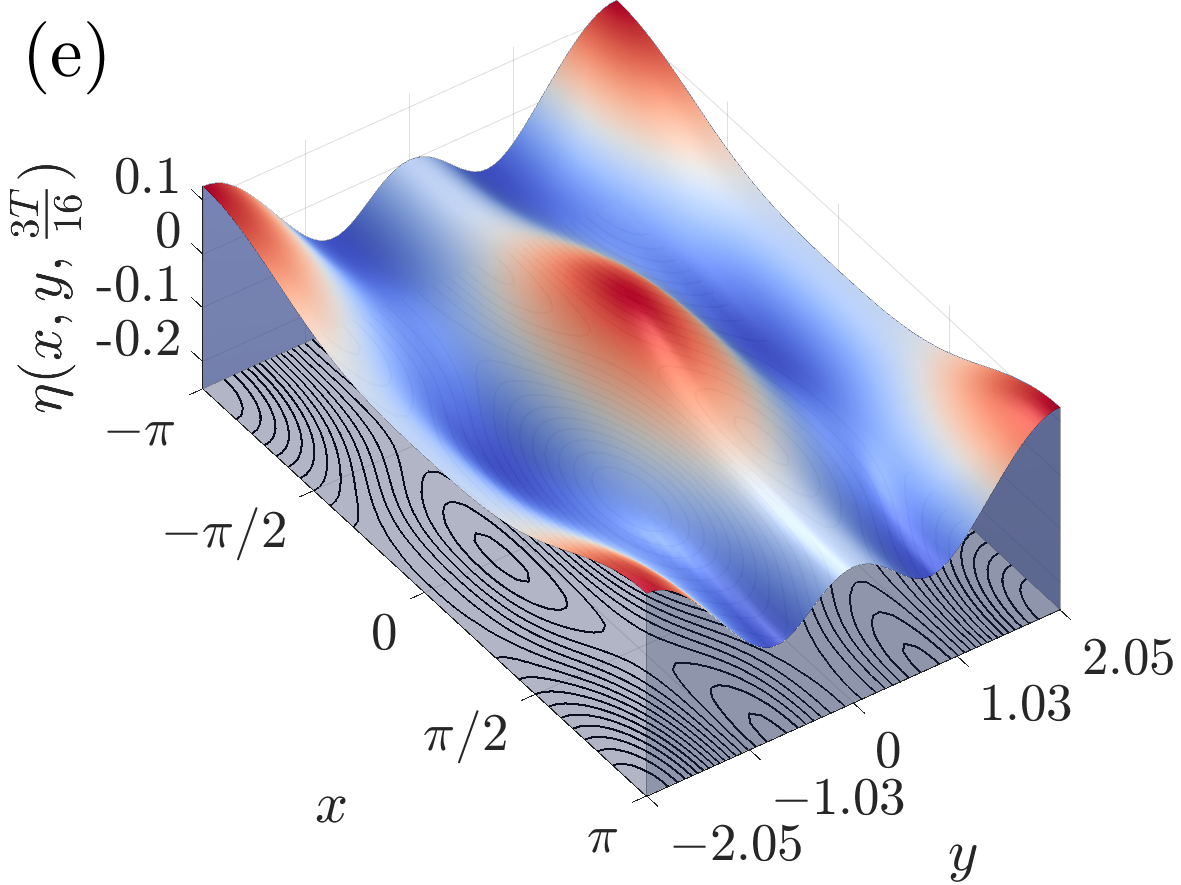}
    \includegraphics[width=0.32\linewidth]{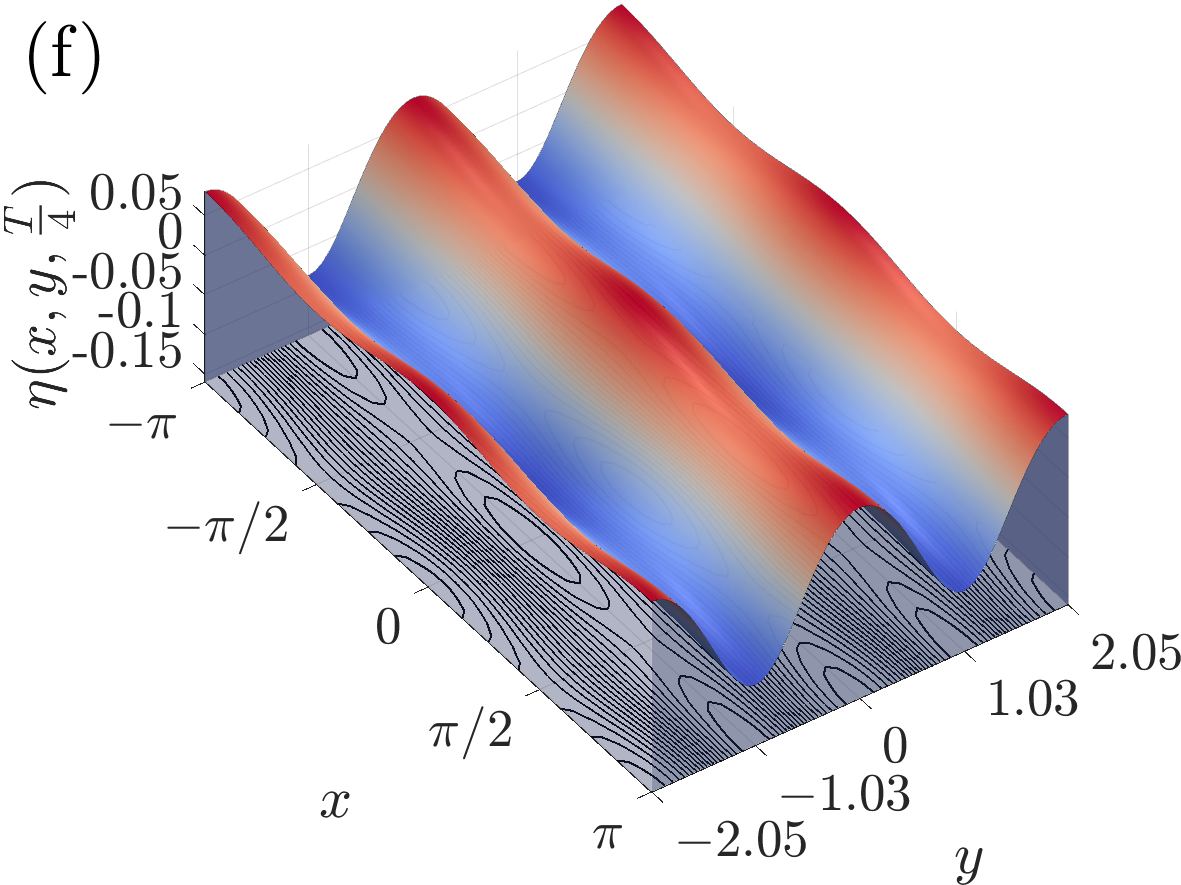}
    \caption{A standing wave with $k = 1,l = 1.5312, H=0.135$ on branch $2$. (a-c) show the top views at $t=0,3T/16$ and $T/4$. The corresponding three-dimensional profiles within a periodic cell (regions surrounded by the dashed lines) are shown in (d) and (e).}
    \label{fig:solution6.55}
\end{figure}
\begin{figure}[h!]
    \centering
    \includegraphics[width=0.45\linewidth]{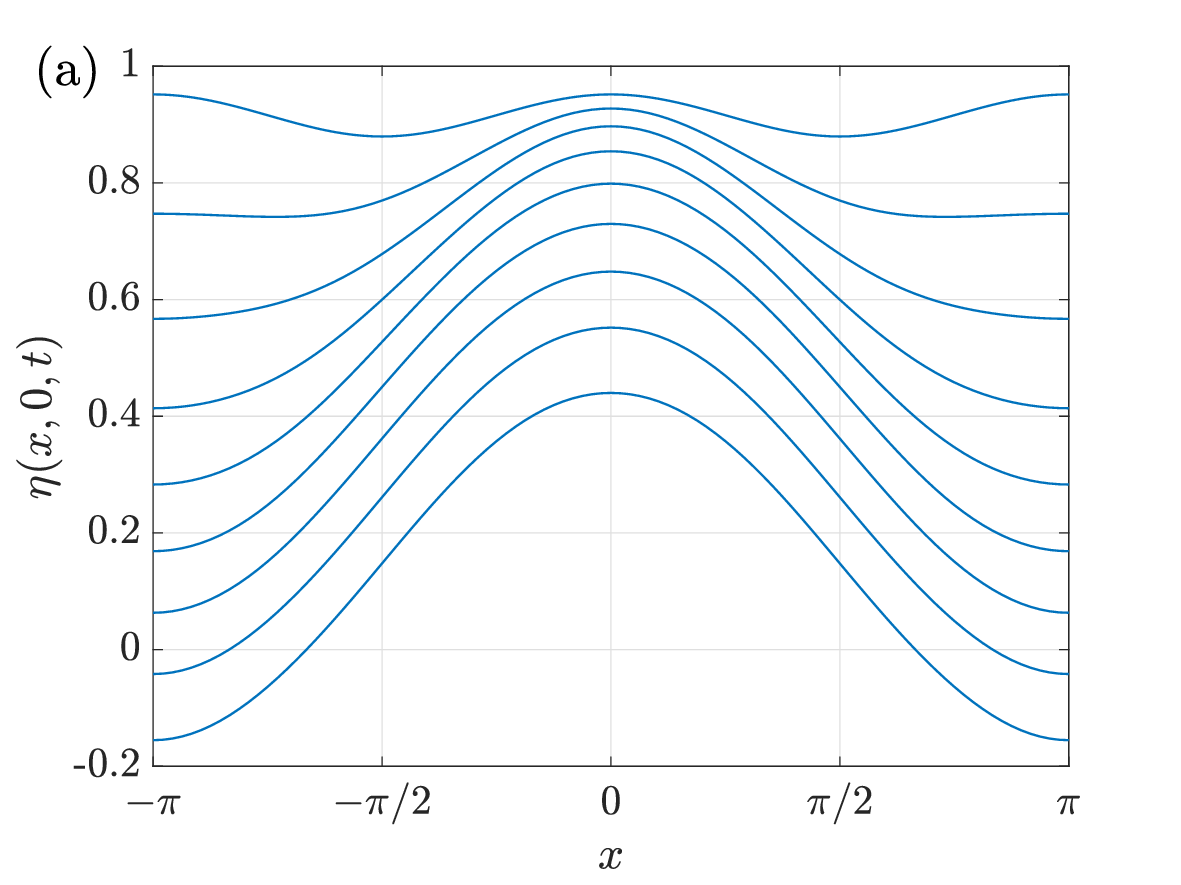}
    \includegraphics[width=0.45\linewidth]{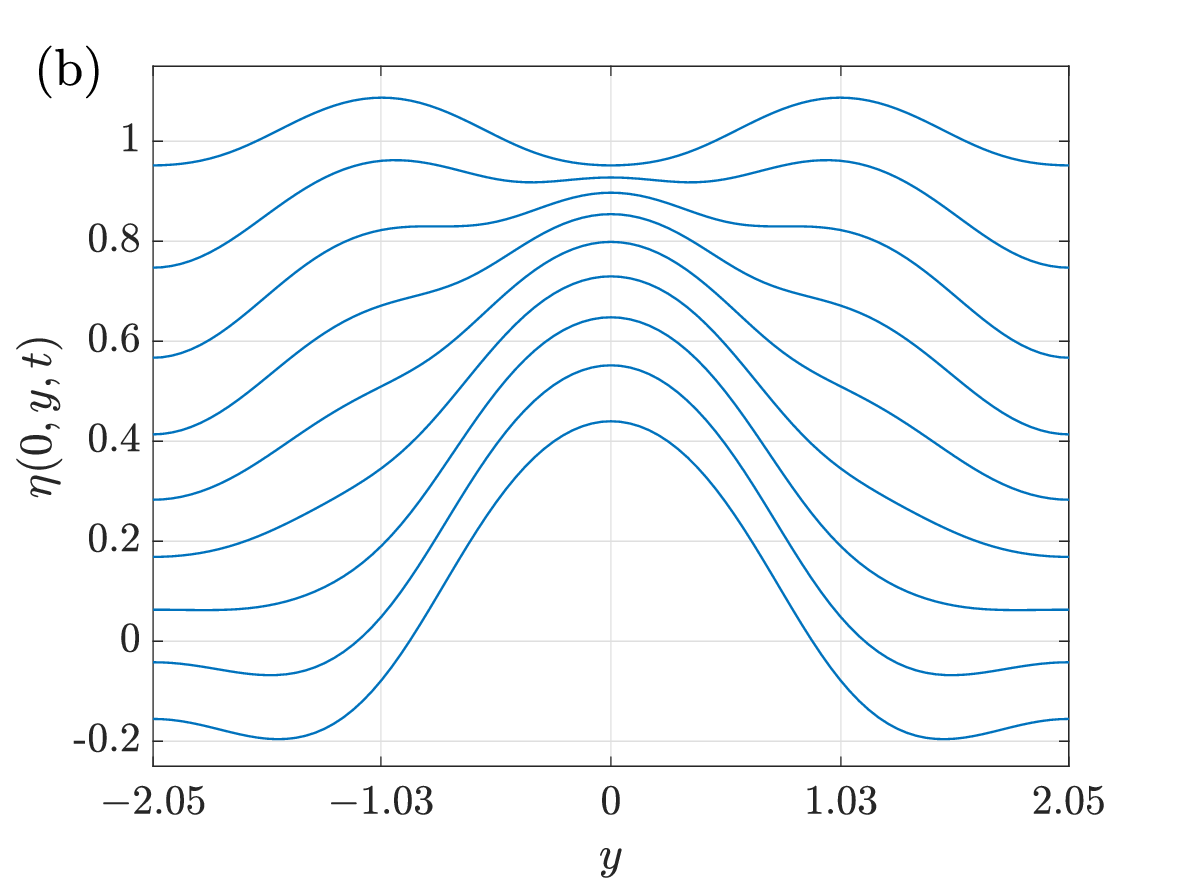}
    \caption{ (a) Cross-sections on the plane $y=0$ for the branch-$1$ solution with $H=0.4$. (b) Cross-sections on the plane $x=0$ for the same solution. From bottom to top, $t = 0, T/32,T/16,\cdots$,$T/4$. }
    \label{fig:solution6.6}
\end{figure}
At larger wave amplitude, branch-$1$ solutions exhibit similar wave profiles owing to the nearly constant value of $a$. Figure \ref{fig:solution6.5} shows the solution with $H=0.4(\epsilon=0.3480),\omega = 2.7182$ on branch $1$. Panels (a-c) correspond to the top views of the solution at $t = 0,3T/16$ and $T/4$, which clearly show the alternative appearance of hexagonal and quasi-two-dimensional patterns. The three-dimensional views of the wave profiles within a periodic cell are plotted in panels (d-f). In contrast, figure \ref{fig:solution6.55} exhibits the top views and wave profiles of a branch-$2$ standing wave with $H=0.135(\epsilon = 0.3402),\omega = 2.7222$. Both solutions are calculated using $96\times 96\times 96$ points. Figure \ref{fig:solution6.6} (a) and (b) show the cross-sections on the planes $y=0$ and $x=0$ for the branch-$1$ solution over a quarter temporal period. The former resembles the non-resonant two-dimensional standing wave profiles, while the latter is similar to the resonant profiles shown in figure \ref{fig:solution5.6}.

\begin{figure}[h!]
    \centering
    \includegraphics[width=0.45\linewidth]{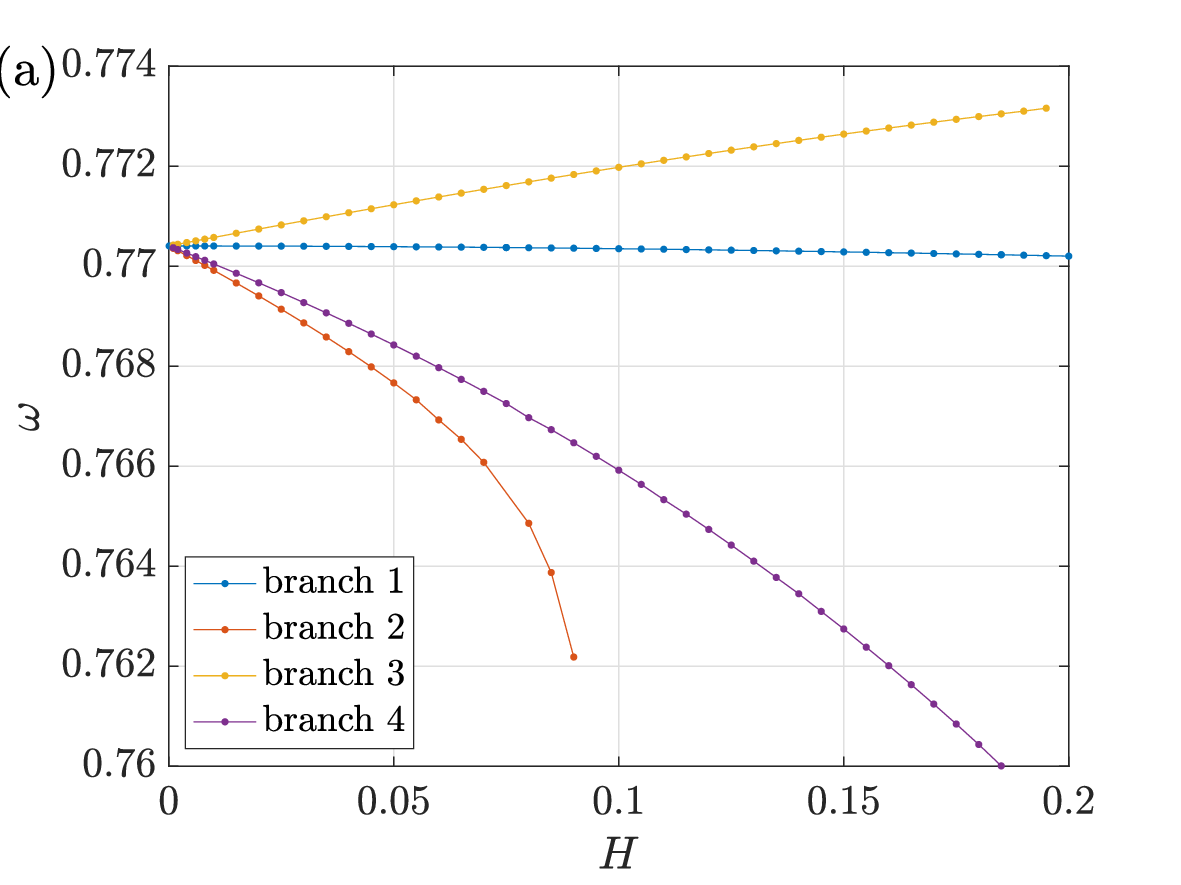}
    \includegraphics[width=0.45\linewidth]{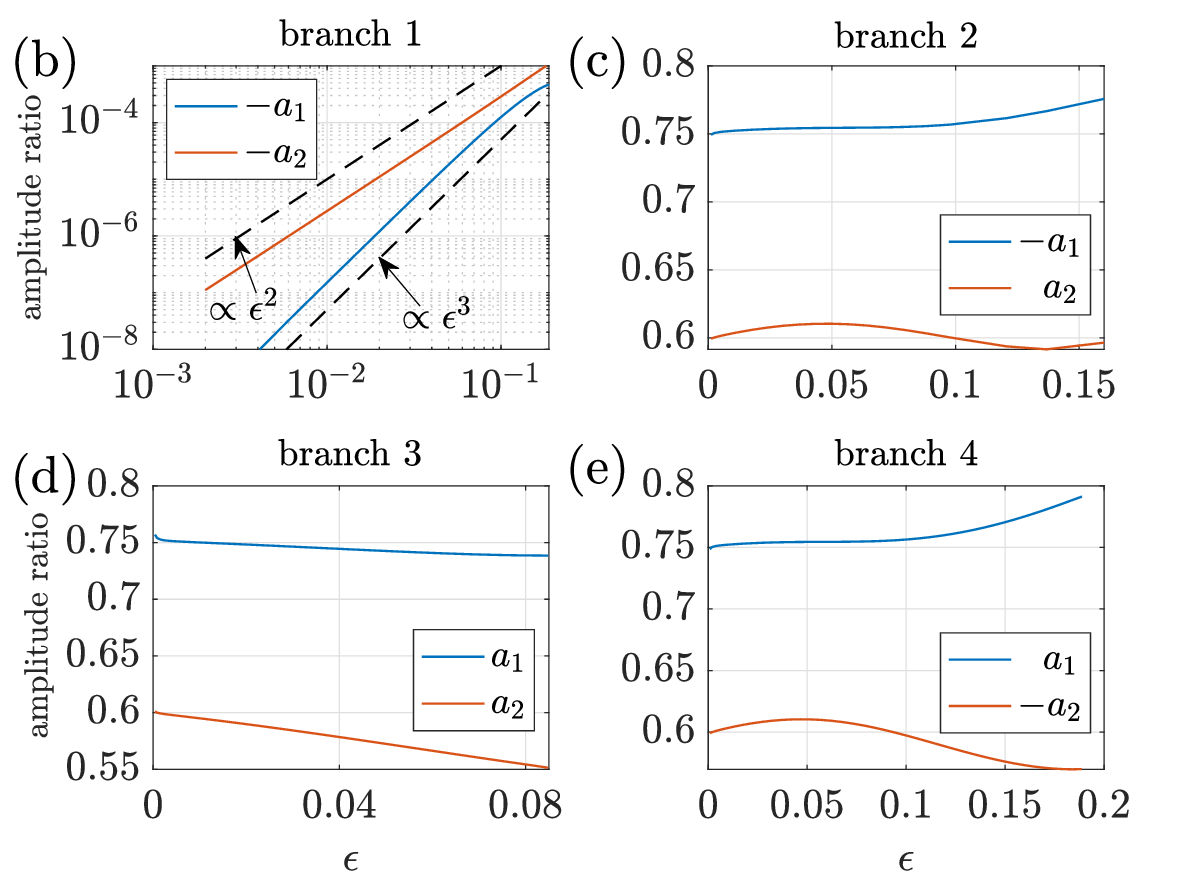}
    \caption{Four bifurcations for standing waves with $k=0.459181,l=0.145664$. (a) $\omega$ versus $H$. (b-e) Amplitude ratio versus $\epsilon$.}
    \label{fig:solution7}
\end{figure}
\begin{figure}[h!]
    \centering
    \includegraphics[width=0.32\linewidth]{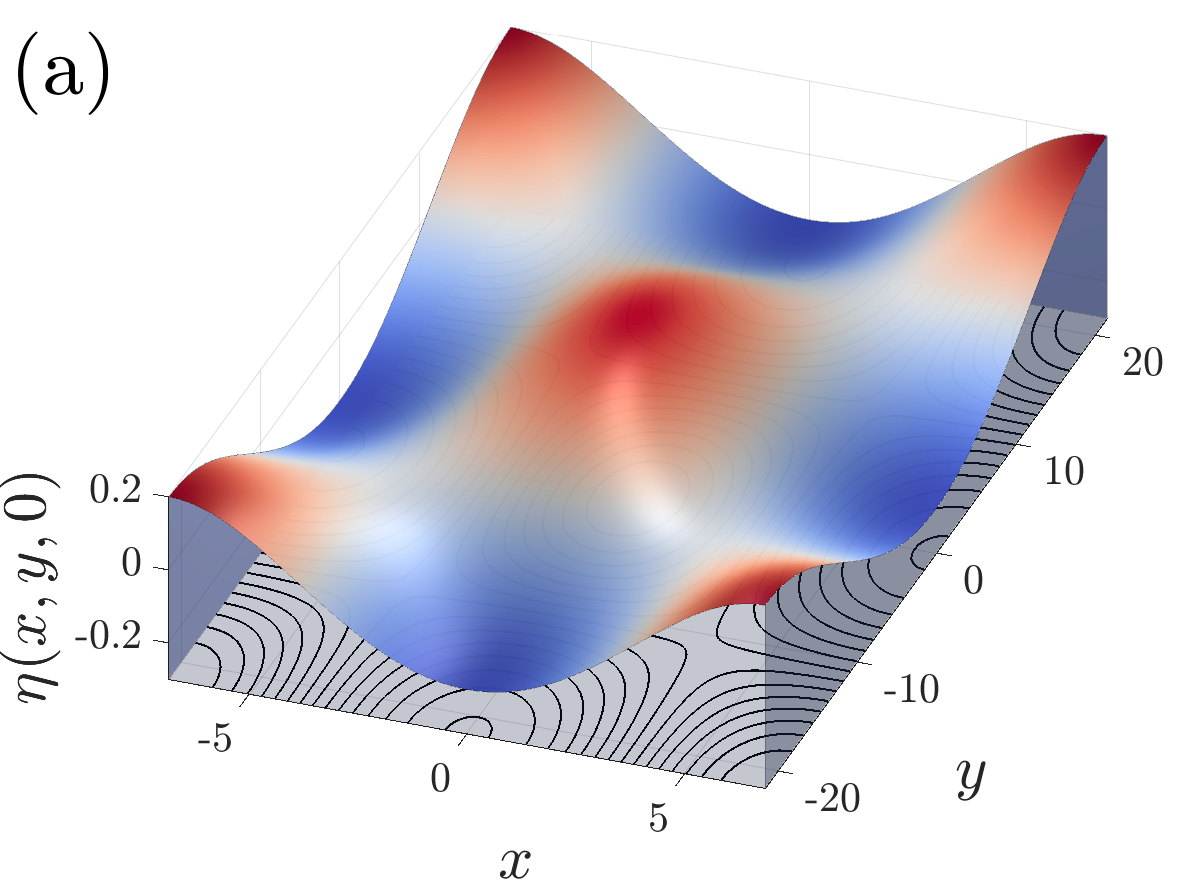}
    \includegraphics[width=0.32\linewidth]{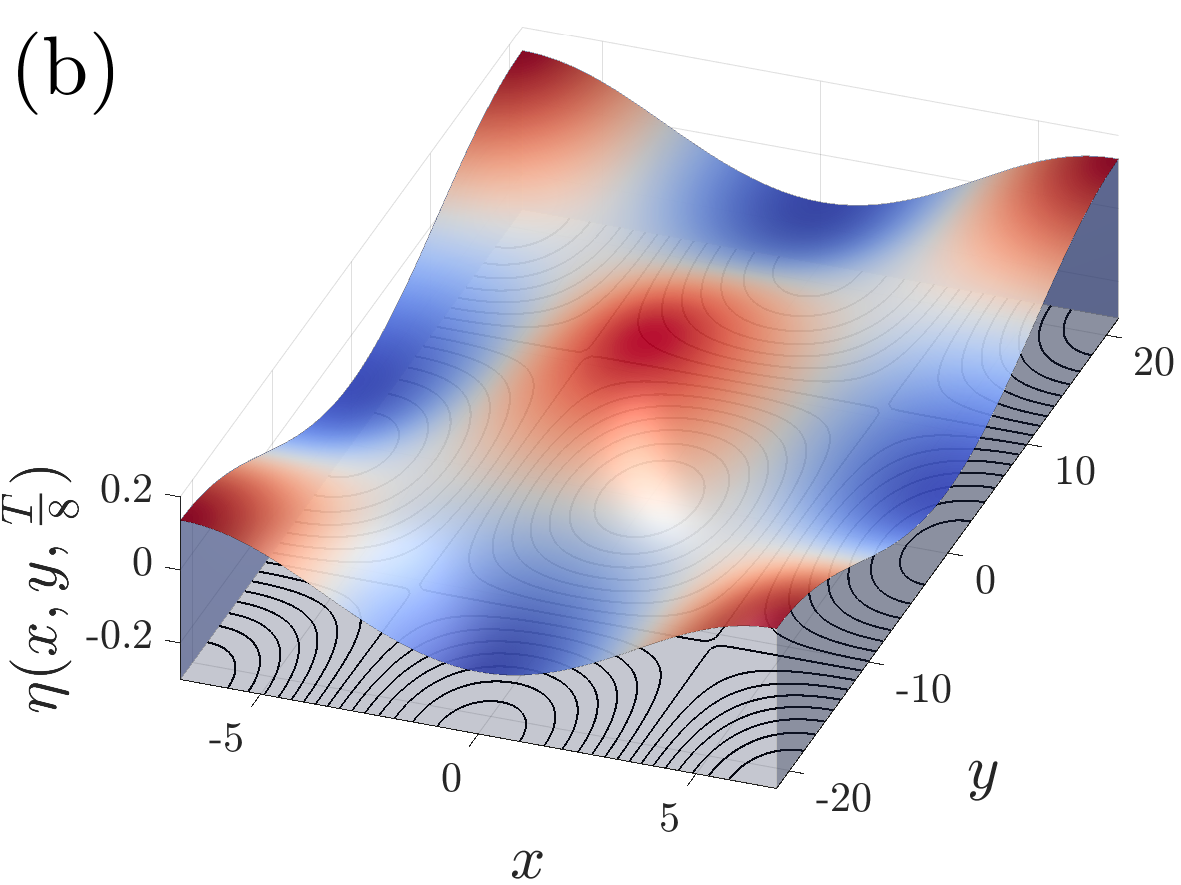}
    \includegraphics[width=0.32\linewidth]{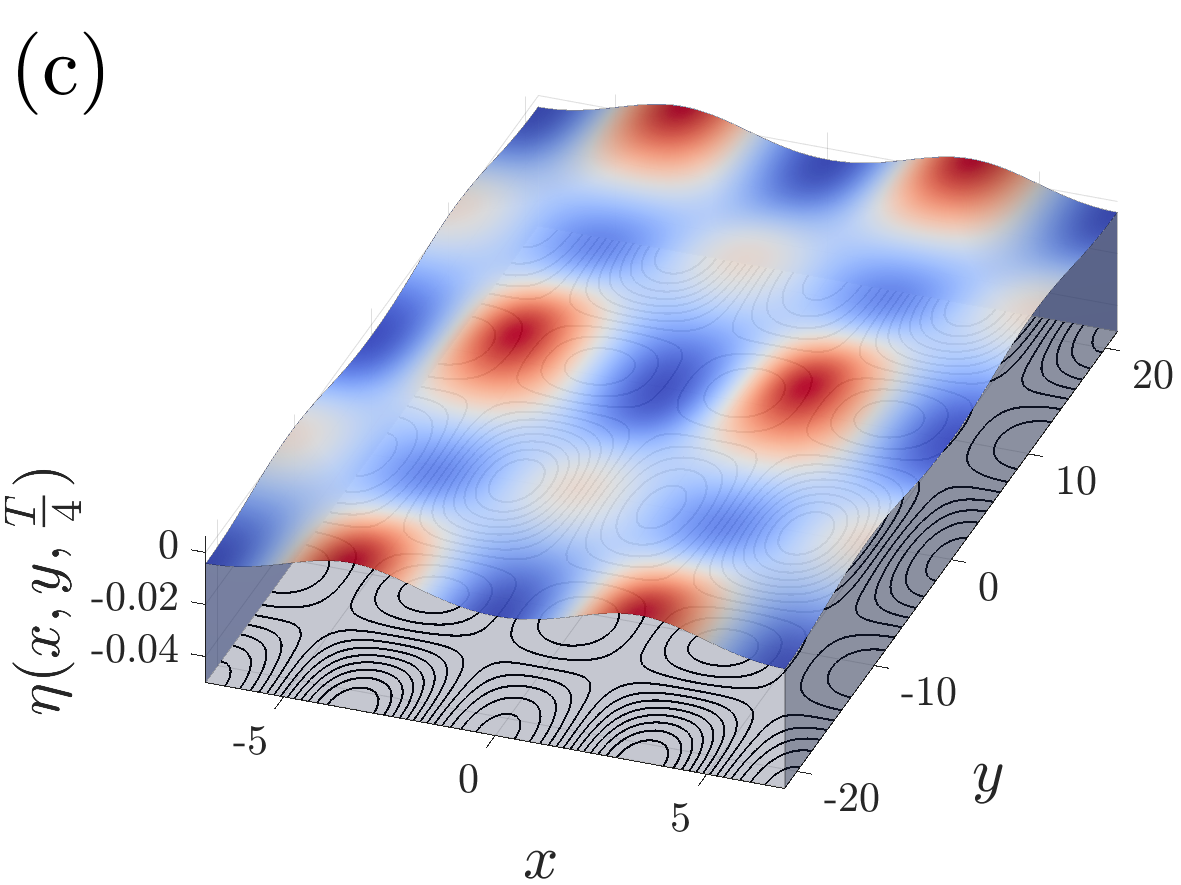}
    \includegraphics[width=0.32\linewidth]{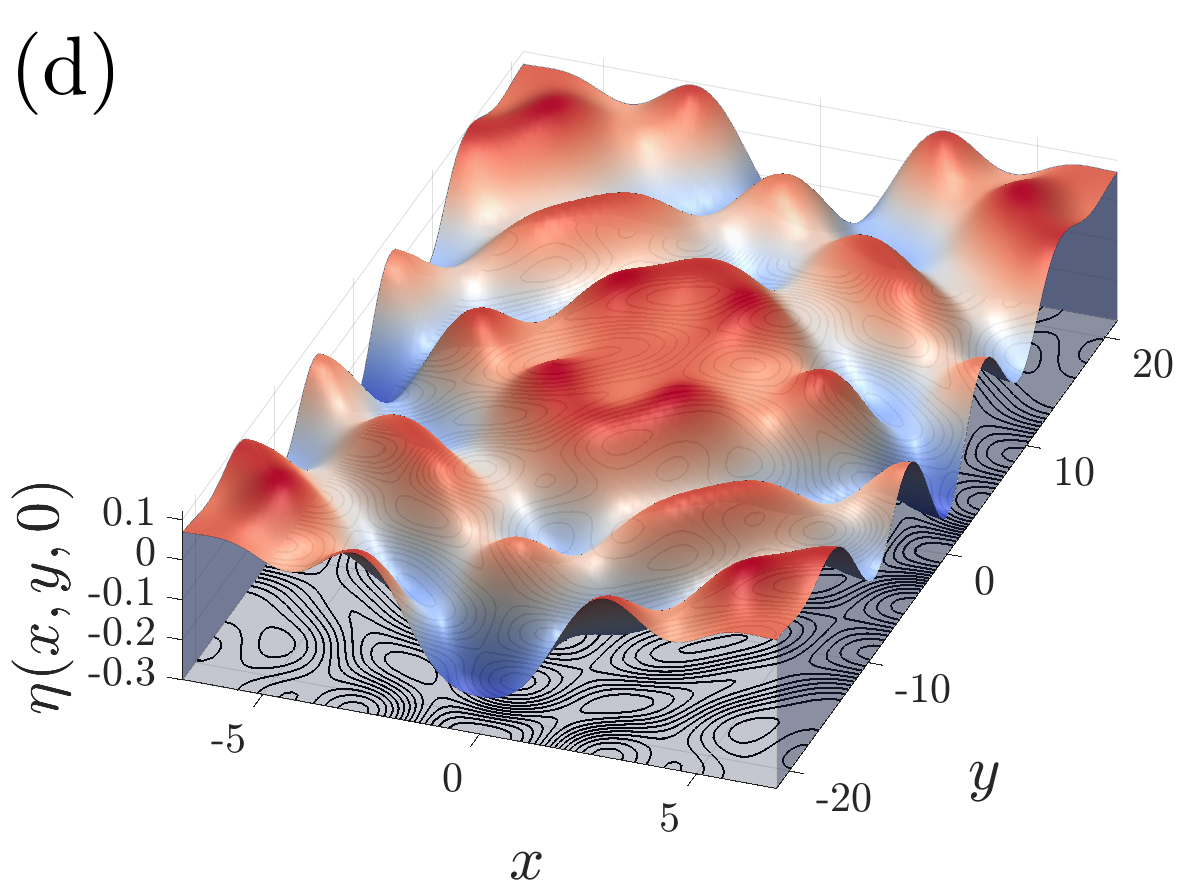}
    \includegraphics[width=0.32\linewidth]{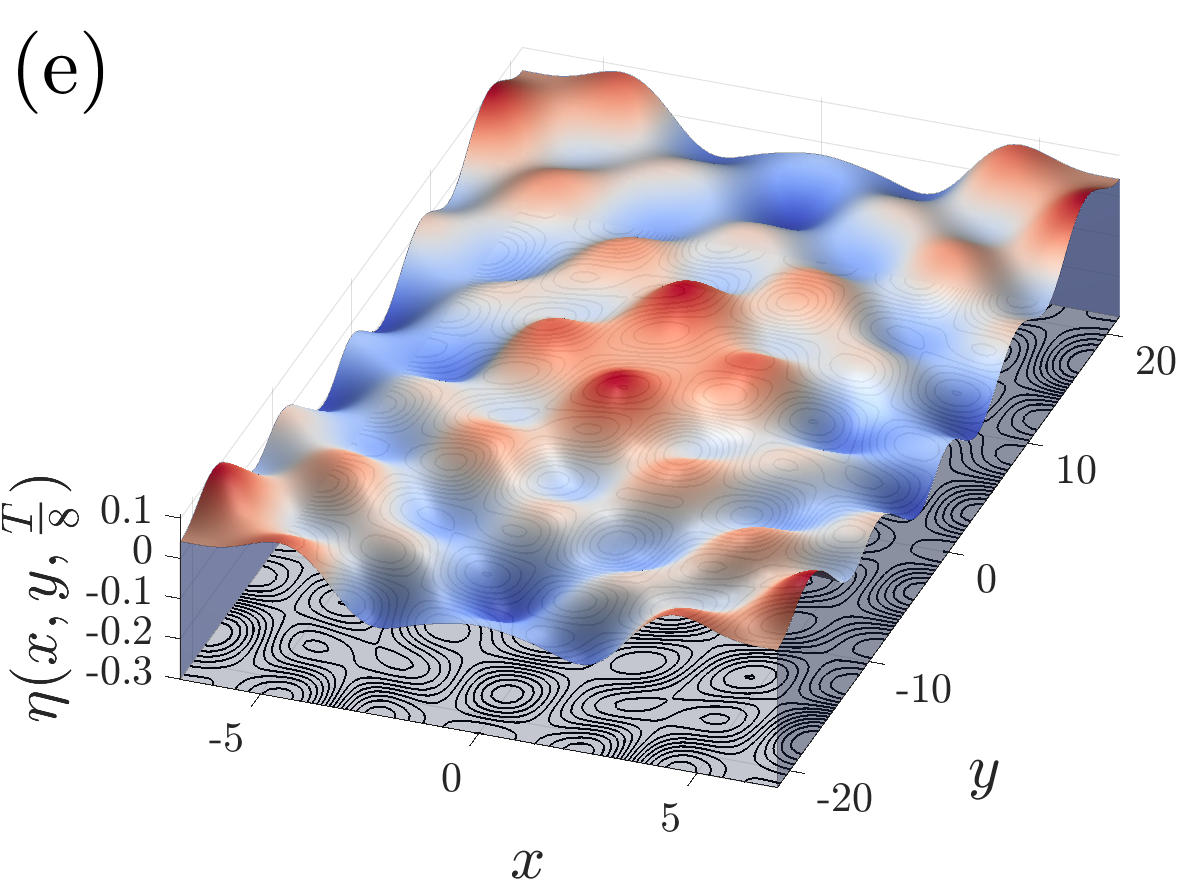}
    \includegraphics[width=0.32\linewidth]{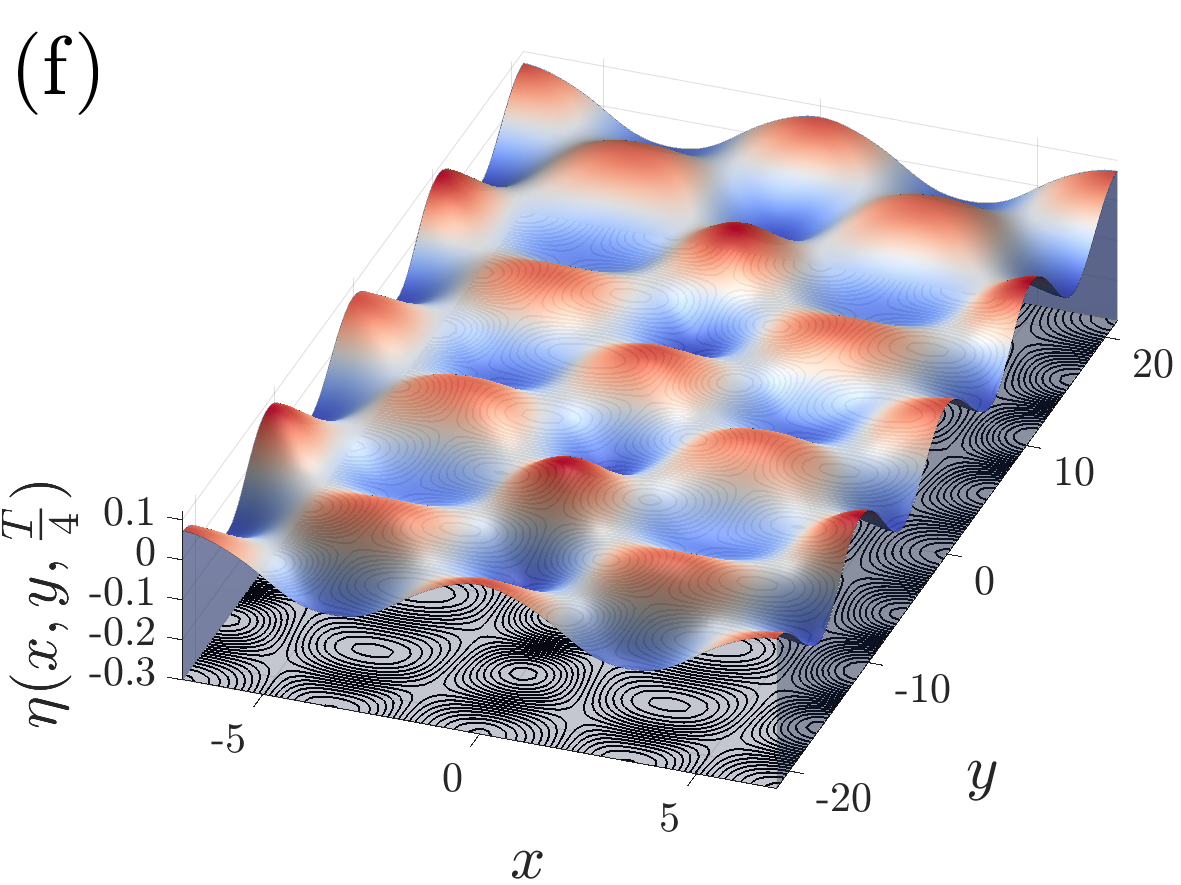}
    \includegraphics[width=0.32\linewidth]{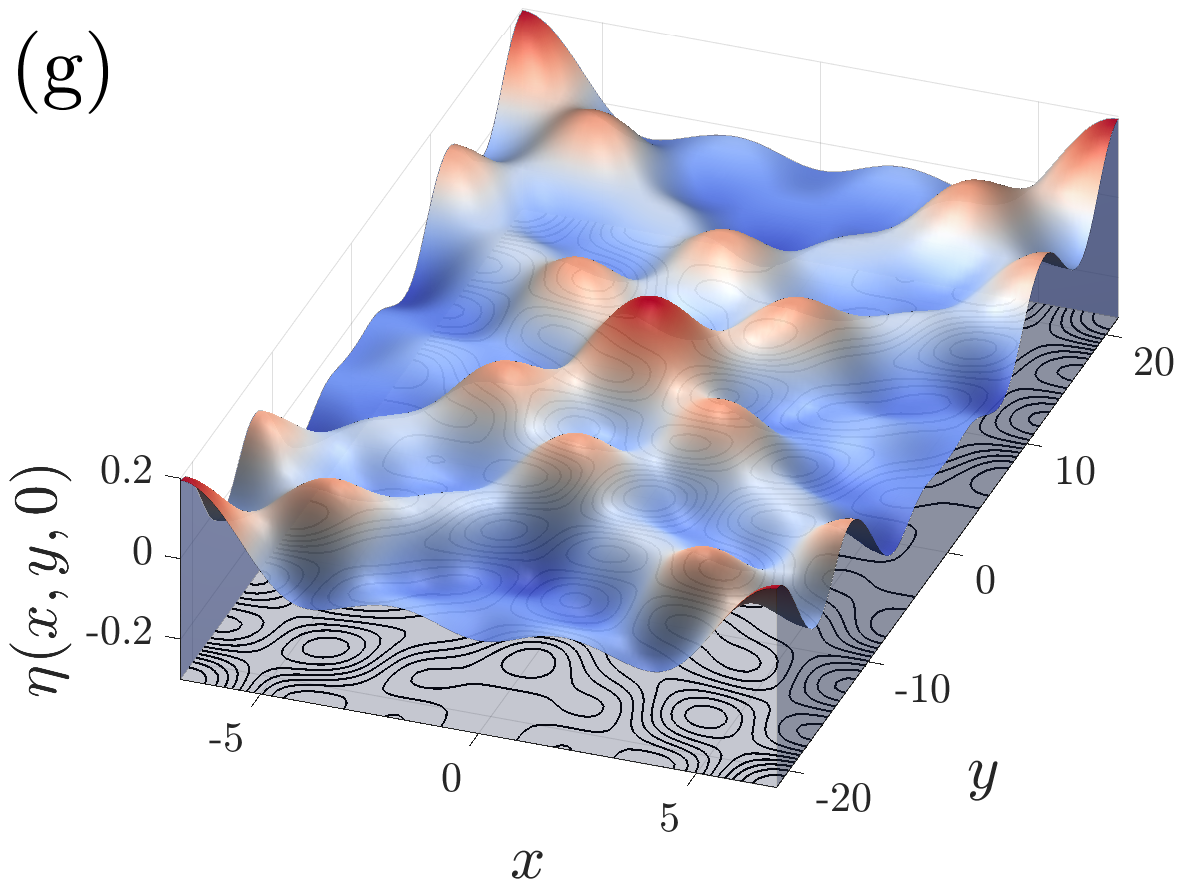}
    \includegraphics[width=0.32\linewidth]{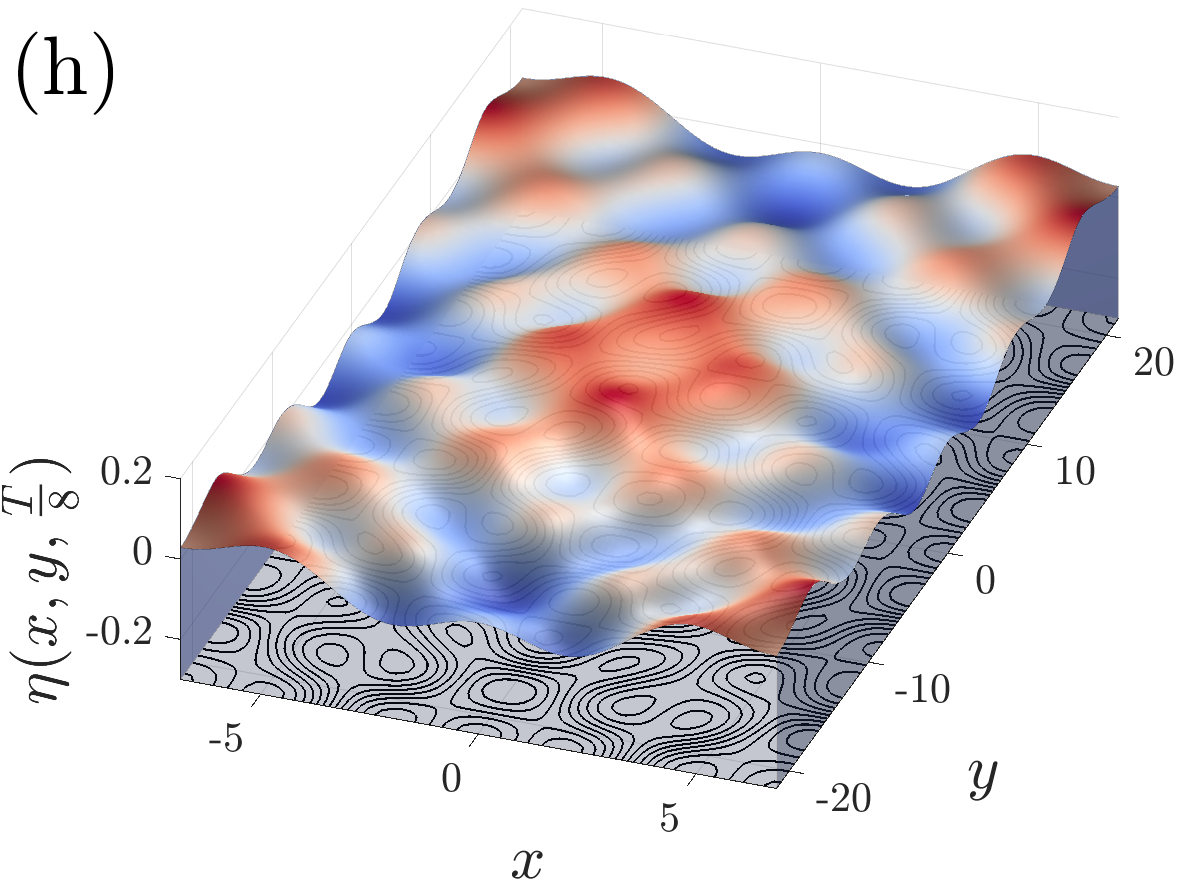}
    \includegraphics[width=0.32\linewidth]{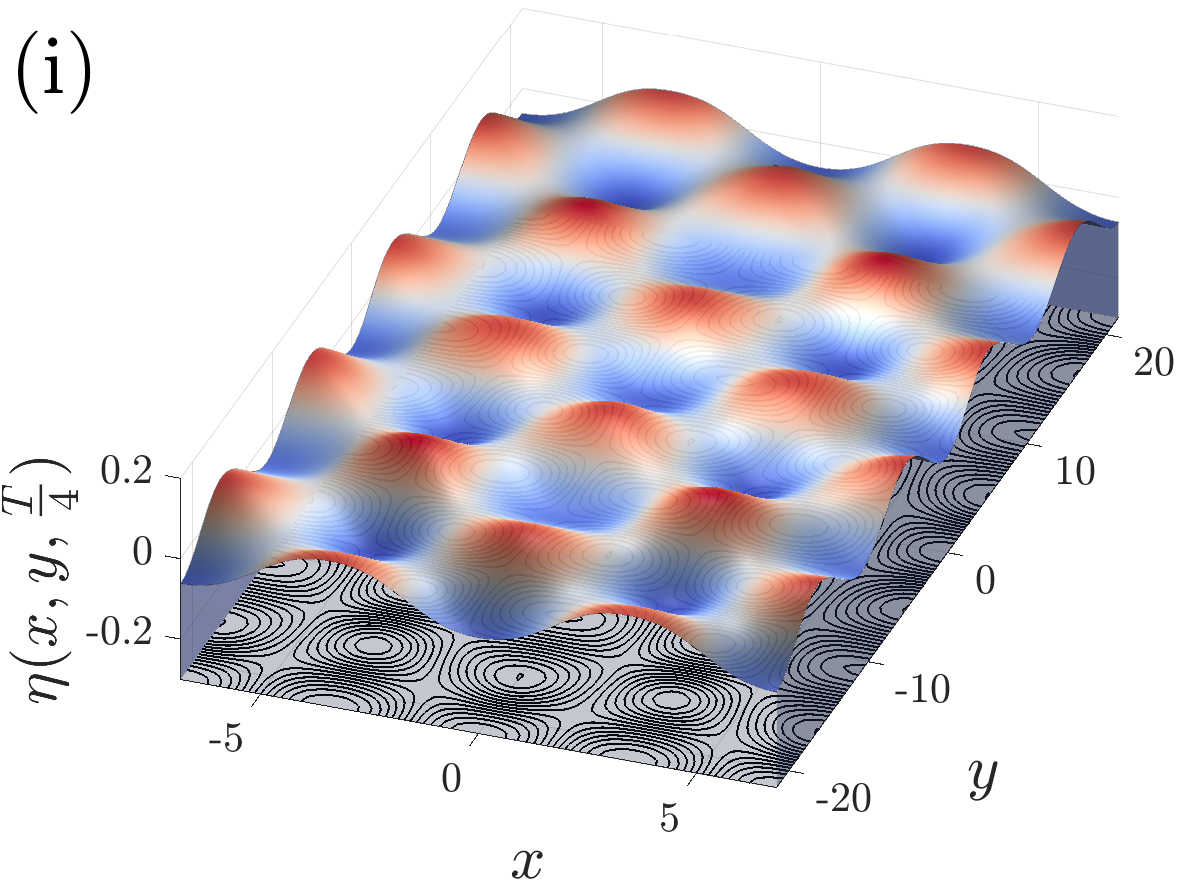}
    \includegraphics[width=0.32\linewidth]{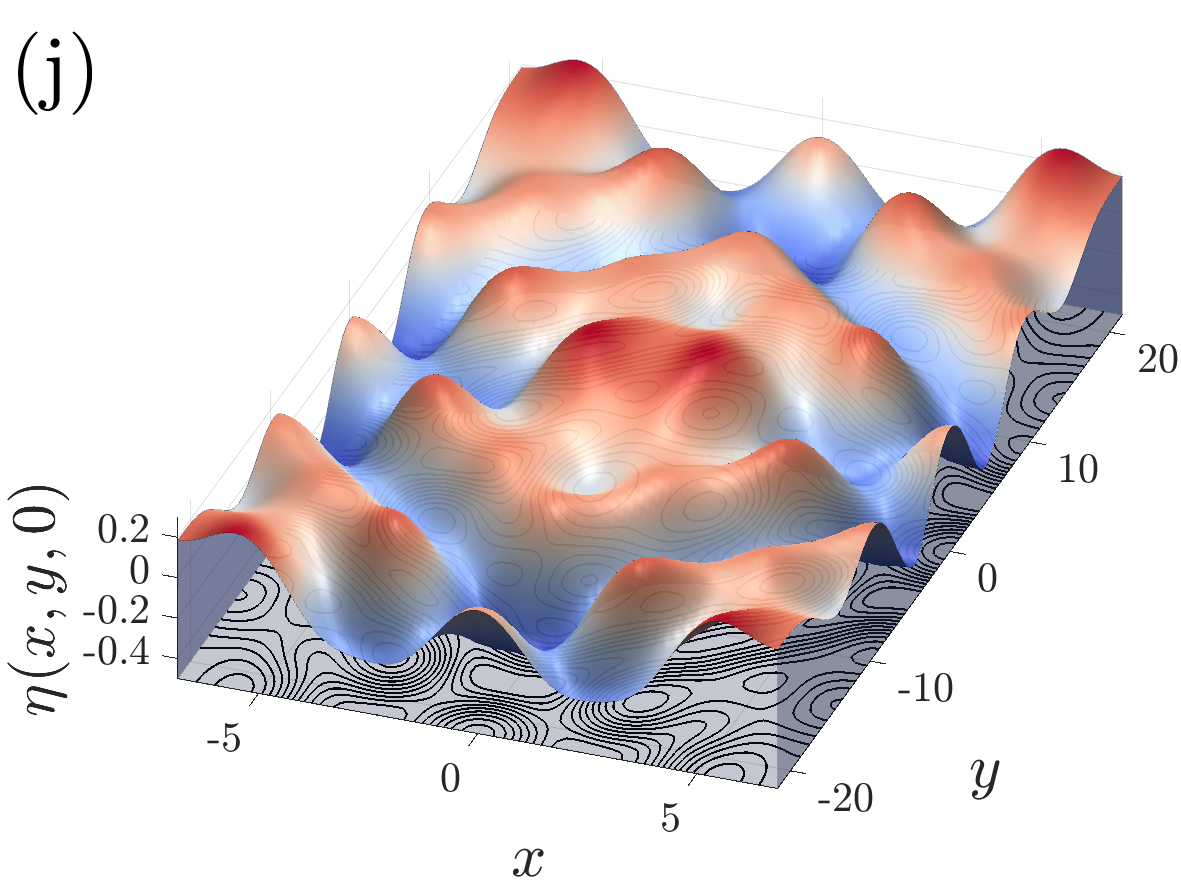}
    \includegraphics[width=0.32\linewidth]{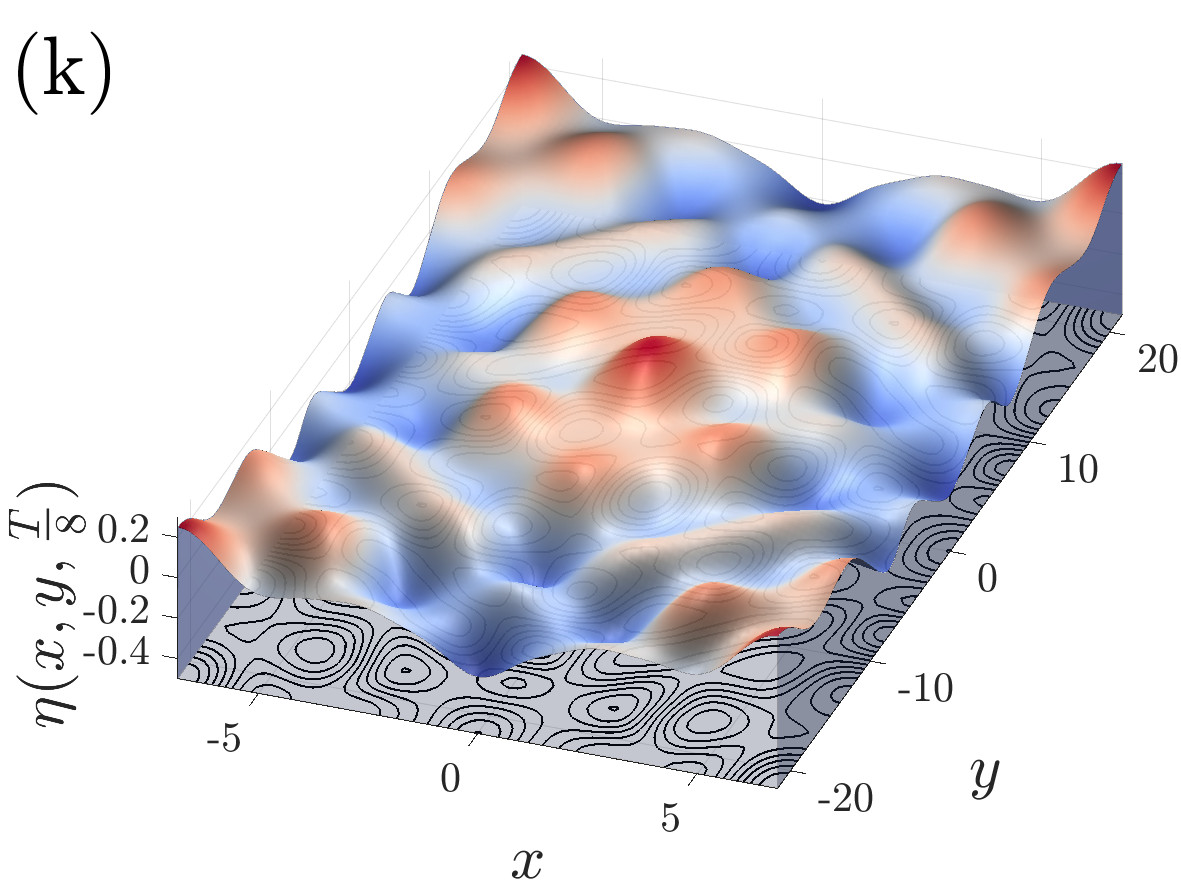}
    \includegraphics[width=0.32\linewidth]{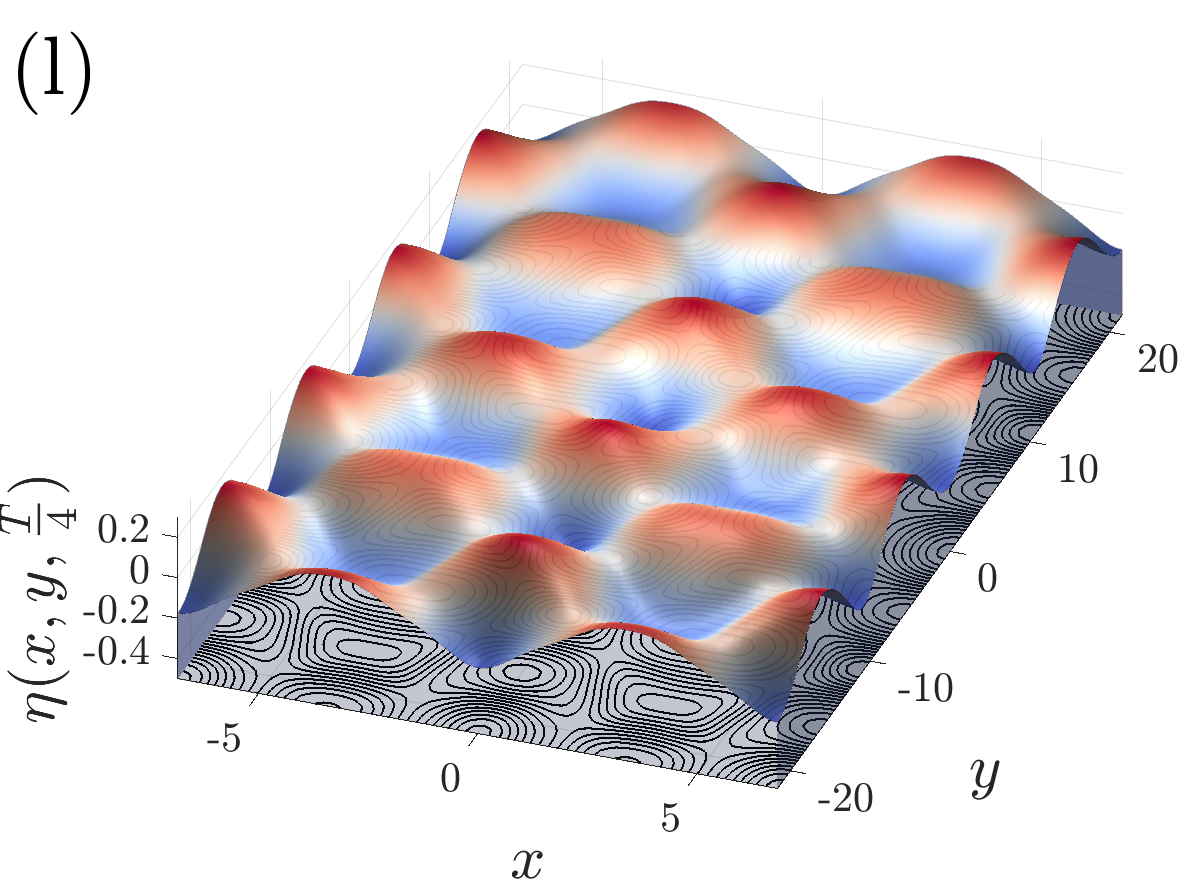}
    \caption{Standing waves with $k=0.459181,l=0.145664$ at $t=0,T/8,T/4$. (a-c) A branch-$1$ solution with $H = 0.2$. (d-f) A branch-$2$ solution with $H = 0.07$. (g-i) A branch-$3$ solution with $H = 0.195$. (j-l) A branch-$4$ solution with $H = 0.185$. For the convenience of visualisation, the figures are stretches in the $x$- and $z$-directions.}
    \label{fig:solution8}
\end{figure}
Finally, we show the standing waves with $k=0.459181,l=0.145664$. In the leading order, $\eta$ can be written as a combination of three components
\begin{align}
    \eta\sim\,& \epsilon \cos(0.459181 x)\cos(0.145664 y)\cos(\omega t)\nonumber\\
    &+ a_1\epsilon \cos(0.918362 x)\cos(0.582656y)\cos(2\omega t)\nonumber\\
    &+ a_2\epsilon \cos(1.377543 x)\cos(0.728320y)\cos(3\omega t) + O(\epsilon^2),\label{stokes3}
\end{align}
It is expected that there are four branches of solutions, corresponding to the combinations of the signs of $a_1$ and $a_2$: $(+,+),(+,-),(-,+)$, and $(-,-)$. Figure \ref{fig:solution7} (a) shows the frequency curves of the four bifurcations, all stemming from the infinitesimal linear solutions at the frequency $\omega = 0.770405$. Branches $1,2$ and $4$ exhibit frequency down-shift, while branch $3$ shows a frequency up-shift. On the other hand, only three branches ($2,3$ and $4$) exhibit resonant standing waves. This can be viewed in panels (b-e) where we plot the amplitude ratios versus $\epsilon$ along the four bifurcations. For branch $1$, it is clear that $a_1\propto -\epsilon^2$ and $a_2\propto -\epsilon^3$ for $\epsilon \ll 1$. Consequently, the small-amplitude standing waves on this branch are non-resonant. In contrast, the amplitude ratios take much greater values on the other three branches. When $\epsilon\rightarrow 0$, $a_1\rightarrow-0.75, a_2\rightarrow 0.6$ on branch $2$, $a_1\rightarrow 0.75, a_2\rightarrow 0.6$ on branch $3$, and $a_1\rightarrow 0.75, a_2\rightarrow -0.6$ on branch $4$. No solution is found for the case $a_1\rightarrow-0.75, a_2\rightarrow -0.6$. Based on our numerical experiments using various initial guesses, it is believed that there are only four solution branches growing from the bifurcation point, excluding the trivia case that a single periodic wave repeats itself multiple times within one computational domain. The profiles at $t=0,T/4$ and $T/2$ of a non-resonant solution and three flower-like resonant solutions are shown in figure \ref{fig:solution8}, where the four rows from top to bottom correspond to the branches $1$ to $4$, respectively.

\section{Temporal evolutions}

In this section, we present some temporal evolutions of the computed standing waves via initial-value calculations. In Fourier space, the quintic model becomes
\begin{align}
    \widehat{\eta}_t &- |\boldsymbol{k}|\widehat{\varphi} = \sum_{i = 1}^{4}\widehat{G_i}\widehat{\varphi},\label{eta_t}\\
    \widehat{\varphi}_t &+ (1+|\boldsymbol{k}|^2)\widehat{\varphi} = \sum_{i = 2}^{5}\widehat{\mathcal N_i} + \mathrm i\boldsymbol{k}\cdot\widehat{\Bigg(\frac{\nabla\eta}{\sqrt{1+|\nabla\eta|^2}}\Bigg)} + |\boldsymbol{k}|^2\widehat{\varphi},\label{phi_t}
\end{align}
where the hat denotes the Fourier transform and $\boldsymbol{k} = (k_1,k_2)$ is the transformed variable. Following Wang \& Milewski \cite{wang2012dynamics}, we introduce
\begin{align}
    \widehat{p} = \widehat{\eta} + \frac{\mathrm i \widehat{\varphi}}{\sqrt{|\boldsymbol{k}|+1/|\boldsymbol{k}|}},\quad \widehat{q} = \widehat{\eta} - \frac{\mathrm i \widehat{\varphi}}{\sqrt{|\boldsymbol{k}|+1/|\boldsymbol{k}|}}.
\end{align}
By combining \eqref{eta_t} and \eqref{phi_t}, we have
\begin{align}
    \widehat{p}_t + \mathrm i \sqrt{|\boldsymbol{k}|+|\boldsymbol{k}|^3} \widehat{p} = \widehat{\mathcal N}_\eta + \frac{\mathrm i}{\sqrt{|\boldsymbol{k}|+1/|\boldsymbol{k}|}}\widehat{\mathcal N}_\varphi, \label{p_t}\\
    \widehat{q}_t - \mathrm i \sqrt{|\boldsymbol{k}|+|\boldsymbol{k}|^3} \widehat{q} = \widehat{\mathcal N}_\eta - \frac{\mathrm i}{\sqrt{|\boldsymbol{k}|+1/|\boldsymbol{k}|}}\widehat{\mathcal N}_\varphi,
\end{align}
where $\widehat{\mathcal N}_\eta$ and $\widehat{\mathcal N}_\varphi$ represent the right-hand sides of \eqref{eta_t} and \eqref{phi_t}, respectively. Using the fact that $\eta$ and $\varphi$ are real, the two equations are essentially the same. Therefore, the problem is reduced to a single evolution equation \eqref{p_t}, which can be temporally integrated by using the fourth-order Runge-Kutta scheme and an integrating-factor method. Once $\widehat{p}$ is obtained, $\widehat{\eta}$ and $\widehat{\varphi}$ can be recovered from
\begin{align}
    \widehat{\eta}(\boldsymbol{k}) = \frac{1}{2}\Big(\widehat{p}(\boldsymbol{k})+\widehat{p}(-\boldsymbol{k})^*\Big),\quad \widehat{\varphi}(\boldsymbol{k}) = \frac{1}{2\mathrm i}\sqrt{|\boldsymbol{k}|+1/|\boldsymbol{k}|}\Big(\widehat{p}(\boldsymbol{k})-\widehat{p}(-\boldsymbol{k})^*\Big),
\end{align}
where $*$ represents complex conjugate. To suppress the aliasing instability, we multiply $\widehat{p}$ by
\begin{align}
    \mathrm e^{-36[(k_1/K_1)^2+((k_2/K_2)^2)]^{18}},
\end{align}
where $K_1 = \pi I/2L_1$ and $K_2 = \pi J/2L_2$, at the end of each Runge-Kutta step. This filter is an extension of those used in \cite{hou1994removing,hou2007computing}, allowing to resolve high-frequency Fourier modes in computation.

\begin{figure}[h!]
    \centering
    \includegraphics[width=0.7\linewidth]{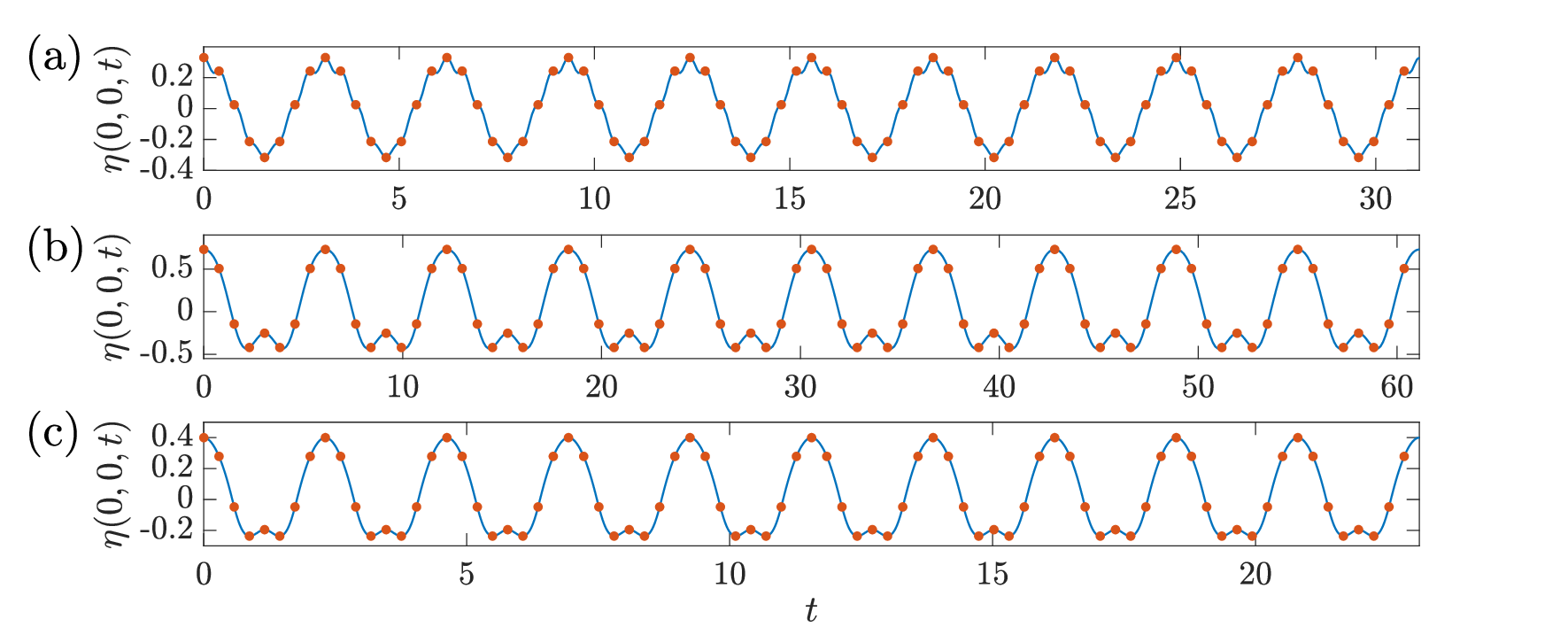}
    \includegraphics[width=0.42\linewidth]{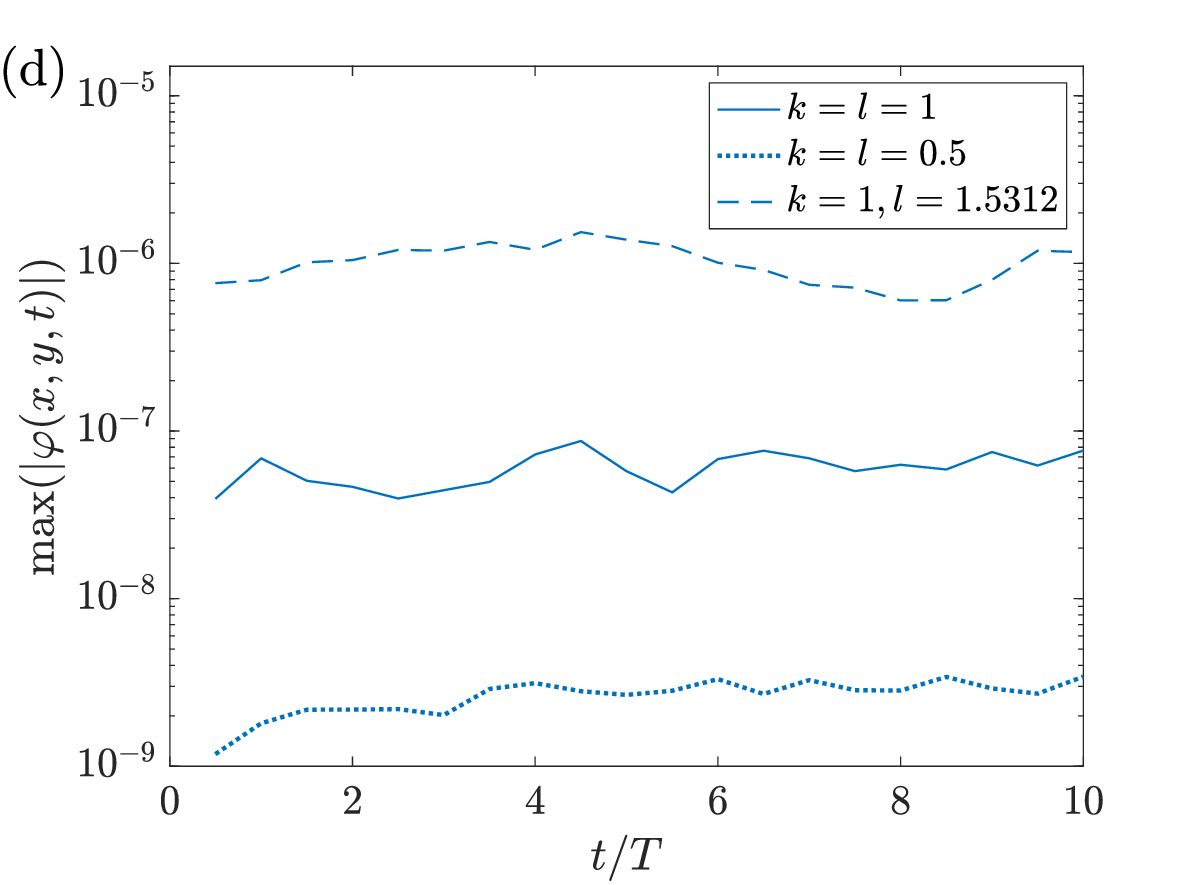}
    \includegraphics[width=0.42\linewidth]{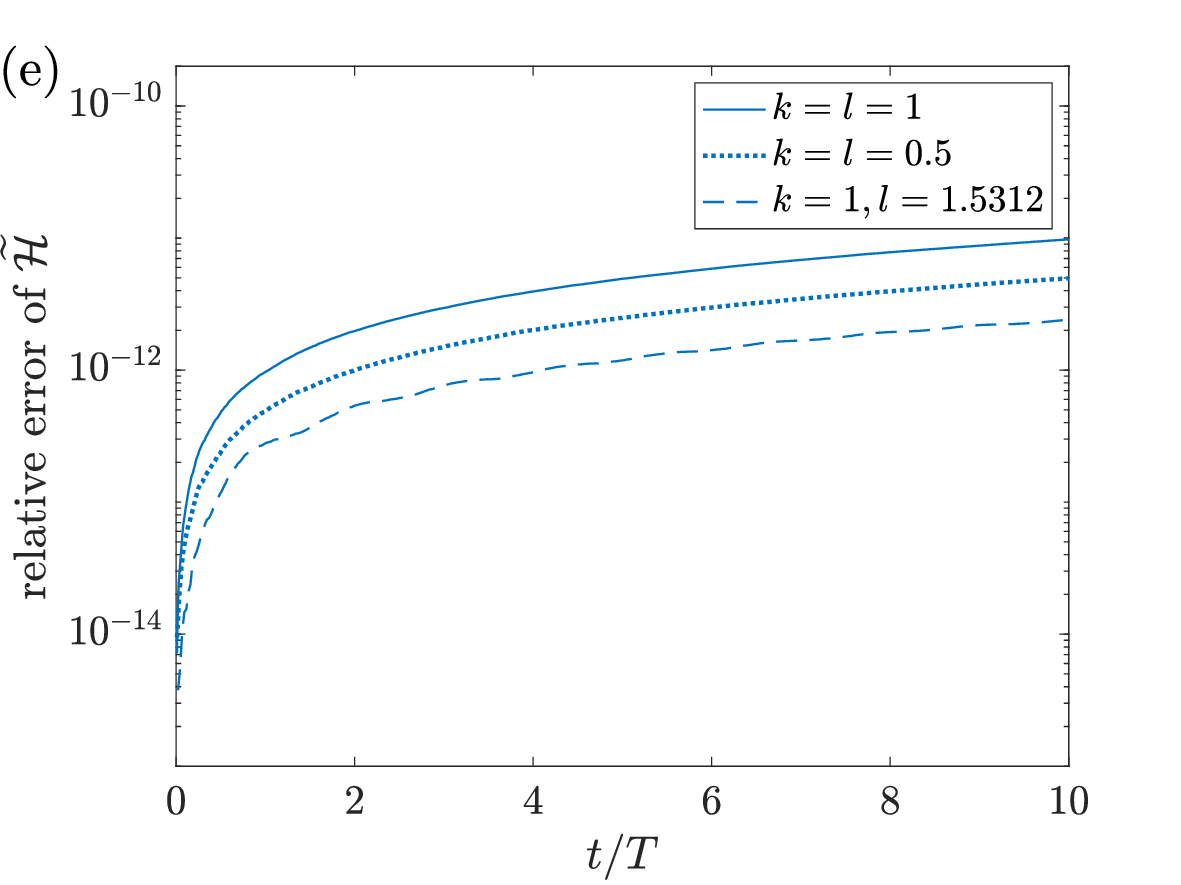}
    \caption{(a-c) Time histories of $\eta(0,0,t)$ over ten periods for the standing waves with $k = l = 1$ (figure \ref{fig:solution2}), with $k = l = 0.5$ on branch $1$ (figure \ref{fig:solution5.5}), and with $k = 1, l = 1.5312$ on branch $1$ (figure \ref{fig:solution6.5}), respectively. The blue curves and red dots denote results of the initial-value and boundary-value calculations. (d) max$(|\varphi(x,y,t/T)|)$ at integer and half temporal periods. (e) Relative energy error in temporal simulations.}
    \label{fig:time simulation1}
\end{figure}
We first validate the numerical accuracy of the computed standing waves. Figure \ref{fig:time simulation1} shows the time histories of $\eta(0,0,t)$ for three different standing waves over ten temporal periods: (a) the \textit{Case II} solution with $k = l = 1,H=0.33$ shown in figure \ref{fig:solution2}, (b) the \textit{Case II} solution solution with $k = l = 0.5,H=0.73$ shown in figure \ref{fig:solution5.5}, and (c) the \textit{Case I} solution with $k = 1, l = 1.5312,H=0.4$ shown in figure \ref{fig:solution6.5}. To avoid numerical instabilities, we uniformly divide each temporal period into $20000$ time steps for the computations in panels (a) and (b), and $40000$ time steps for the one in panel (c). The blue lines and red dots denote the numerical results of the initial-value and the boundary-value approaches, respectively. The excellent agreement among the three comparisons confirms the high accuracy of our spatio-temporal collocation method. As a second examination, we check the amplitude of $\varphi(x,y,t)$ at each integer and half temporal period, which should be zero theoretically. Panel (d) shows three curve of $\text{max}(|\varphi(x,y,t)|)$ which fluctuate around $10^{-7}, 10^{-9}$, and $10^{-6}$, corresponding to the simulations in panels (a), (b), and (c), respectively. The different residual levels primarily reflect the deviation from exact standing waves in the given initial conditions, therefore relating to the resolutions used in the boundary-value calculations: $128\times 128\times 128$ for the solid and dotted lines, and $96\times 96\times 96$ for the dashed line. On the other hand, the numerical errors arising from the time integration, spatial discretisation, and Fourier filter are almost negligible, as demonstrated by panel (e) where the relative error of $\widetilde{\mathcal {H}}$, the Hamiltonian of the quintic model, is shown.

\begin{figure}[h!]
    \centering
    \includegraphics[width=0.7\linewidth]{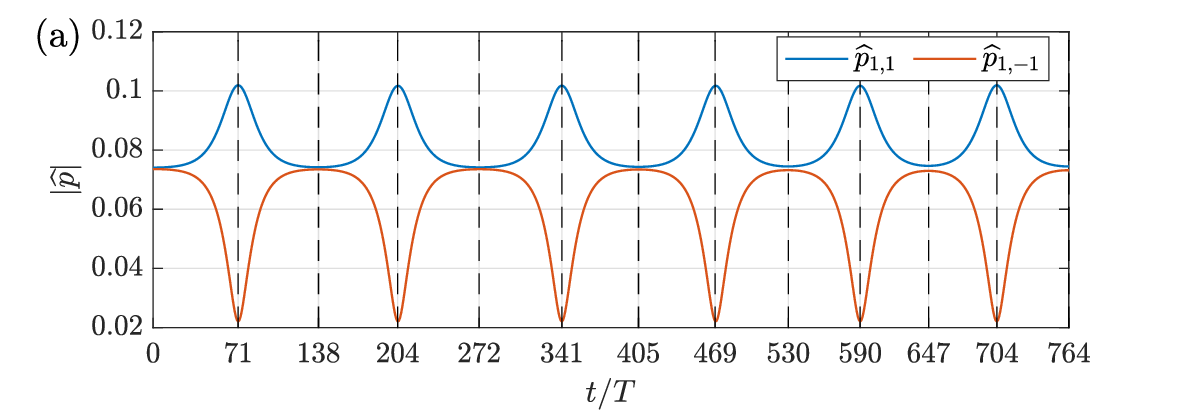}
    
    \includegraphics[width=0.32\linewidth]{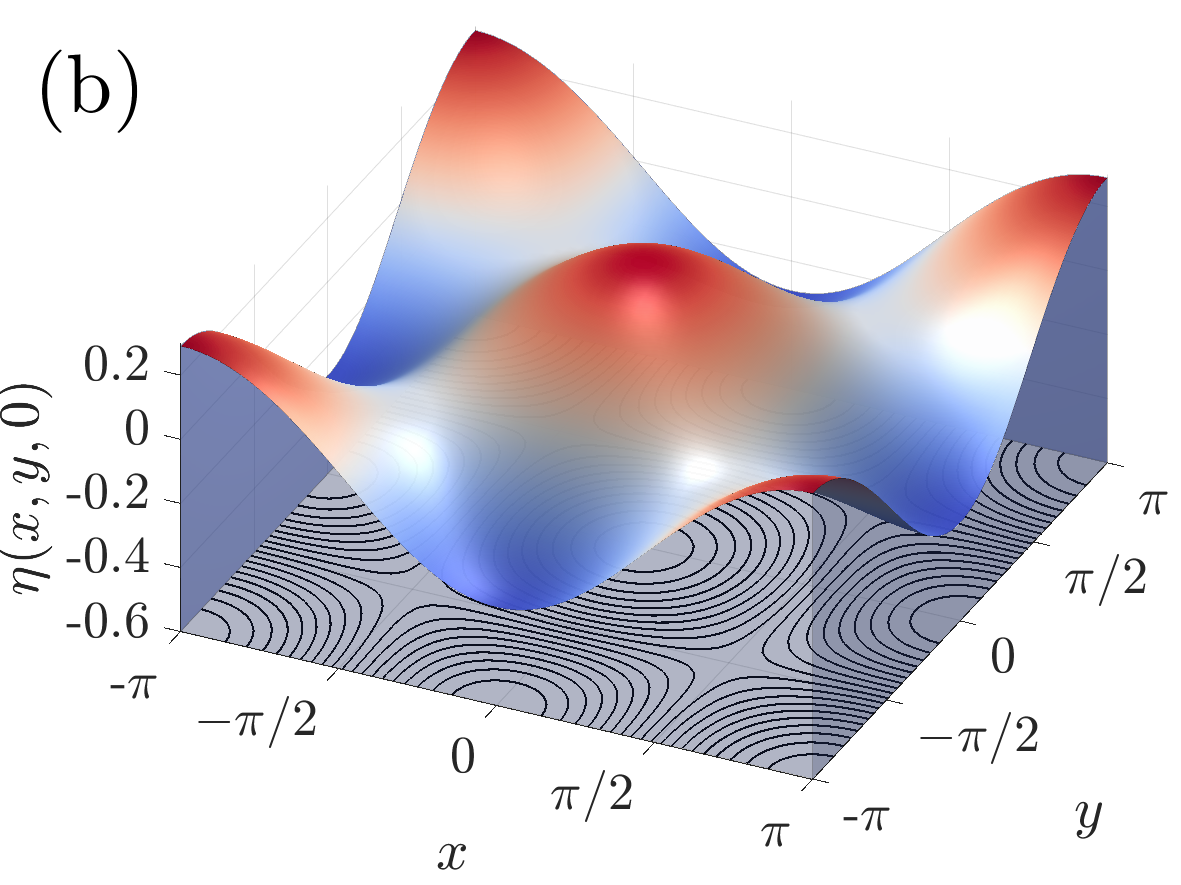}
    \includegraphics[width=0.32\linewidth]{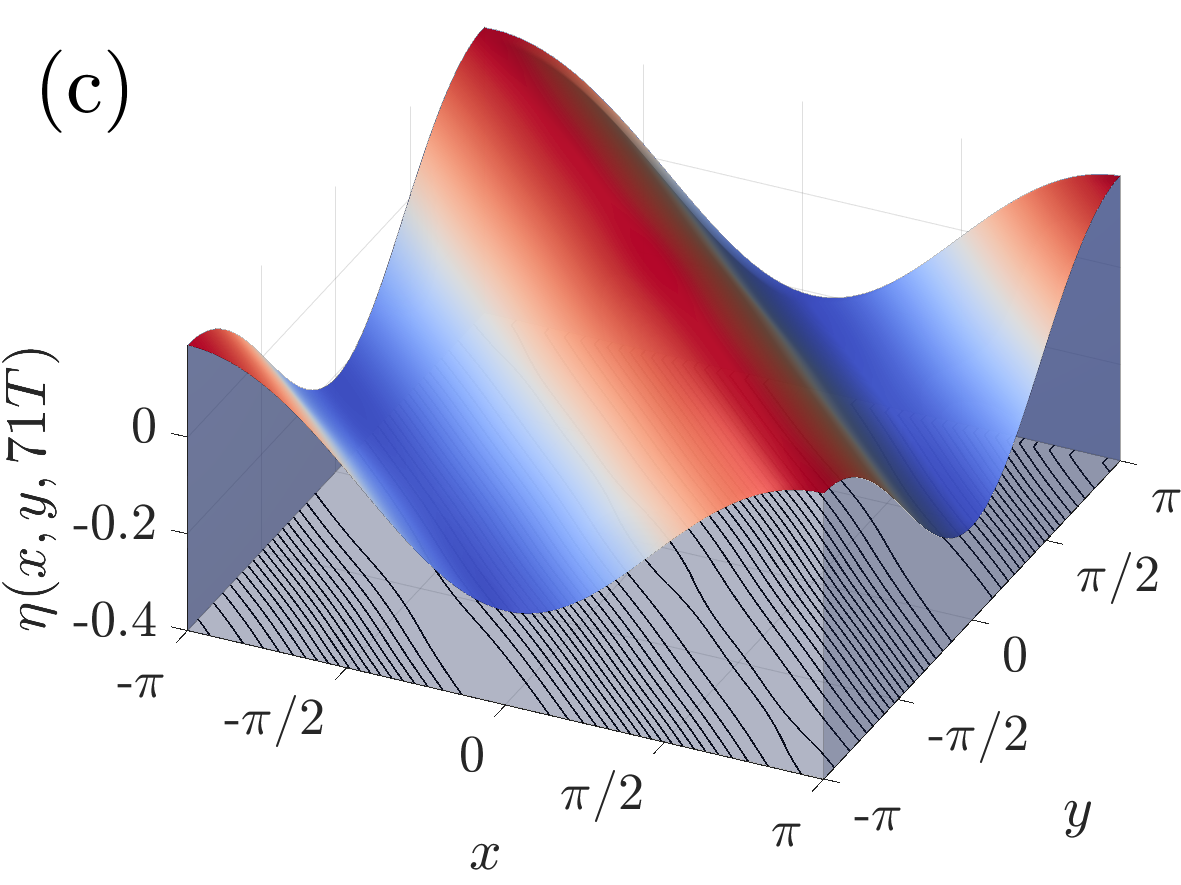}
    
    \includegraphics[width=0.32\linewidth]{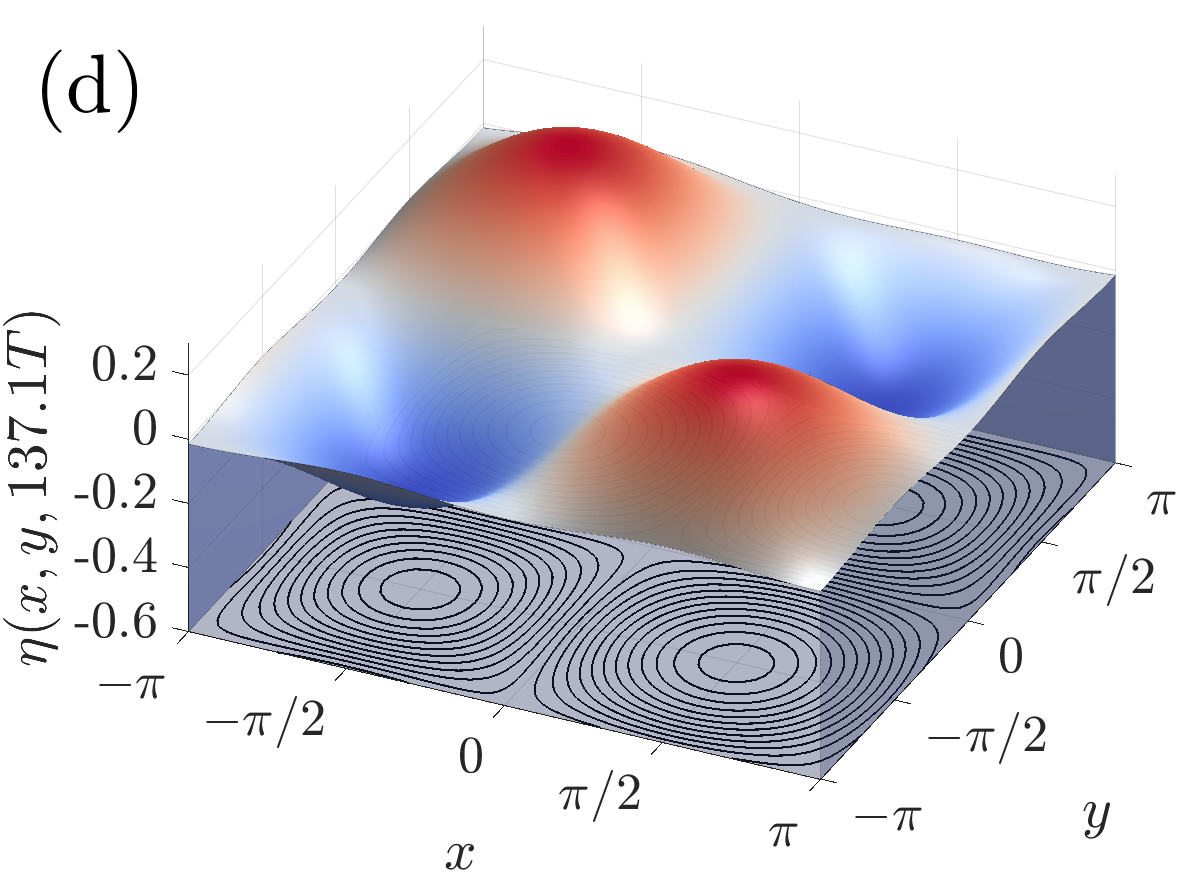}
    \includegraphics[width=0.32\linewidth]{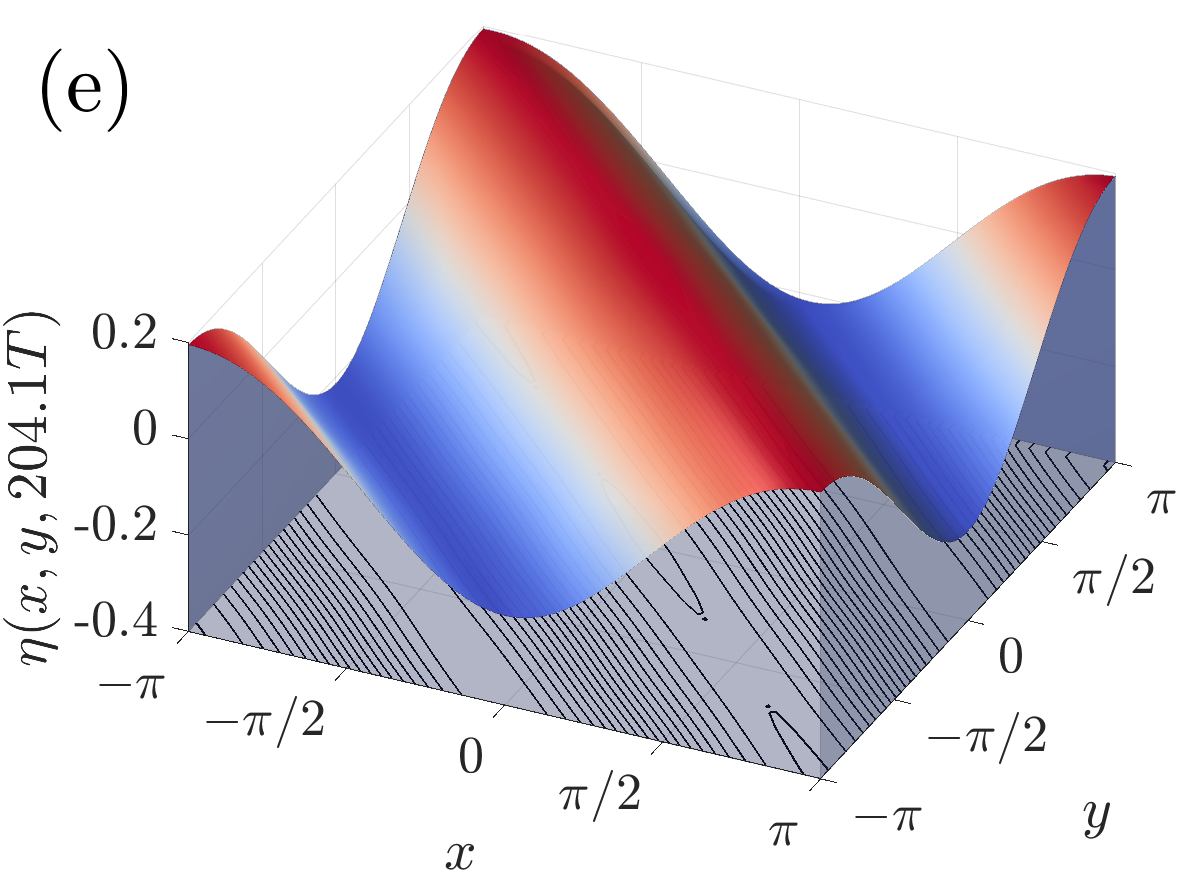}
    \includegraphics[width=0.32\linewidth]{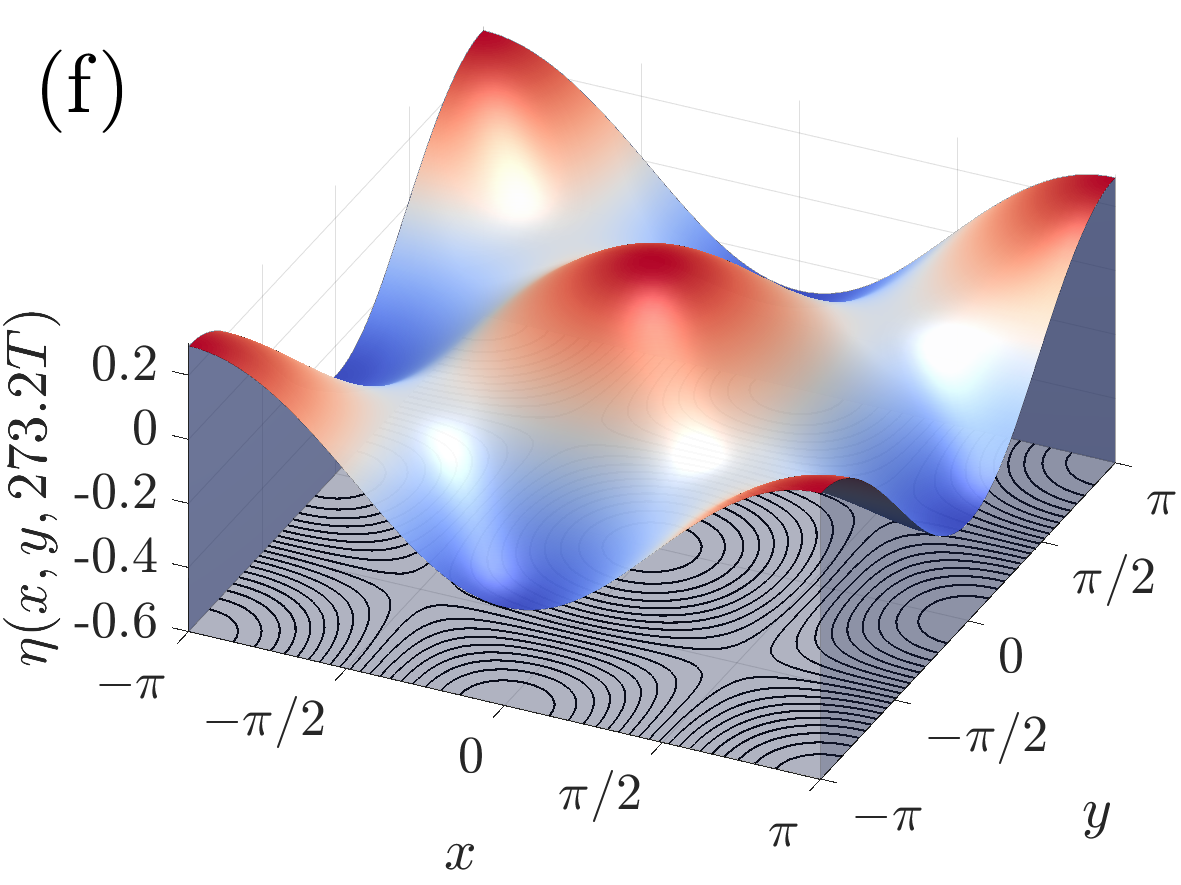}
    \caption{Temporal evolution of a non-resonant \textit{Case II} standing wave $(k=l=1,H=0.29)$ perturbed by $0.001\cos(x+y)$. (a) Time histories of $|\widehat{p}_{1,1}|$ and $|\widehat{p}_{-1,1}|$. The black dashed lines label the instants when the curves reach maximums and minimums. (b-f) Representative surface profiles at $t=0,72T,137.1T,204.1T$ and $273.2T$.}
    \label{fig:recurrence1}
\end{figure}
\begin{figure}[h!]
    \centering
    \includegraphics[width=0.7\linewidth]{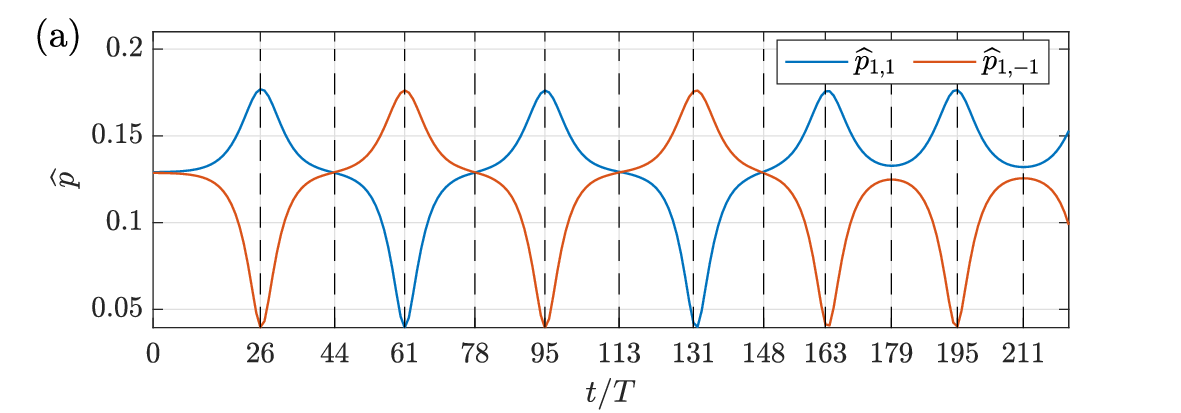}
    \includegraphics[width=0.32\linewidth]{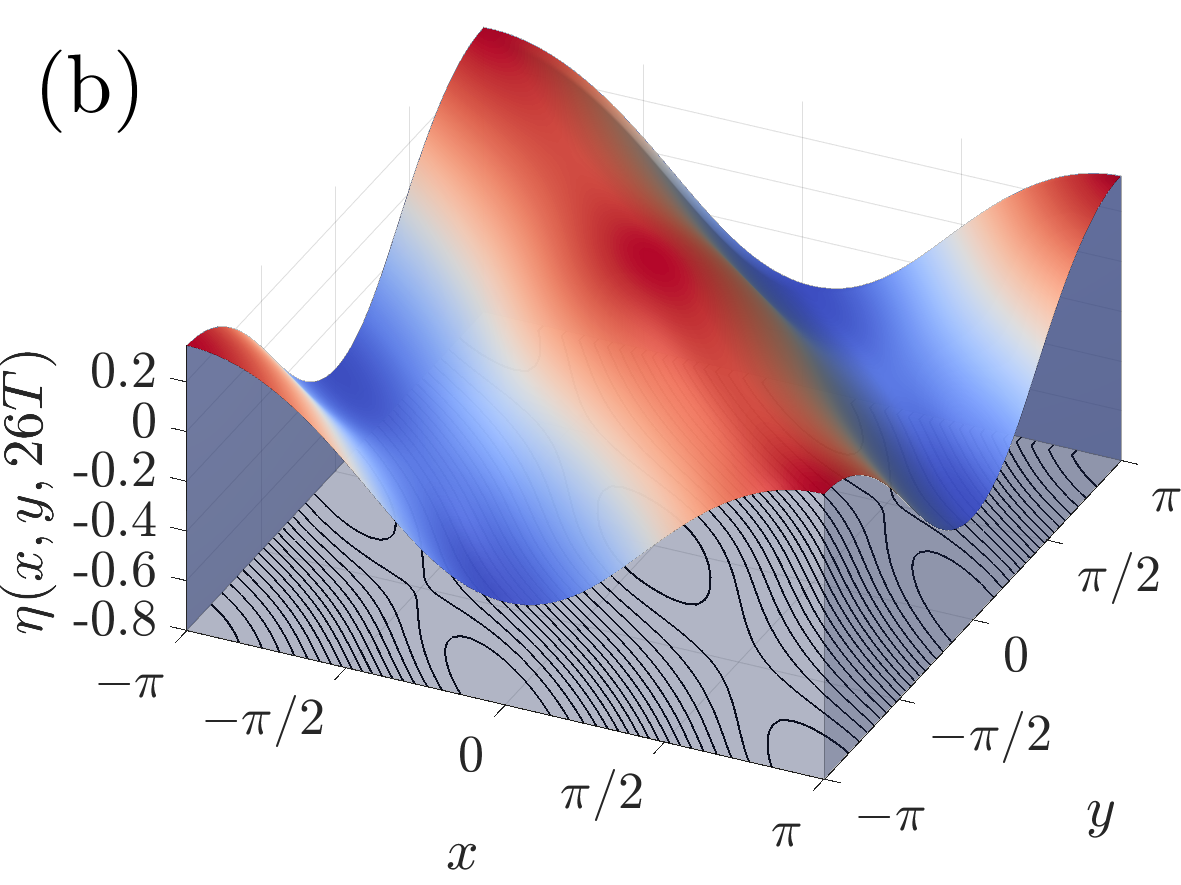}
    \includegraphics[width=0.32\linewidth]{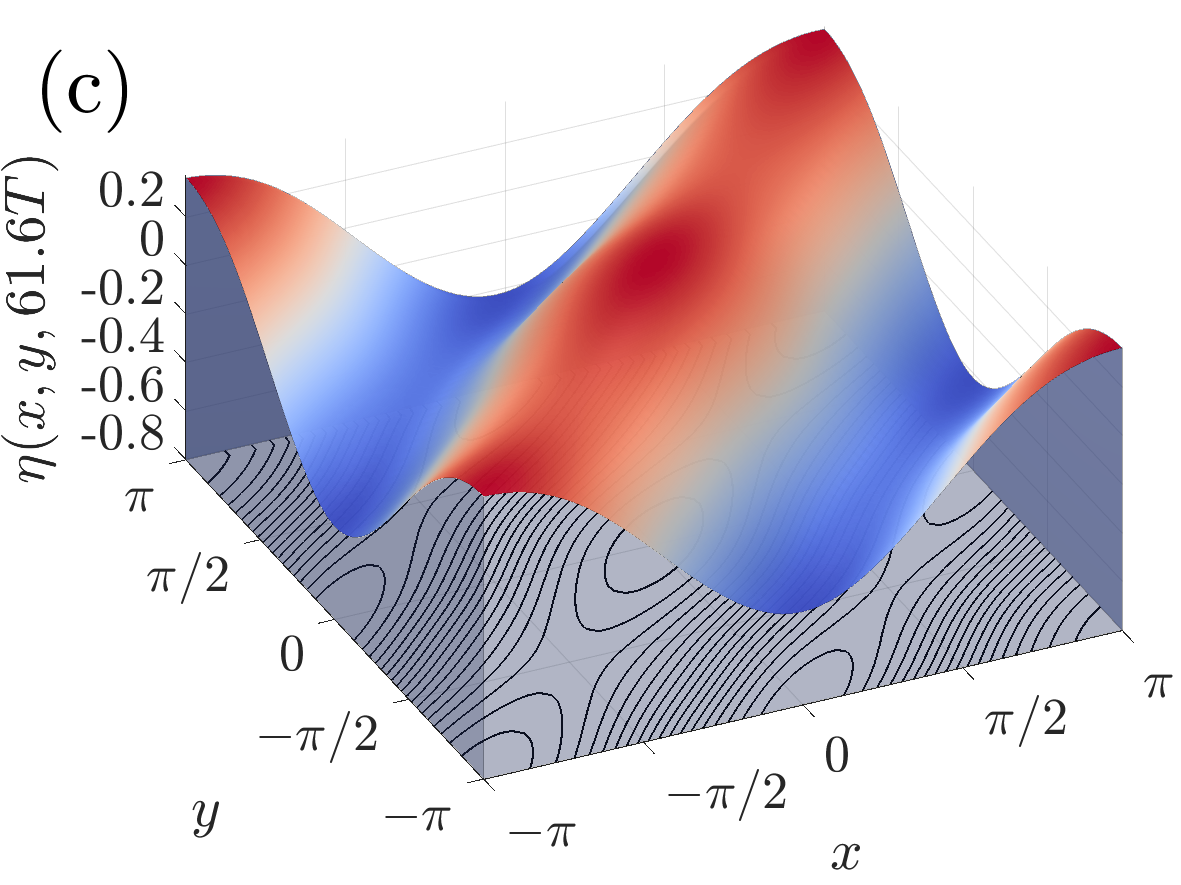}
    \caption{Temporal evolution of a non-resonant \textit{Case II} standing wave $(k=l=1,H=0.5)$ perturbed by $0.001\cos(x+y)$. (a) Time histories of $|\widehat{p}_{1,1}|$ and $|\widehat{p}_{1,-1}|$. The black dashed lines label the instants when the curves intersect and reach maximums and minimums. (b-c) Surface profiles at $t=26T$ and $61.6T$.}
    \label{fig:recurrence2}
\end{figure}
Next we examine the stability of three-dimensional standing waves. Zhu \textit{et al.} \cite{zhu2003three} have demonstrated that two-dimensional gravity standing waves, say, in $x$-direction with a wavenumber $\kappa$, are linearly unstable to harmonic disturbances $\mathrm e^{\mathrm i(\alpha x + \beta y)}$ if $\alpha^2+\beta^2\approx \kappa^2$. From a point of view of resonance, this condition implies that the disturbances have approximately the same frequency as the base standing waves, thereby supporting energy transfer between them. Over long periods, they showed that the instability leads to nearly cyclic return to the initial condition. Suppose a similar instability mechanism exists in gravity-capillary case, then the leading-order decomposition \eqref{decom1} implies that the three-dimensional standing waves are likely unstable to certain harmonic disturbances as well. We start to examine this idea by considering a non-resonant \textit{Case II} standing wave with $k=l=1, H = 0.29$, as shown in \ref{fig:recurrence1} (b). To satisfy the periodic boundary condition, we initially perturb this solution by a small disturbance $0.001\cos(x+y)$, which corresponds to two complex harmonics $0.0005\mathrm e^{\pm\mathrm i(x+y)}$. Figure \ref{fig:recurrence1} (a) shows the time histories of $|\widehat{p}_{1,1}|$ and $|\widehat{p}_{1,-1}|$ at integer temporal periods, which can be regarded as ensemble averages measuring the typical amplitudes of $\cos(x+y)$ and $\cos(x-y)$ modes. These two curves clearly demonstrates that energy is transferred between the two modes in a quasi-periodic fashion with an approximate period of $70 T$. Therefore, the base standing wave is unstable to the initial perturbation. Panels (c-f) show four representative surface profiles at $t=71T,137.1T,204.1T$ and $273.2T$. Together with the initial profile in (b), these form a near cyclic return to the initial standing wave. For base standing waves with larger amplitude and the same initial disturbance, we observe similar cyclic returns occurring on relatively shorter temporal periods. Furthermore, there could be two alternatively appearing quasi-two-dimensional modes. A temporal evolution initiated with a $\textit{Case II}$ standing waves for $k=l=1,H=0.5$ is shown in figure \ref{fig:recurrence2}. Panel (a) presents the curves of $|\widehat{p}_{1,1}|$ and $|\widehat{p}_{1,-1}|$ taken at each integer period, and panels (b) and (c) correspond to the surface profiles at $t = 26 T$ and $61.6 T$ when the quasi-two-dimensional modes appear.

\begin{figure}[h!]
    \centering
    \includegraphics[width=0.6\linewidth]{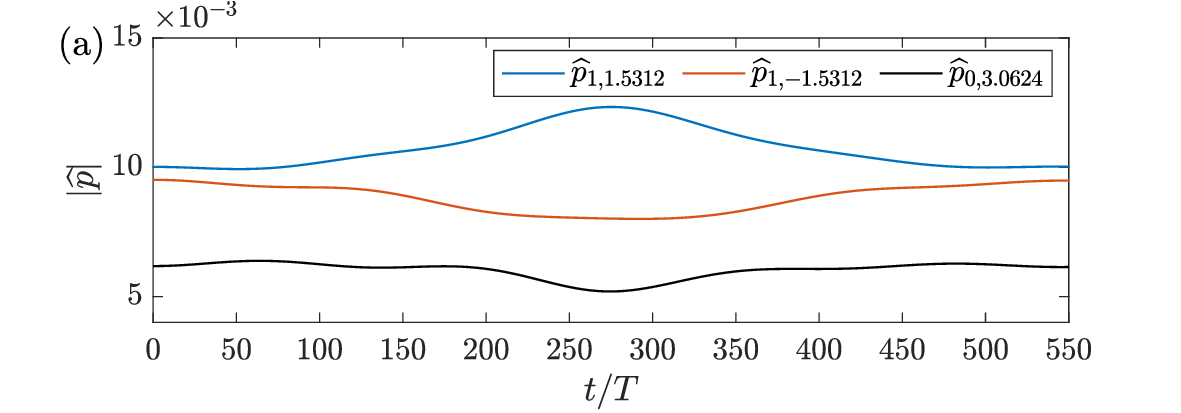}

    \includegraphics[width=0.45\linewidth]{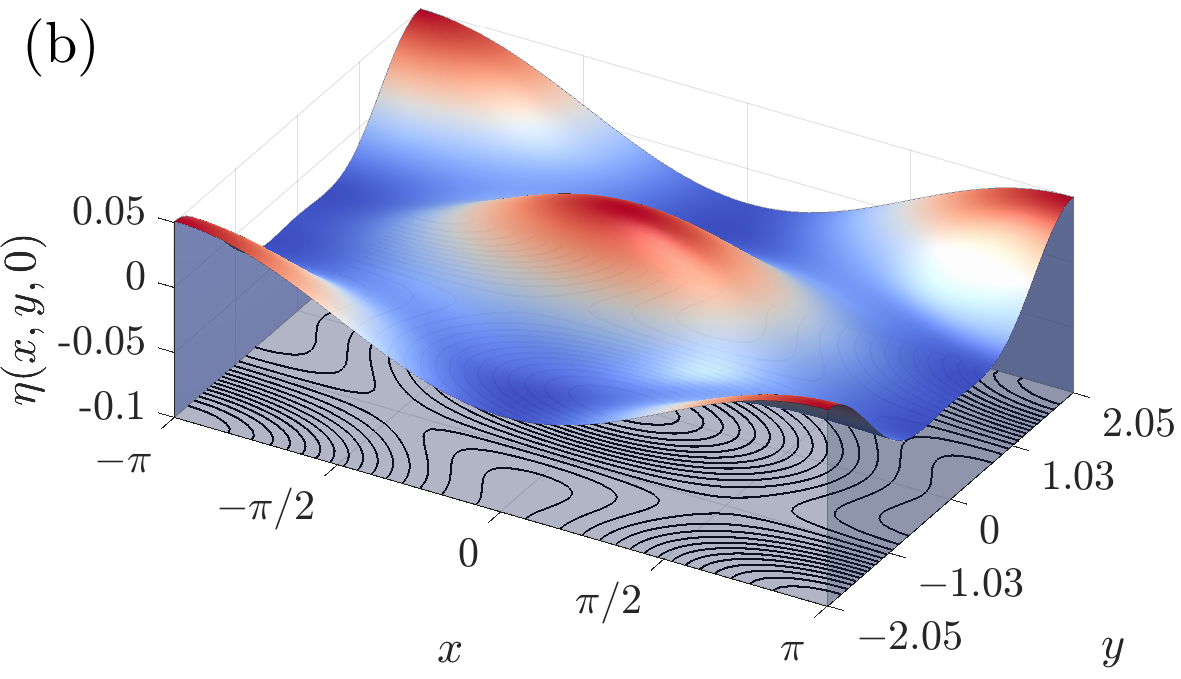}
    \includegraphics[width=0.45\linewidth]{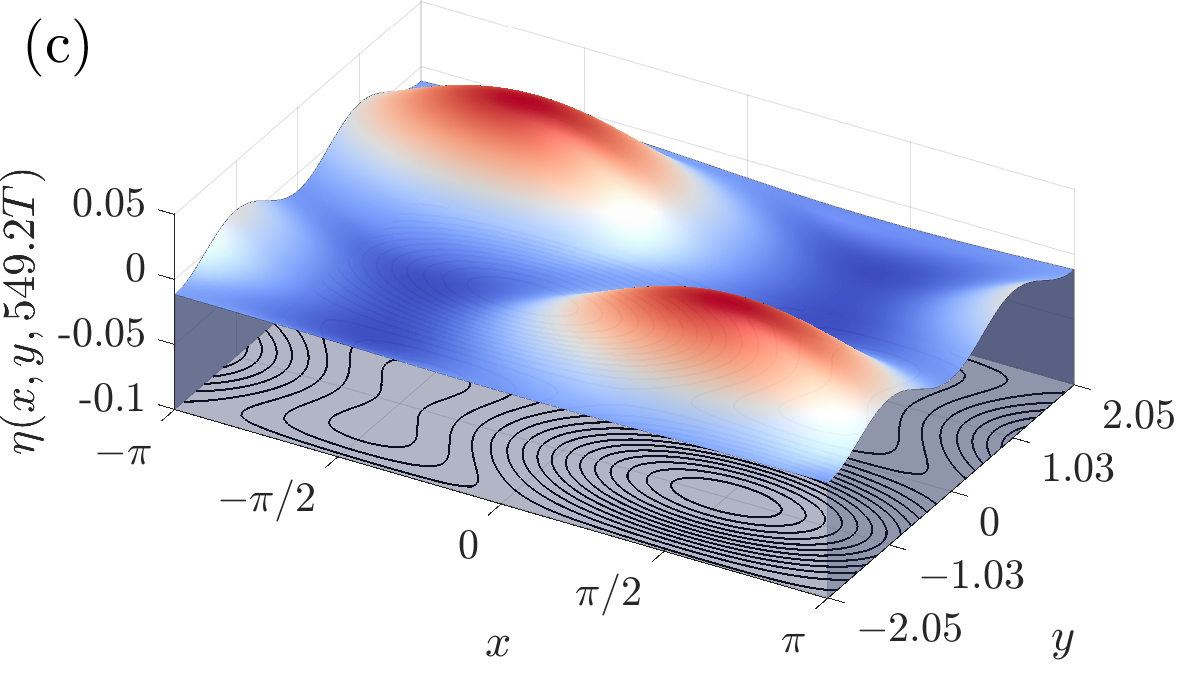}
    \caption{Temporal evolution of a branch-$1$ standing wave $(k=1,l=1.5312,H=0.05)$ perturbed by $0.001\cos(x+1.5312y)$. (a) Time histories of $|\widehat{p}_{1,1.5312}|$, $|\widehat{p}_{1,-1.5312}|$, and $|\widehat{p}_{0,3.0624}|$. (b-c) Surface profiles at $t=0$ and $453.2T$.}
    \label{fig:recurrence3}
\end{figure}
In general, consider three-dimensional standing waves dominated by $\cos(kx)\cos(ly)\cos(\omega t)+\cos(lx)\cos(ky)\cos(\omega t)$ and perturbed by $\delta \cos(kx\pm ly), \delta \ll 1$. It is found that such an instability is universal and the resulting energy-transfer rate is related to the frequencies of the base standing wave and the one used as initial perturbation. For two dimensional standing waves with wavenumber $\kappa = \sqrt{k^2+l^2}$, Concus \cite{concus1962standing} showed that the frequency $\omega$ is a monotonically decreasing function of wave amplitude unless $\kappa\in(\kappa_1,\kappa_2)$. For infinite deep water, $\kappa_1 = 0.5154$ and $\kappa_2 = \sqrt{2}/2$ (see figure 1 in \cite{concus1962standing}). 
Therefore, if $\kappa$ lie outside this specific interval and the frequency curves of the base standing waves have down-shifts close to the bifurcation point, the base standing waves are likely unstable. Otherwise, there is no overlap between the the frequency curves of the base stadning waves and the perturbations, admitting no significant energy transfer. To confirm this, we consider the branch-$1$ standing waves with $k=1,l=1.5312$. The leading Fourier components can be decomposed into
\begin{align}
    \eta\sim \frac{\epsilon}{2} \Big(\cos(x+1.5312y) + \cos(x-1.5312y)\Big)\cos(\omega t) + a\epsilon \cos(3.0624 y)\cos(2\omega t) + O(\epsilon^2).
\end{align}
We first consider the initial disturbance $0.001\cos(x+1.5312y)$ whose oscillation frequency is close to $2.8188$. As shown in figure \ref{fig:bif2}, the frequency curve of the base standing wave is slightly above this value for $H<0.12$. Figure \ref{fig:recurrence3} exhibits the temporal evolution of the standing wave with $H=0.05$ whose frequency is $\omega = 2.8224$. Panel (a) shows the time histories of $|\widehat{p}_{1,1.5312}|, |\widehat{p}_{1,-1.5312}|$, and $|\widehat{p}_{0,3.0624}|$, which are taken at integer temporal periods and represent the typical amplitudes of $\cos(x+1.5312y)$, $\cos(x-1.5312y)$, and $\cos(3.0624y)$ modes. Energy transfer among the three modes is weak and only has influences over long periods. Note that the evolution also demonstrates a nearly periodic return to the initial state (with a phase shift) with an approximate period of $550T$, as shown by the surface profiles in panels (b) and (c). 
\begin{figure}[h!]
    \centering
    \includegraphics[width=0.8\linewidth]{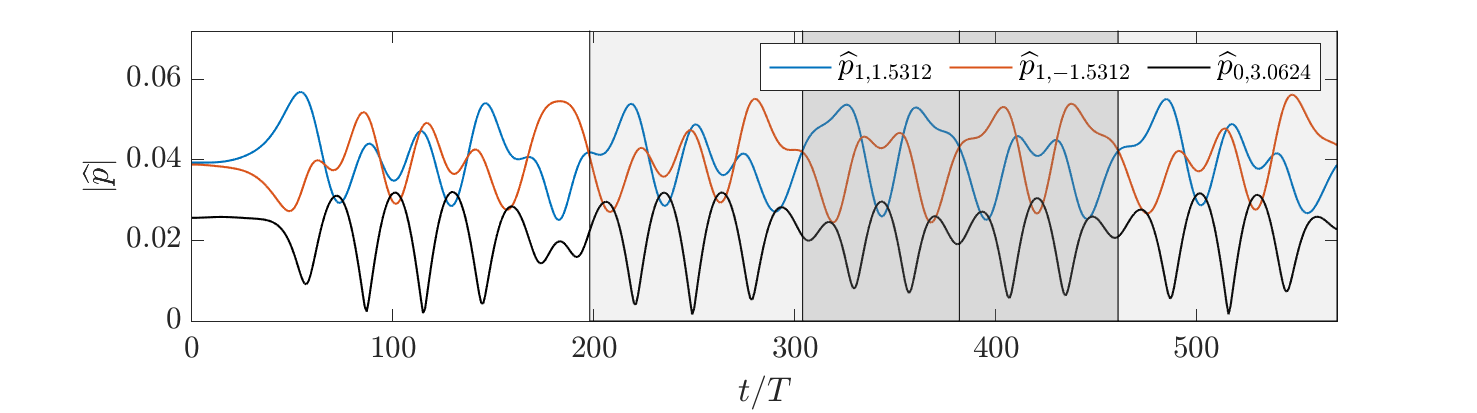}
    \caption{Time histories of $|\widehat{p}_{1,1.5312}|$, $|\widehat{p}_{1,-1.5312}|$, and $|\widehat{p}_{0,3.0624}|$ in the temporal evolution of a branch-$1$ standing wave $(k=1,l=1.5312,H=0.2)$ perturbed by $0.001\cos(x+1.5312y)$. The two light shaded regions show similar patterns, so are the two dark shaded regions.}
    \label{fig:recurrence4}
\end{figure}
In contrast, figure \ref{fig:recurrence4} shows the evolution of the perturbed standing wave on the same branch with $H=0.2$ and $\omega = 2.8051$. The energy transfer among the three modes is violent, leading to a more complex temporal evolution. Note that there still exist some fundamental, quasi-periodic patterns, as shown by the four shaded regions.

\begin{figure}[h!]
    \centering
    \includegraphics[width=0.6\linewidth]{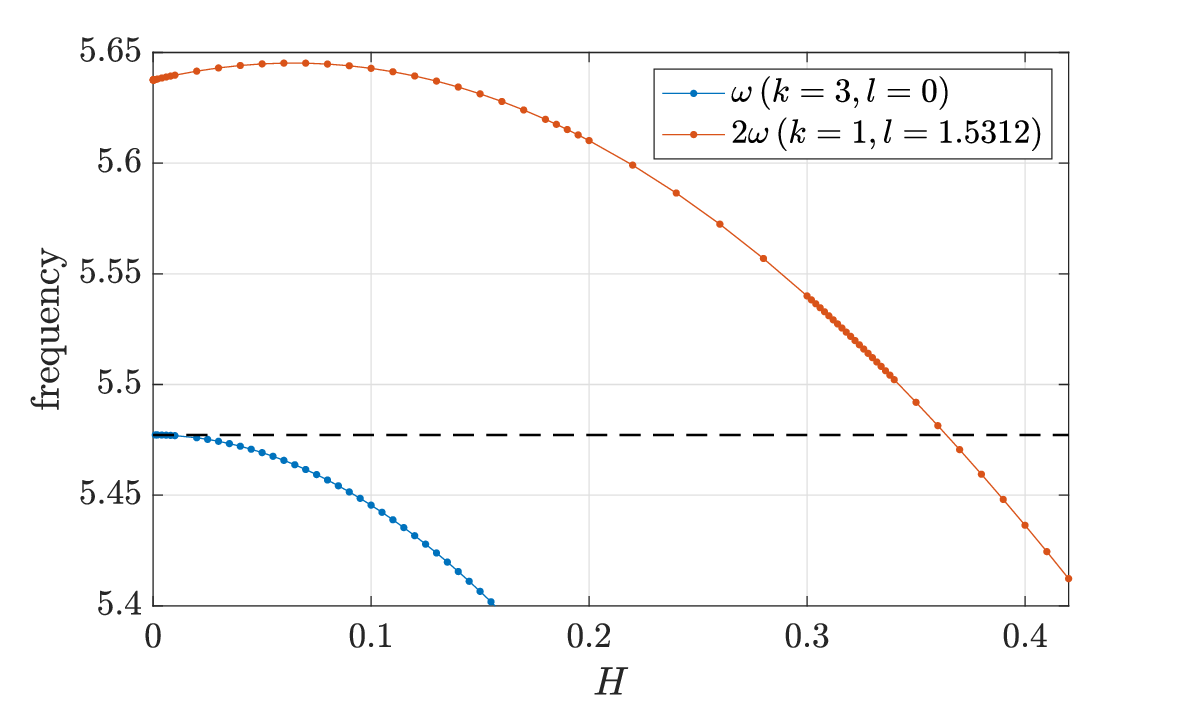}
    \caption{Frequency curves of the branch-$1$ standing waves with $k = 1, l = 1.5312$, and of the standing waves with $k = 3,l=0$.}
    \label{fig:omega_downshift}
\end{figure}
Next we perturb the standing waves on the same solution branch by imposing an initial disturbance $0.001\cos(3 x)$. Although this mode does not dominate the base standing wave, its linear frequency is close to that of $\cos(3.0624 y)\cos(2\omega t)$, one of the fundamental Fourier components. Figure \ref{fig:omega_downshift} reproduces the nonlinear frequency curves of the three-dimensional base standing waves and the two-dimensional standing waves used as the perturbation. As clearly shown, there is no overlap of the two frequency curves until $H \ge 0.364$ for the three-dimensional base standing waves, thereby suggesting that the prescribed perturbation only causes instability for certain finite-amplitude base standing waves.
\begin{figure}[h!]
    \centering
    \includegraphics[width=0.9\linewidth]{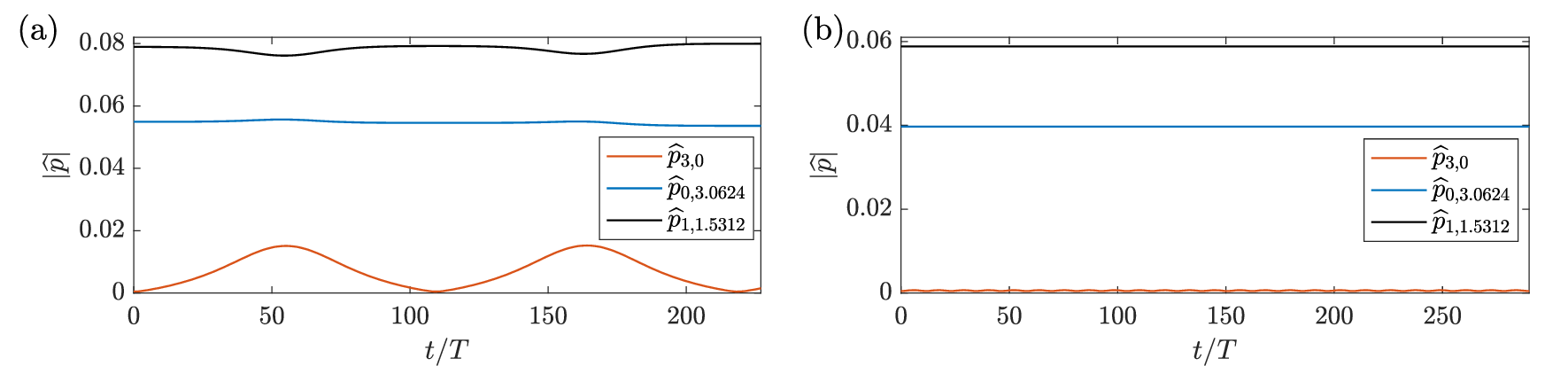}
    \caption{Time histories of $|\widehat{p}_{3,0}|,|\widehat{p}_{0,3.0624}|$ and $|\widehat{p}_{1,1.5312}|$ for different branch-$1$ standing waves with an initial disturbance $0.001\cos(3x)$. (a) $H=0.4$. (b) $H=0.3$.}
    \label{fig:energytransfer}
\end{figure}
This explains the observed energy transfer in figure \ref{fig:energytransfer} (a), where the base standing wave has wave amplitude $H = 0.4$ and frequency $\omega = 2.7182$. In contrast, panel (b) shows the evolution of the standing wave with $H = 0.3,\omega = 2.7700$ and perturbed by the same disturbance initially. As can be clearly seen, $|\widehat{p}_{3,0}|$ only slightly fluctuates around its initial state, indicating that no significant energy transfer occurs. All solutions are computed using $96\times96\times96$ grid points.

\section{Conclusions}
We have numerically investigated three-dimensional gravity-capillary standing waves in deep water. The full water-wave problem is reformulated using Zakharov's Hamiltonian approach and the Craig-Sulem expansion of the DNO, leading to the cubic- and quintic-truncated models. They provide accurate approximations to the non-local relationship between the surface velocity potential and the normal velocity, thereby avoiding directly solving Laplace's equation. We have developed a spatio-temporal collocation method to compute triply periodic (in two spatial directions and one temporal direction) solutions. This approach is based on boundary-value calculation, thereby avoiding the numerical stiffness associated with surface tension and the numerical instabilities arising in initial-value calculation. We considered the symmetric standing waves in square or rectangular basins. By using the time-reversal argument, we revealed the multiple spatio-temporal symmetries of the system. In square basins, solutions are invariant after reflections about the $x$-axis, $y$-axis, the two diagonals, and the four straight lines connecting the midpoints of adjacent edges. In rectangular basins, solutions are even functions with respect to the $x$-axis and $y$-axis. Along the temporal direction, we showed that solutions are symmetric about the quarter period, and there are additional spatial symmetries emerging at this instant. By fully exploiting these properties, we have significantly reduced the number of unknowns by a factor of 64 or 32, allowing efficient computations via Newton's method and the Fourier pseudo-spectral method.

We showed that the cubic and quintic models align closely with the full-potential formulation in both two and three dimensions. Typical non-resonant gravity-capillary standing waves exhibit round crests and troughs, in contrast to the pyramidal shape of pure gravity standing waves. When resonances occur, standing waves exhibit complicated and rapidly oscillating surface patterns, and their bifurcations could break into multiple, disjoint solution branches. In particular, we considered the resonant standing waves that support the three-wave resonance, generalising the concept of classical Wilton ripples occurring in two-dimensional gravity-capillary travelling waves. For both collinear and non-collinear resonant triads, we identified the existence of multiple solution branches bifurcating from the linear solutions, along with various resonant standing waves featuring square, hexagonal, and more complex flower-like surface patterns. We examined the temporal periodicity of the computed standing waves via initial-value calculations, validating the excellent numerical accuracy of the spatio-temporal collocation method. Using the same approach, we also studied the stability of three-dimensional standing waves, and showed that they could be unstable to certain harmonic disturbances whose frequencies are close to those of the base standing waves. Over long temporal periods, such an instability usually leads to the quasi-periodic return to the initial state with a possible phase shift, akin to those reported in \cite{bryant1995water,zhu2003three}. 

In the future, we plan to exploit other problems regarding time-periodic water waves, including the stability and long-term dynamics of standing waves, Faraday waves, and breathers. These problems require highly accurate numerical computations, hence a major challenge is to overcome the computer-memory limitation in our spatio-temporal collocation method by using some matrix-free methods. Two promising approaches are the preconditioned Newton-Krylov method which has been successfully applied to several water-waves problems \cite{pethiyagoda2014jacobian,ctugulan2022three}, and the adjoint-based variational method which has been developed to find the time-periodic solutions of the Kuramoto–Sivashinsky equation \cite{azimi2022constructing,ashtari2023jacobian}, the Navier-Stokes equations \cite{parker2022variational}, the Benjamin-Ono equation \cite{ambrose2010computationb}, and the vortex-sheet problem \cite{ambrose2010computationa}. Another challenge is to compute the time-periodic solutions of the full water-wave equations by combining the spatio-temporal collocation method and the three-dimensional boundary-integral method. This approach will allow computations in the strongly nonlinear regime where the quintic and other higher-order truncated models would fail or become numerically inefficient. Recent advances in developing fast solvers of three-dimensional Laplace's equation using the Ewald summation technique \cite{ambrose2013small,duan2000ewald,fructus2005efficient} have shed light on this direction.

\section*{Appendix 1}\label{sec:appendix1}
$G(\eta)$ has a convergent Taylor series
\begin{align}
    G(\eta) = \sum_{i=0}^{\infty} G_i(\eta),
\end{align}
provided $\eta$ smaller than a certain constant \cite{coifman1985nonlinear,craig2000traveling}. The zeroth-order term is given by 
\begin{align}
    G_0 = |D|\tanh(|D|h),
\end{align}
where $D = -\mathrm i\nabla$, $|D| = (-\Delta)^{1/2}$, and $h$ is the mean water depth. Following \cite{craig2002traveling}, the recursive expansions of $G_i(\eta)$ are given as: for $i = 2r>0$
\begin{align}
    G_{2r}(\eta) =& \frac{1}{(2r)!} G_0(|D|^2)^{r-1}D\cdot \eta^{2r}D - \sum_{s = 0}^{r-1}\frac{1}{(2(r-s))!}(|D|^2)^{r-s}\eta^{2(r-s)}G_{2s}(\eta)\nonumber\\
    &- \sum_{s = 0}^{r-1} \frac{1}{(2(r-s)-1)!}G_0(|D|^2)^{r-s-1}\eta^{2(r-s)-1}G_{2s+1}(\eta),
\end{align}
and, for $i = 2r-1>0$
\begin{align}
    G_{2r-1}(\eta) =& \frac{1}{(2r-1)!} (|D|^2)^{r-1}D\cdot \eta^{2r-1}D - \sum_{s = 0}^{r-1}\frac{1}{(2(r-s)-1)!}G_0(|D|^2)^{r-s-1}\eta^{2(r-s)-1}G_{2s}(\eta)\nonumber\\
    &- \sum_{s = 0}^{r-2} \frac{1}{(2(r-s-1))!}(|D|^2)^{r-s-1}\eta^{2(r-s-1)}G_{2s+1}(\eta).
\end{align}
By truncating $G(\eta)$ up to $G_4(\eta)$, substituting into the Hamiltonian \eqref{Hamiltonian}, and taking variational derivatives with respect to $\varphi$ and $\eta$, we obtain the quintic model
\begin{align}
    \eta_t & = \sum_{i=0}^{4} G_i(\eta)\varphi,\label{quintic1}\\
    \varphi_t &= \sum_{i=2}^{5}\mathcal N_i(\eta,\varphi) - \eta + \nabla\cdot \Bigg( \frac{\nabla\eta}{\sqrt{1+|\nabla\eta|^2}}\Bigg),\label{quintic2}
\end{align}
where
\begin{align}
    \mathcal N_2 =& \frac{1}{2}\Big((G_0\varphi)^2 - |\nabla\varphi|^2\Big),\\
    \mathcal N_3 =& \frac{1}{2}\Big(2(G_0\varphi)(G_1\varphi) + 2(G_0\varphi)(\nabla\varphi\cdot \nabla \eta)\Big),\\
    \mathcal N_4 =& \frac{1}{2}\Big(2(G_0\varphi)(G_2\varphi) + (G_1\varphi)^2 + 2(G_1\varphi)(\nabla \varphi\cdot \nabla \eta) + (\nabla \varphi\cdot \nabla \eta)^2 - |\nabla\eta|^2(G_0\varphi)^2\Big),\\
    \mathcal N_5 =& \frac{1}{2}\Big(2(G_0\varphi)(G_3\varphi) + 2(G_1\varphi)(G_2\varphi) + 2(G_2\varphi)(\nabla \varphi\cdot \nabla \eta) - 2|\nabla\eta|^2(G_0\varphi)(G_1\varphi)\nonumber\\
    &- 2|\nabla \eta|^2(G_0\varphi)(\nabla\varphi\cdot\nabla\eta)\Big).
\end{align}
Similarly, the cubic model is derived by truncating the $G(\eta)$ to $G_2(\eta)$ and taking the variational derivative. The final form can be obtained by keeping the summations in \eqref{quintic1} and \eqref{quintic2} up to $G_2$ and $\mathcal N_3$, respectively.

\section*{Appendix 2}\label{sec:appendix2}
There are different boundary-integral formulations in two-dimensional water waves, based on the choices of integral relations and parameterisations. Because we are not handling overturning waves, and for the convenience of comparisons with the cubic and quintic models, we employ a $x$-coordinate-parameterised Cauchy's integral formula \cite{guan2025time}. Using the DNO, the kinematic and dynamic boundary conditions are expressed as
\begin{align}
    \eta_t &= G(\eta)\varphi,\label{BC1}\\
    \varphi_t &= \frac{1}{2}\frac{(\varphi_x\eta_x+G(\eta)\varphi)^2}{1+\eta_x^2} -  \frac{1}{2}\varphi_x^2 -\eta+\frac{\eta_{xx}}{(1+\eta_x^2)^{3/2}}.\label{BC2}
\end{align}
Here $G(\eta)\varphi = \phi_{\boldsymbol{n}}\sqrt{1+\eta_x^2}$, and $\phi_{\boldsymbol{n}}$ is solved from the following Fredholm integral equation of the second kind
\begin{align}
    \phi_{\boldsymbol{n}}(x_0) =&- \frac{k}{2\pi}\text{PV}\int_{-\pi/k}^{\pi/k} \text{Im}\Big(\mathrm e^{\mathrm i\theta(x_0)} \cot[k(z(x)-z(x_0))/2]\Big)\phi_{\boldsymbol{n}}(x)\sqrt{1+\eta_x^2(x)}\,\mathrm dx\nonumber\\
    & -\frac{k}{2\pi}\text{PV}\int_{-\pi/k}^{\pi/k} \text{Re}\Big(\mathrm e^{\mathrm i\theta(x_0)} \cot[k(z(x)-z(x_0))/2]\Big)\varphi_x(x)\,\mathrm dx,
\end{align}
where $k$ denotes the wavenumber, PV represents Cauchy principal value, $z = x+\mathrm i \eta$, and $\theta$ is the angle between the tangent of the curve and the $x$-axis. Following a two-dimensional spatio-temporal method similar to that outlined in {\S} 3, standing-wave solutions can be calculated by solving a doubly periodic boundary-value problem.

\bibliographystyle{unsrt} 
\bibliography{references}

\end{document}